\documentclass{aastex62}

\received{September ??, 2019}
\revised{November ??, 2019}
\accepted{December ??, 2019}
\submitjournal{ApJ}

\shorttitle{Silicate emission of type 1 AGNs}
\shortauthors{Mart\'inez-Paredes et al.}

\begin{document}

\title{Modelling the strongest silicate emission features of local type 1 AGN}

\correspondingauthor{M. Mart\'inez-Paredes}
\email{mariellauriga@kasi.re.kr}

\author[0000-0002-0786-7307]{M. Mart\'inez-Paredes}
\affil{Korea Astronomy and Space Science Institute 776, Daedeokdae-ro, Yuseong-gu, Daejeon, Republic of Korea (34055)}

\author{O. Gonz\'alez-Mart\'in}
\affil{Instituto de Radioastronom\'ia y Astrof\'isica UNAM 
Apartado Postal 3-72 (Xangari), 58089 Morelia, Michoac\'an, Mexico}  

\author{D. Esparza-Arredondo}
\affil{Instituto de Radioastronom\'ia y Astrof\'isica UNAM 
Apartado Postal 3-72 (Xangari), 58089 Morelia, Michoac\'an, Mexico}  


\author{M. Kim}
\affil{Korea Astronomy and Space Science Institute 776, Daedeokdae-ro, Yuseong-gu, Daejeon, Republic of Korea (34055)}

\author{A. Alonso-Herrero}
\affiliation{Centro de Astrobiolog\'ia, CSIC-INTA, ESAC Campus, E-28692 Villanueva de la Cañada, Madrid, Spain}

\author{Y. Krongold}
\affil{Instituto de Astronom\'{\i}a UNAM, M\'exico, CDMX., C.P. 04510}  

\author{T. Hoang}
\affil{Korea Astronomy and Space Science Institute 776, Daedeokdae-ro, Yuseong-gu, Daejeon, Republic of Korea (34055)}

\author{C. Ramos Almeida}
\affiliation{Instituto de Astrof\'isica de Canarias (IAC),E-38205 La Laguna, Tenerife, Spain}
\affiliation{Departamento de Astrof\'isica, Universidad de La Laguna (ULL), E-38206 La Laguna, Tenerife, Spain}

\author{I. Aretxaga}
\affiliation{Instituto Nacional de Astrof\'isica, \'Optica y Electr\'onica (INAOE), Luis Enrrique Erro 1, Sta. Ma. Tonantzintla, Puebla, Mexico}

\author{D. Dultzin}
\affil{Instituto de Astronom\'{\i}a UNAM, M\'exico, CDMX., C.P. 04510}  

\author{J. Hodgson}
\affil{Korea Astronomy and Space Science Institute 776, Daedeokdae-ro, Yuseong-gu, Daejeon, Republic of Korea (34055)}



\begin{abstract}

We measure 
the 10 and $18\mu$m silicate features in a sample of 67 local ($z<0.1$) type 1 active galactic nuclei (AGN) with available {\it Spitzer} spectra 
dominated by non-stellar processes.
We find that the 
$10\mu$m silicate feature peaks at $10.3^{+0.7}_{-0.9}\mu$m with a strength (Si$_{p}$ = ln f$_{p}$(spectrum)/f$_{p}$(continuum)) of $0.11^{+0.15}_{-0.36}$, while the  
$18\mu$m one
peaks at $17.3^{+0.4}_{-0.7}\mu$m with a 
strength of $0.14^{+0.06}_{-0.06}$. We select from this sample sources with the strongest 10$\mu$m silicate strength ($\sigma_{Si_{10\mu m}}>0.28$, 10 objects). We carry out a detailed modeling of the IRS/{\it Spitzer}  
spectra
by comparing several models that assume different geometries and dust composition: a smooth torus model,  
two clumpy torus models, 
a two-phase medium torus model, 
and a disk+outflow 
clumpy model. We find that the silicate features are well
modeled by the clumpy model of Nenkova et al. 2008, and among all models
those including outflows and complex dust composition 
are the best (Hoenig et al. 2017). 
We note that even in 
AGN-dominated galaxies
it is usually necessary to add stellar contributions to reproduce
the emission at the shortest wavelengths. 
\end{abstract}



\section{Introduction} 

A dusty torus surrounds the central 
engine  of active galactic nuclei (AGN) on 
a scale of 
few pc \citep[e.g.,][]{Krolik_Begelman88, Antonucci93, Robson95, Peterson97}. It shines at infrared (IR) 
wavelengths between 1 to 1000 $\mu$m 
peaking at around 20 $\mu$m \citep[e.g,][]{Sanders89, Elvis94}. 
This emission is the result of the IR re-radiation of optical-UV light generated around the central black hole (BH) that has been absorbed by the dusty torus
\citep[][]{Neugebauer79}.
The main observational 
components that describe it
are the slope of the spectral energy 
distribution (SED) between 1-8 $\mu$m, and the 
strength of the silicate features around 10$
\mu$m and 18$\mu$m, that are produced within the warm dust of the torus that the AGN directly illuminates. 
The silicate features have been 
observed mostly in emission with the mid-infrared (MIR) infrared spectrometer (IRS) {\it Spitzer} in 
type 1 AGN \citep[][]{Siebenmorgen05,Hao05, Hatziminaoglou15}. In these AGN the 10$\mu$m silicate feature is broader and peaks at much 
longer wavelengths ($\sim10.0-11.5$ $\mu$m) than 
the ``normal'' silicate emission feature of the 
Galactic inter-stellar medium (ISM) 
\citep[e.g,][]{Hao05, Siebenmorgen05, 
Sturm05,Li08}. This suggests different silicate 
compositions, such as a different proportion and/or size of grains
\citep[e.g.,][]{Shi06, Li08}. 

\citet{Martinez-Paredes17} used the starburst-subtracted IRS/{\it Spitzer} spectra of 20 quasi stellar objects (QSOs) between $\sim7.5$ 
to 15 $\mu$m plus the unresolved near-infrared (NIR) emission to 
constrain the parameters of the dusty torus using 
the CLUMPY models of \citet{Nenkova08a, Nenkova08b}. They 
noted that in most cases the spectral range around 8$\mu$m is underestimated by the {\sc CLUMPY} models and 
that
trying to fit this part of the spectrum resulted 
in a bad fit of the silicate feature at 10$\mu$m. Furthermore, in most cases the 
peak of the 10$\mu$m silicate feature was 
shifted 
from the model location.

Considering that the shape and peak of the 
silicate features are strongly correlated with 
the properties of the dust \citep[e.g.,][]{Draine07, Sirocky08}, in this work we use five of the most popular torus models to investigate how well they reproduce the strongest silicate emission features observed in type 1 AGN. 
Recently, \citet{Gonzalez-Martin19a} compared these models and found that they can be distinguished according to the continuum slopes and silicate strengths, and in a second work \citep[][]{Gonzalez-Martin19b} they used  MIR spectroscopy data of type 1 and type 2 AGNs in the Swift/BAT survey to investigate how these models reproduce their spectral continuum.
We focus our study on 
type 1 AGN with strong silicate emission features because: 1) for type 2 AGN, the 
inclusion of a fraction ($\sim50\%$) of 
silicate grains, either based on \citet{Draine03} or \citet{Ossenkopf92} are 
enough to reproduce the observed silicate 
features in absorption \citep[see e.g.,][]{RamosAlmeida11, Alonso-Herrero11, Martinez-Paredes15}; and 2) the silicate features are well above the continuum in the low resolution {\it Spitzer} spectra, which 
support the interpretation
that the silicate features are due to AGN activity.

In this paper we aim at exploring how accurately smooth, clumpy, two-phase, and outflow torus models reproduce the IRS/{\it Spitzer} spectra of these objects, and analyzing the results in terms of 
physical differences
like the geometry and dust composition assumed. The first model that we consider is the smooth torus model of \cite[][Fritz06 hereafter]{Fritz06}, 
which assumes a continuous 
distribution of dust composed of graphite and silicate grains in almost equal percentages. The 
second model is the CLUMPY torus model of \citep[][Nenkova08]{Nenkova08a,Nenkova08b}, which assumes a standard ISM dust composition of $53\%$ silicates and $47\%$ graphite with sizes between 0.025 and 0.25 $\mu$m. The third torus model is clumpy \citep[][Hoenig10]{Hoenig10} and assumes three dust components that 
include the standard ISM composition, the standard ISM composition with larger 
grains (0.1 - 1.0 $\mu$m), and a larger fraction of graphite 
grains ($30\%$ silicates and $70\%$ graphite, with sizes between 0.05 and 0.25 $\mu$m). The fourth model \citep[][Hoenig17]{Hoenig17} 
has two components: a dusty clumpy disk that 
takes into account the emission from the 
hotter dust close to the central engine; and a 
hollow cone formed by a wind of clumpy dusty clouds elongated 
toward the polar direction. This model assumes that the wind is originated close to the sublimation zone of the dusty disk. Because of that, the dust in the wind has similar properties to the dust in the sublimation zone.

The paper is organized as 
follows: in Section 2 we present our sample 
selection and data; in Section 3 we describe 
the models; in Section 4 we present the analysis; and 
in Section 5 the discussion. Our summary and  
conclusions are presented in Section 6. We 
adopt the following cosmology: H$_{0}$ = 73 km 
s$^{-1}$ Mpc$^{-1}$, $\Omega_{M}=0.30$ and $\Omega_{\Lambda}=0.70$.

\begin{figure}
\centering
\includegraphics[scale=0.39]{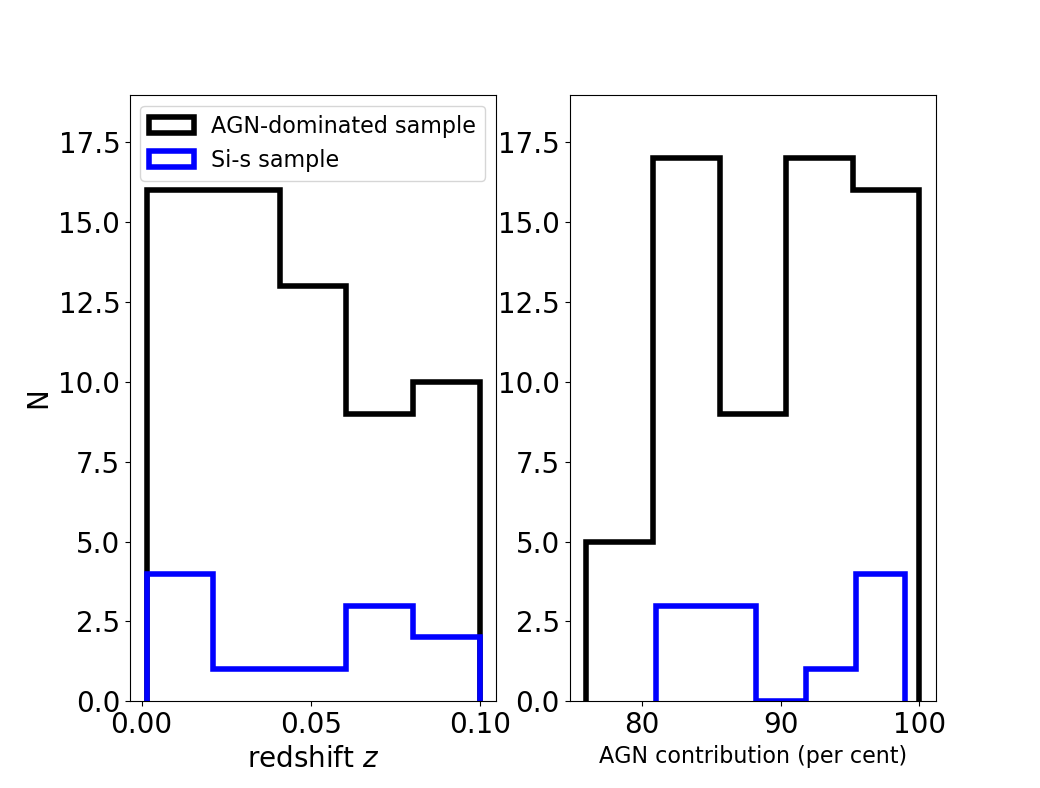} 
\caption{Properties of the sample.
{\bf Left:} distribution of redshifts. 
{\bf Right:} distribution of 
AGN contributions to the IRS/{\it Spitzer} spectra. The black line 
represents the full sample,
and the blue line 
is a subsample defined
in section 2.1.}

\label{sample}
\end{figure}

\section{The sample and data}

\subsection{The silicate dominated local AGN}
We use the latest version of the AGN catalog of \cite{Veron_Veron10} and 
the sample of low luminosity type 1 AGN in \citet{Mason12} to select AGN that have
redshifts $z<0.1$ and available 2D low-resolution 
IRS/{\rm Spitzer} spectra in the CASSIS database \citep[v6.,][]{Lebouteiller11}. 
We select those objects 
with spectra extracted as point-sources that cover the 
$5-35\mu$m spectral range. The first criterion allows us to probe the MIR emission from the central region of the AGN, and the second 
one allows us to study both the 10 and 18$\mu$m silicate features. 
As an exception, we include NGC~3998, a strong silicate emitter  \citep[e.g.,][]{Mason12}, which has a low-resolution spectrum in the $\sim$7.5-14.5$\mu$m range and a high-resolution spectrum in the $\sim$14-35$\mu$m range. 
In order to ensure that the emission in the IRS/{\it Spitzer} spectra is mostly dominated by dust heated 
by the AGN, we decompose them 
 into their stellar, 
ISM and AGN emission (dust heated by AGN) 
components at $\sim5-15\,\mu$m, using the spectral decomposition tool {\sc DeblendIRS}  \citep[][]{Hernan-Caballero15}. 
For more details on the spectral 
decomposition  
we refer the reader to Appendix A.  
We select only those objects (67) for which the spectral 
contribution due to AGN is $>80\%$ (see Figure~\ref{sample}). Hereafter, we refer to the spectra of these objects as the AGN-dominated IRS/{\it Spitzer} spectra. Note that we refer here to the original spectra, not to the component obtained from the decomposition. 

\subsection{The IRS/{\it Spitzer} spectra}
We obtained the reduced low resolution 
($R\sim60-127$) \emph{IRS/Spitzer} spectra from the 
CASSIS database \citep[v6.,][]{Lebouteiller11}. The 
spectra include the SL1 ($\lambda\sim7.4-14.5\,\mu$m) 
and SL2 ($\lambda\sim5.2-7.7\,\mu$m) modules with a 
slit-width of 3.6 arcsec, and the LL1 ($\lambda\sim19.9-39.9\mu$m) and LL2 ($\lambda\sim13.9-21.3\,\mu$m) modules with a slit-width 
of 10.5 arcsec \citep[][]{Werner04, Houck04}. We use 
the final stitched spectra between $5-35$ $\mu$m. The 
different module spectra were stitched by scaling the 
LL and SL1 flux modules to the shortest module SL2 
flux. 

\subsection{Measuring the silicate emission features}
\label{measuring_SS}
To measure the silicate emission features we start by interpolating a local continuum between both sides of the emission line complexes. 
We inspect each spectrum visually 
and choose three bands in which to measure the continuum. Band 1 is located at the blue extreme of the 10$\mu$m silicate feature, 
while band 3 is located at the red extreme of 
the 18$\mu$m silicate feature, and band 2 is located 
between the two silicates features. 
Band 1, 2 , and 3 are located within the $7.8-8.5$, $13.5-14.0$ and $20.0-21.0\,\mu$m ranges.

\begin{figure}
\centering
\includegraphics[scale=0.45]{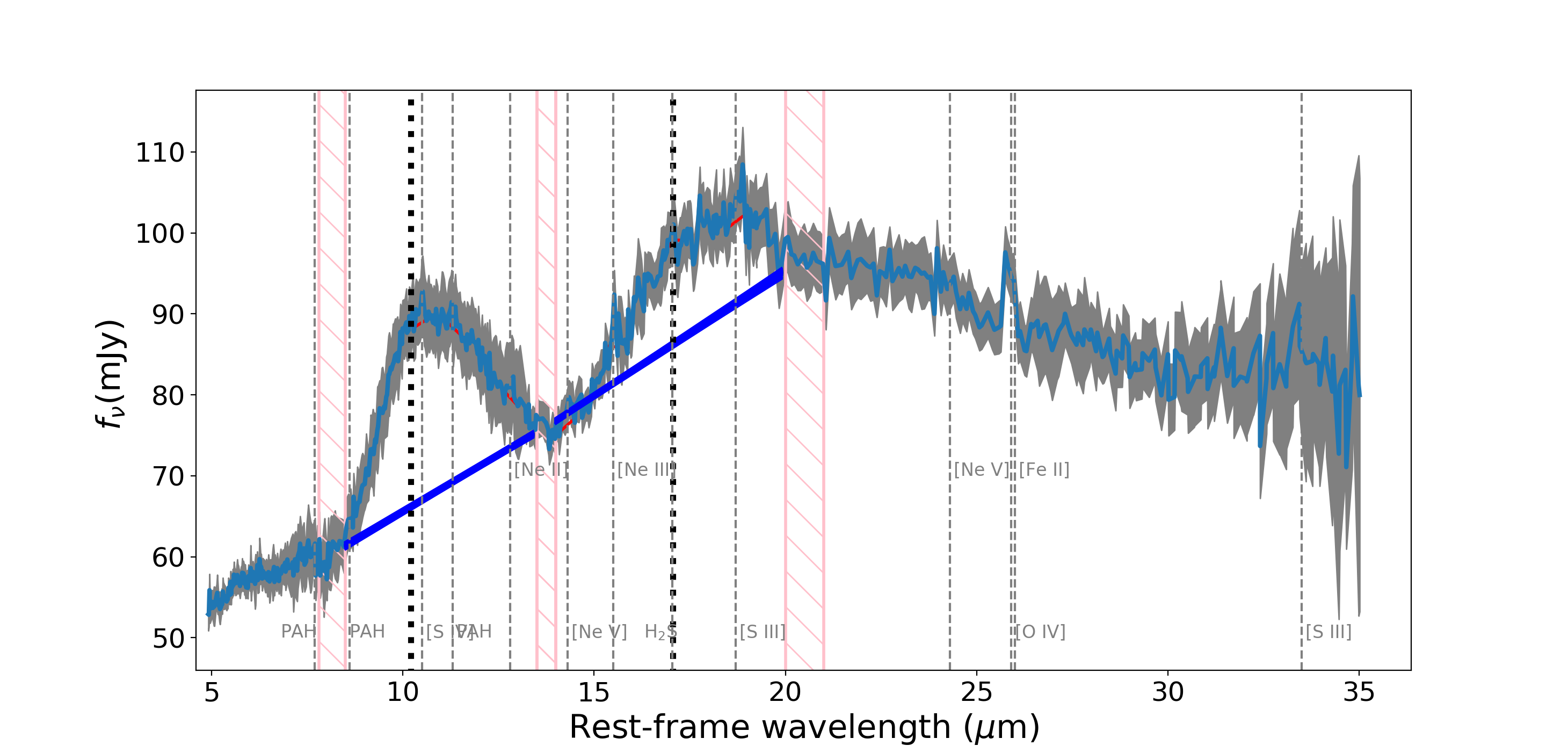}
\caption{IRS/{\it Spitzer} spectrum of PG~2214+139 (light blue solid line). The red line is the local continuum that follows the broad features of the IRS/{\it Spitzer} spectrum. 
The blue solid lines are the bootstrapped local continua and the vertical pink dashed bars are the bands used to fit the continua around the features. The vertical black dashed-lines indicate the wavelength where the silicate strength is measured. The vertical grey dashed-lines mark other  emission lines.\label{fig:silicates_examp_irs}}
\end{figure}

In order to better define uncertainties in the continuum definition, which might be affected by the presence of Polycyclic Aromatic Hydrocarbon (PAH) molecular emission lines around 7$\mu$m, and in some cases by the artificial ``teardrop''\footnote{An excess of emission present in the 2D SL1 spectrum of some objects.} feature around 14$\mu$m, we trace
fiducial mean values of each continuum band by 
bootstrapping on the measured fluxes (hatched pink 
regions in Figure~\ref{fig:silicates_examp_irs}). We 
generate 100 continuum values, considering the 
uncertainties, between band 1 and 2, and between band 
2 and 3. We randomly associate shorter and longer 
wavelength mean continuum values to generate linear 
continua below the silicate features. We note that fitting a spline continuum gives similar results. The dark blue solid lines in Figure~\ref{fig:silicates_examp_irs} are bootstrapped local continua derived from the continuum bands and they define the regions where we measure the strength of the silicate features. 

Considering that the peak of 
the silicate features vary from object to object,
we choose to measure the strength of the features at the wavelengths where they peak. However, for simplicity we still call them the 10 and 18$\mu$m silicate features.
The silicate strength is, hence, defined as the silicate peak relative to the continuum, at the wavelength where the silicate feature peaks \citep[see e.g.,][]{Hao07}, 
according to the equation:
\begin{equation}
Si_{p}=\ln\frac{f_{p}\,({\rm spectrum})}{f_{p}\,({\rm continuum})}.
\end{equation}
In Table~\ref{tab:silicates_IRS} we list the mean and the 68 percent intervals for the objects with the largest silicate feature strengths. In Appendix B (Table~\ref{tab:silicates_IRS_all}) we list the silicate emission strengths for the full sample.
On average, we find that type 1 AGN have a 10$\mu$m silicate strength $Si_{10.3\mu\text{m}}=0.13^{+0.15}_{-0.37}$ that peaks at $10.3^{+0.7}_{-0.9}\mu$m, and a 18$\mu$m silicate strength $Si_{17.3\mu\text{m}}=0.14^{+0.06}_{-0.06}$ that peaks at 17.3$\mu$m.

\begin{table*}
\caption{{\bf Basic properties of objects in the Si-s sample and their silicate feature strengths measured from the IRS/{\it Spitzer} spectra}. Columns 1 and 2 list the name and type of activity,  column 3 gives the redshift, and columns 4 and 5 list the logarithm of the intrinsic hard X-ray luminosity and the corresponding reference. 
Columns 6 and 7 list the wavelength where the $10\mu$m silicate emission feature peaks, and the $10\mu$m silicate strength, and columns 8 and 9 are are the wavelength and strength of 
the $18\mu$m silicate emission feature. 
\label{tab:silicates_IRS}}
\begin{center}
\resizebox{18cm}{!}{
\begin{tabular}{l|cccccccc}
				\hline
Name & Activity$^{a}$&  z &$log L_{X (2-10 {\rm keV})}$ & Ref.& $\lambda_{p}$& Si$_{10\mu\text{m}}$ &$\lambda_{p}$ & Si$_{18\mu\text{m}}$\\
     &               &    & erg s$^{-1}$  & &($\mu$m)&   & ($\mu$m) & \\
\hline
NGC7213 &LINER  & 0.0058 & 42.2 & 6&  $10.7\pm0.1$ & $0.52\pm0.05$&    $17.3\pm0.1$ & $0.23\pm0.04$ \\
PG2304+042 &Sy1.2 & 0.0420 & 43.4$^{*}$ & 9&  $10.3\pm0.1$ & $0.30\pm0.06$ & $17.3\pm0.1$ & $0.33\pm0.08$ \\
PKS0518-45 &LINER & 0.0420 & 44.9& 8&  $10.4\pm0.2$ &$0.31\pm0.05$&$17.6\pm0.1$ & $0.18\pm0.04$ \\
PG0844+349 & Sy1/QSO & 0.0640 &43.7 & 1&  $10.3\pm0.1$ & $0.33\pm0.06$ & $16.7\pm0.1$ & $0.18\pm0.04$ \\
PG1351+640 & Sy1.5/QSO & 0.0882 & 43.1 & 6& $9.8\pm0.1$ & $0.52\pm0.03$ & $18.4\pm0.1$ & $0.14\pm0.03$ \\
PG2214+139 &Sy1.0/QSO & 0.0658 & 43.8 & 1& $10.2\pm0.1$ & $0.29\pm0.04$ & $17.8\pm0.2$ & $0.14\pm0.04$ \\
PG0804+761 &Sy1/QSO & 0.1000 & 44.5 & 1&  $9.9\pm0.1$ & $0.36\pm0.03$ & $17.3\pm0.1$ & $0.11\pm0.04$ \\
OQ208 &Sy1.5 & 0.0766 & 40.8 & 3 &  $10.2\pm0.1$ & $0.34\pm0.05$ & $16.5\pm0.1$     & $0.12\pm0.04$ \\
NGC4258 & LINER$^{b}$ & 0.0015 & 40.9 & 5&  $11.0\pm0.1$ & $0.29\pm0.05$ & $17.3\pm0.1$  &   $0.17\pm0.04$  \\
NGC3998 & LINER & 0.0035 & 41.2 & 4&  $10.7\pm0.1$ & $0.36\pm0.01$ & $16.3\pm0.1^{*}$ & $0.23\pm0.02$ \\ 
\hline
\hline
\end{tabular}}\\
Note.-$^{*}$This value needs to be used carefully, since we are using the high angular resolution spectrum to fix the band 3 and measure the $18\mu$m silicate feature strength. References:$^{a}$NED and \citet{Veron_Veron10}, $^{b}$\citet{Mason12}. $^{*}$Estimated from (15-150)+keV X-ray luminosity assuming a spectral power law with an index $\alpha=1.8$. References for hard X-ray luminosity: 1: \citet{Zhou2010}; 2: \citet{Sambruna11}; 3: \citet{Ueda05}; 4: \citet{Younes11}; 5: \citet{Cappi06}; 6: \citet{Bianchi09}; 7:\citet{Brightman_Nandra11}; 8: \citet{Winter12}; 9: \citet{Tueller10}; 10: \citet{Cusumano10}.
\end{center}
\end{table*}

\begin{figure}
\centering
\includegraphics[scale=0.39]{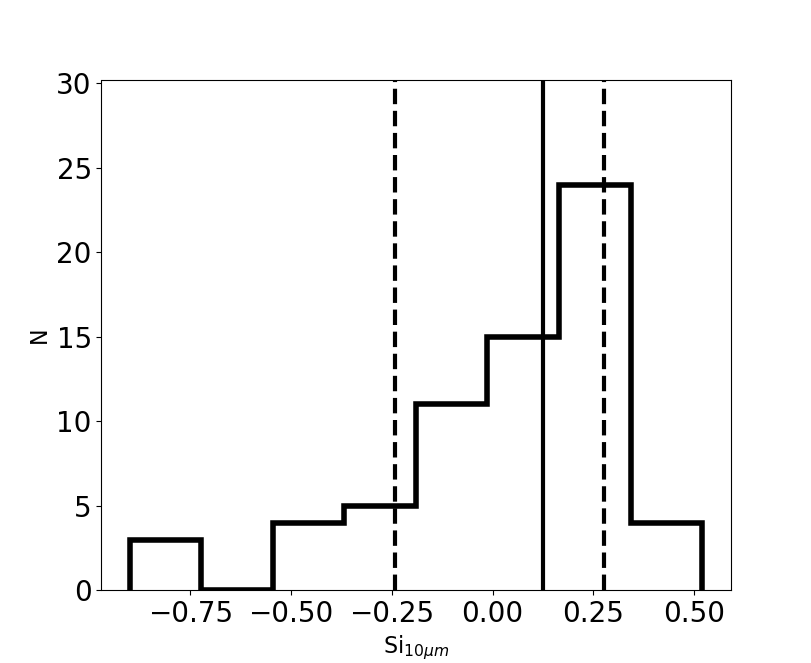} 
\caption{Distribution of the $10\mu$m silicate strength feature 
(see Table 1).
The solid and dashed black lines indicate the mean and its 1$\sigma$ confidence interval.
 \label{sil_select}}
\end{figure}

In most cases, our measurements, which were performed using the AGN-dominated {\it Spitzer} spectra, produce similar results to previous works \citep{Hao05, Thompson09, Sirocky08, Mendoza-Castrejon15} (see Tables~\ref{tab:silicates_IRS_all}). 
Additionally, we compare our measurements of the silicate strengths with the values obtained using {\sc deblendIRS} and find that they are similar within the uncertainties (see Figure~\ref{fig:comparison} in Appendix A).

\begin{table*}
\caption{{\bf Dust properties of the models}. \label{tab:dust_model_comparison}
}
\resizebox{16cm}{!}{
\begin{tabular}{l|ccccc}
\hline
Model name & Geometry & Dust composition & Scattering and absorption & \multicolumn{2}{c}{Grain sizes$^{*}$($\mu$m)}   \\
      & &   & coefficients & Graphites& Silicates \\
\hline
Fritz06 & Smooth torus &  Graphite and Silicates & graphite and silicates & 0.005-0.25  & 0.025-0.25\\
  &  &   &  \citet{Laor_Draine93}   &  & \\        
  & &   &                          & &  \\
Nenkova08& Clumpy torus& Standard ISM & graphites, \citet{Draine03}    & \multicolumn{2}{c}{(0.005-0.01)-0.25}  \\
 &  &   & silicates, \citet{Ossenkopf92}  & &   \\
   & &   &                          &  & \\
   & &   &                          &  & \\
Hoenig10& Clumpy torus &  Standard ISM     & graphites, \citet{Draine03}    & \multicolumn{2}{c}{(0.005-0.01)-0.25}   \\
      &       &+standard ISM with large grains      & silicates, \citet{Draine03}   &  & 0.1-1.0 \\
            &       & +graphite dominated dust      &  \citet{Ossenkopf92}   & & 0.05-0.25  \\
      &        &($70\%$ graphite and $30\%$ silicates)      &    &  & \\      
   & &   &                          & & \\     
Hoenig17& Disk+Outflow & Standard ISM      & like Hoenig10    &   \multicolumn{2}{c}{(0.005-0.01)-0.25} \\
  & & +standard ISM with large grains  &                          & & \\
Stalevski16&Two phase media torus& Graphite and Silicates     &\citet{Laor_Draine93}  &0.005-0.25 & 0.005-0.25 \\
      &       &       &  \citet{Li_draine01}    &  & \\
\hline
\hline
\end{tabular}}
\\
Note. The standard ISM is composed by $47\%$ of graphite and $53\%$ silicates. $^{*}$Assuming MNR distribution \citep[][]{Mathis77}. The numbers inside the parenthesis indicate the minimal range of sizes for the smaller grains.
\end{table*}

\subsection{Sample selection}
We build our final sample by selecting only those objects that show the 
strongest silicate emission features (see Table~\ref{tab:silicates_IRS}). From the original sample of 67 local type 1 AGNs we select those objects with the largest 10$\mu$m silicate strength ($\sigma_{Si_{10\mu\text{m}}}>0.28$, see Figure~\ref{sil_select}). The final sample is 
composed of 10 objects: six Seyfert (Sy) galaxies, and four 
low-ionization nuclear emission-line region (LINER)  galaxies. Four of the 
Seyfert galaxies are also classified as PG QSOs. Note that we are not selecting the 
sample according to the type of AGN. We list the type only to give 
information about their basic properties. 

Hereafter, we 
will refer to this sample as the type 1 AGN 
strong Silicate selected sample (Si-s). The sample spans a range of 
hard (2-10 keV) X-ray unobscured luminosities between $\sim10^{41}$~erg s$^{-1}$ and $\sim10^{45}$~erg 
s$^{-1}$. 

\section{Dusty torus models}
\begin{figure}
\begin{center}
 \includegraphics[scale =0.52]{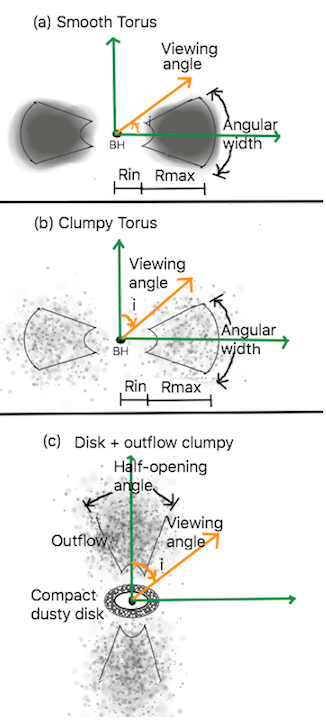}\\
  \caption{{\bf Cartoons of torus and disk+outflow models}: a) smooth models \citep{Fritz06}, b) clumpy models \citep{Nenkova08a,Nenkova08b, Hoenig10} and, c) disk + outflow clumpy models \citep{Hoenig17}. {\bf For smooth models}: viewing angle $i$ (degrees) $=[0-90]$, angular width $\Theta$ (degrees) $=[20-60]$, parameter of the dust distribution $\gamma =[0-6]$, and the outer-to-inner radius ratio $Y=R_{\rm out}/R_{\rm in}=[10-150]$. {\bf For clumpy models}: viewing angle $i$ (degrees) $=[0-90]$, angular width $\sigma$ (degrees) $=[15-70]$, number of clouds along the equatorial line ${N_{0}=[1-15]}$ (for Nenkova08) and $[2.5-10]$ (for Hoenig10), parameter of radial distribution ($\propto r^{-q}$, in Nenkova08) $q\rm{=[0-2.5]}$, and parameter of radial distribution ($\propto r^{a}$, in Hoenig10) $a=[-2.5-0]$, the outer-to-inner radius ratio $Y=R_{\rm out}/R_{\rm in}=[5-100]$ (for Nenkova08), and ${R_{\rm max}=170R_{\rm in}}$ (for Hoenig10). {\bf For disk+outflow clumpy models}: viewing angle $i$ (degrees) $=[0-90]$, number of clouds along the equatorial line ${N_{0}=[2.5-10]}$, half-opening angle of the outflow $\rm{\sigma_{\Theta}}$ (degrees) $=[30-45]$, and angular width of the disk (degrees) $\rm{\Theta_{w}=[7.0-15]}$. For a more detailed comparison between the parameters of the models see 
  \citet[][]{Gonzalez-Martin19a, Gonzalez-Martin19b}.}
 \label{fig:torus_models}
 \end{center}
\end{figure}
There are two types of torus models 
widely used in the literature: 
smooth torus models and clumpy torus models. Smooth models 
assume a continuous distribution of the dust in the 
torus 
\citep[][]{Pier_Krolik92, Granato97, Efstathiou_Rowan-Robinson95, Fritz06}, while clumpy models assume a distribution of dusty clouds or clumps in a toroidal structure \citep[][]{Dullemond_vanBemme05, Nenkova08a, Nenkova08b, Hoenig10, Hoenig17}.

For simplicity, smooth models assumed a continuous distribution of the dust in the torus, so the temperature of the dust decreases monotonically with the distance from the central BH. According to this type of model only type 1 
AGN produce $10\mu$m silicate features in emission and type 2 AGN produce silicate features in 
absorption, implying  
an edge-on orientation with respect to the observer. In type 2 AGN, 
therefore, the outer region of the torus hides the emission produced by the hotter and 
warmer dust in the inner part of the torus. However, 
the silicate depth that this model predicts is larger 
than observed in type 2s, and it fails to explain 
silicate features in emission in type 2 QSOs \citep[e.g.,][]{Sturm06}. 

Conversely, clumpy models \citep[e.g.,][]{Nenkova08a, Nenkova08b} never produce
very deep absorption silicate features, and they predict 
silicate features in emission in type 1 and type 2 AGN for a large combination of parameters \citep[see e.g.,][]{Nikutta09}. A third type of model combines clumpy and smooth properties to produce a two-phase medium dusty torus model \citep[][]{Stalev12, Stalev16}. This model assumes a distribution of dusty clumps with constant and high density, embedded in a smooth dusty component of low density. 
 
It was motivated by observational evidence that suggests that dust around AGN seems to  
have a multiphase filamentary structure \citep[][]{Wada09, Wada12}.

A more recent model proposes a geometry composed by a compact and geometrically thin disk in the equatorial region plus an extended elongated polar structure of clumpy dust, that is cospatial with the outflow of the AGN \citep{Hoenig17}. This geometry  
is inferred from  interferometric 
observations of nearby Seyfert galaxies, 
where the bulk of the MIR emission originates from the polar region rather than the equatorial plane \citep[e.g.,][]{Raban09, Hoenig12,Hoenig13,  Tristram14, LopezGonzaga16}.

In this paper, we use the AGN-dominated IRS/{\it Spitzer} spectra of our Si-s sample of type 1 AGN to investigate which models are  able to better reproduce the peak and shape of the strongest silicate emission features observed. 
In the following subsections we briefly describe the main geometrical and physical properties of the models and Table~\ref{tab:dust_model_comparison} summarizes the different geometries and dust compositions assumed for each model.

\subsection{Smooth torus model of \citet{Fritz06}}

This model is represented by a flared disc delimited by the inner and outer torus radius \citep[e.g.,][]{Efstathiou_Rowan-Robinson95}. The 
inner radius ($R_{\rm min}$) is defined by the sublimation 
temperature of dust grains (1500 K) under the strong radiation 
produced by the central engine. The size of the torus 
is determined by the outer radius ($Y=R_{\rm out}/R_{\rm in}$, where $Y$ is a free parameter), and the angular 
width of the torus $\Theta$ (see the cartoon (a) in 
Figure~\ref{fig:torus_models}). This model assumes that 
the dust is composed of graphite grains (53 per cent) 
with sizes from 0.005 to 0.25$\mu$m, and silicate 
grains (47 per cent) with sizes from 0.025 to 0.25$\mu$m. The grain sizes are distributed according to \citet{Mathis77}. For the two species they used the scattering and 
absorption coefficients given by \citet{Laor_Draine93}. The dust is illuminated by an 
isotropic central point-like emitting source, which is 
represented by a broken power law of the form $\lambda L_{\lambda}\propto\lambda^{\alpha}$, with $\alpha=1.2$ if $0.001<\lambda<0.03\,\mu$m,  $\alpha=0$ if $0.03<\lambda<0.125\,\mu$m, and $\alpha=0.5$ if $0.125<\lambda<20.0\,\mu$m. 
Other parameters of the model are
the viewing angle $i$, 
the polar ($\gamma$) and radial ($\beta$) indices of 
the gas density distribution $\rho(r,\Theta)\propto r^{\beta}e^{-\gamma\times \cos(\Theta)}$ within the 
torus, and the optical depth $\tau_{9.7\,\mu {\rm m}}$. For 
a more complete description of this model see \citet{Fritz06}.

\subsection{Clumpy torus models of \citet{Nenkova08a,Nenkova08b}}
This type of model assumes a central point-like emitting 
source surrounded by a toroidal distribution of 
clouds. The emission of the central source is 
characterized by a broken power law of the form $\lambda f_{\lambda}\propto\lambda^{1.2}$ for $\lambda\leq0.01\mu$m, $\propto\lambda^{-0.5}$ for $0.1\mu {\rm m}\leq\lambda\leq1\mu {\rm m}$, $\propto\lambda^{-3}$ 
for $\lambda\geq1\mu$m, and a constant power-law 
between 0.01 and 0.1$\mu$m. Due to 
its clumpy nature, 
the central source can directly heat the dust in the inner 
region of the torus 
and the dust located at several 
sublimation radii from the central source. This model 
assumes standard Galactic spherical dust grains 
(standard ISM) composed of graphites \citep[$47\%$,][]{Draine03} and silicates \citep[$53\%$,][]{Ossenkopf92}, with a power-law distribution of grain 
sizes ($\propto a^{-3.5}$), where $a_{\rm min}=0.05-0.01\mu$m and $a_{\rm max}=0.25\mu$m, 
respectively. This model has six 
free parameters: the viewing angle $i$, the number of 
clouds along the equatorial ray $N_{0}$, the angular 
width $\sigma$, the radial extend $Y=R_{\rm outer}/R_{\rm inner}$, the index of the radial distribution of 
clouds $q$, and the optical depth per cloud $\tau_{V}$ (see cartoon (b) in Figure~\ref{fig:torus_models}). 
For a more complete description of this model see \citet{Nenkova08a, Nenkova08b}.

\subsection{Clumpy torus models of \citet{Hoenig10}}
In this case the toroidal distribution of dusty clumps 
surrounds a central point-like emitting source, that 
is described by a broken power-law of the form $\lambda f_{\lambda}\propto\lambda$ for $\lambda<0.03\mu$m, $\propto\lambda^{-3}$ for $\lambda>3\mu$m, constant for $\lambda$ between 0.03$\mu$m and $0.3\mu$m, and $\propto\lambda^{-4/3}$ for $\lambda$ between 0.3 and 3$\mu$m. They modeled the torus 
following a 3D Monte Carlo radiative transfer 
simulation. These kinds of simulations fail to 
properly sample optically thick surface regions with 
enough grid cells so that each cell is optically thin, which results in 
underestimating the emission temperature and resulting 
in a smaller number of model clouds with respect to \citet{Nenkova08a,Nenkova08b}. The dust is composed of three components:
a standard ISM component, a standard ISM component with larger grains ($a_{\rm min}=0.1\mu$m and $a_{\rm max}=1.0\mu$m),
and 
intermediate to larger grains ($a_{\rm min}=0.05\mu$m and $a_{\rm max}=0.25\mu$m) with  $70\%$ 
graphite and $30\%$ silicates. The free parameters 
that describe this model are the viewing angle $i$, 
the number of clouds along the equatorial ray $N_{0}$, 
the half-opening angle $\Theta$, the index of the 
radial distribution $a$, and the optical depth $\tau_{V}$ (see cartoon (b) in 
Figure~\ref{fig:torus_models}). For a more complete 
description of this model see \citet{Hoenig10}.

\subsection{Clumpy disk+outflow models of \citet{Hoenig17}}
This type of model is based on the parametrization of the 
clumpy torus model of \citet{Hoenig10}, but 
instead of adding a blackbody component to take into 
account the NIR emission, they include a set of 
different sublimation temperatures for silicate and 
graphite dust that results in more emission from graphite located in the inner edge of the torus. In 
this way, when the dust is heated to temperatures $>1200$ K smaller silicate grains are destroyed leaving 
only graphite grains that can 
be heated up to 1900 K. In the innermost radius only, 
grains with a size between 0.075 and 1$\mu$m survive.

The clouds are distributed according to a radial 
power-law $\propto r^{a}$, where $a$ is the power law 
index and $r$ the distance from the black hole in units of 
the sublimation radius $r_{\rm sub}$. 
They also add a polar outflow, 
modeled as a hollow cone that can be 
characterized by the radial distribution of dust 
clouds in the wind $a_{\rm w}$, the half-opening angle of 
the wind ($\Theta_{\rm w}$), and the angular width ($\sigma_{\Theta}$). Other parameters are the number of 
clouds along the equatorial ray $N_{0}$, and the scale 
height in the vertical Gaussian distribution of 
clouds $h$ in the disk, (see cartoon (c) in 
Figure~\ref{fig:torus_models}). For a more complete 
description of this model see \citet{Hoenig17}.

\subsection{Two-phase medium dusty torus models of \citet{Stalev16}}
This kind of models assumes a distribution of high density dusty clumps embedded in a low density smooth dusty component. This assumption produces both weaker silicate features and a pronounced NIR emission. They assume that the accretion disk in the nucleus radiates 
as a broken power-law of the form $\lambda L_{\lambda}\propto\lambda^{\alpha}$, where $\alpha=1.2$ for a spectral range of $0.001\le\lambda\le0.01$ $\mu$m, $\alpha=0$ for $0.01<\lambda\le0.1$ $\mu$m, $\alpha=-0.5$ for $0.1<\lambda\le5$ $\mu$m, and $\alpha=-3$ for $5<\lambda\le50$ $\mu$m. The dust is distributed 
following a law that allows a density gradient along the radial ($r$) and polar ($\theta$) directions, 
inside a flare disc defined by the inner ($R_{\rm in}$), outer radii ($R_{\rm out}$) and half opening angle. The inner radius is defined by the sublimation temperature of 1500 K for an average dust grain size of 0.05 $\mu$m. They assumed a standard ISM dust composition with optical properties from \citet{Laor_Draine93} and, \citet{Li_draine01}.
For a more complete 
description of this model see \citet{Stalev12, Stalev16}. 

\begin{figure*}
\includegraphics[width=1.\columnwidth]{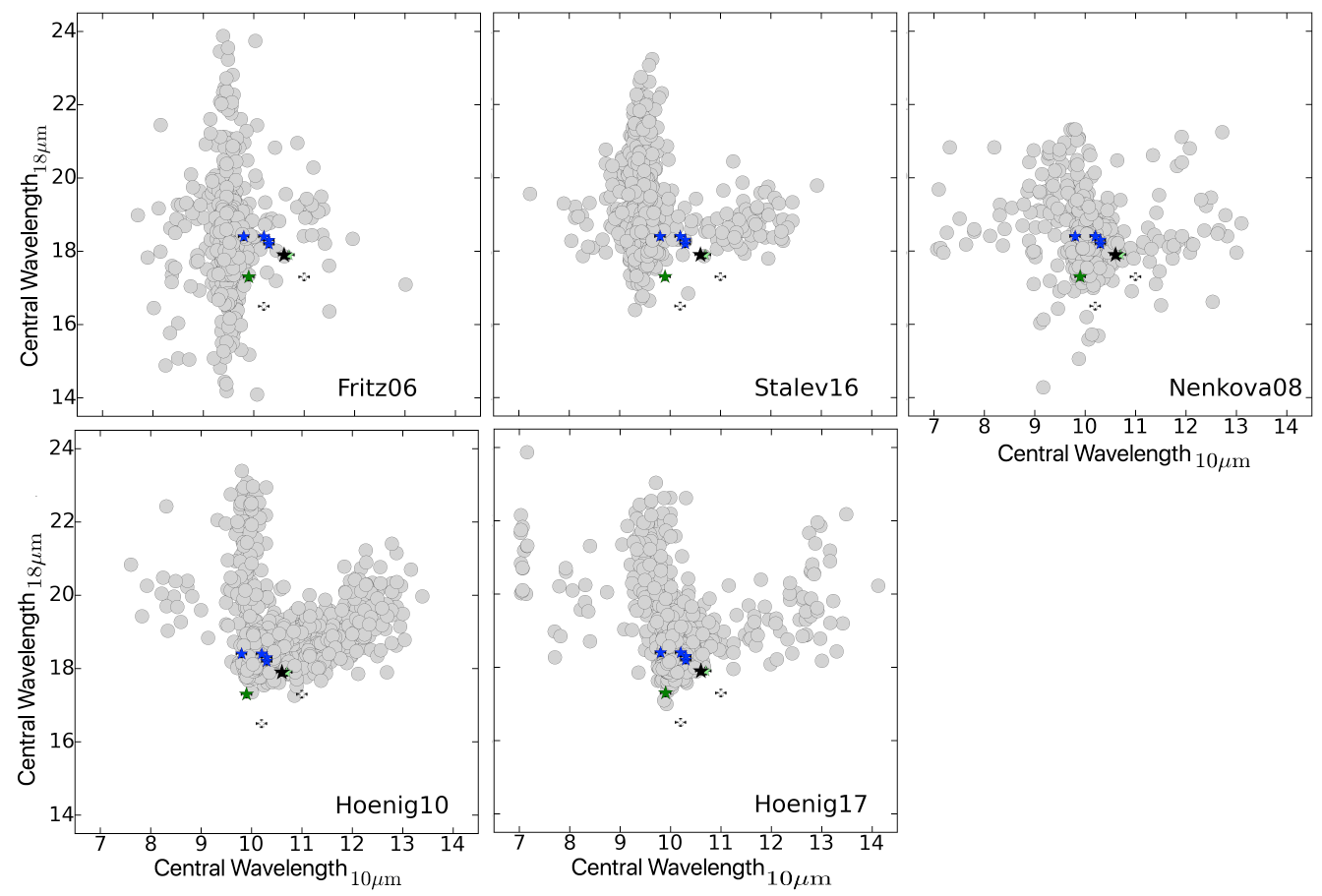}
\caption{{\bf Central wavelength of the synthetic and observed peaks of the 10 and 18$\mu$m silicate features}. Grey circles represent the values measured in the models, while the stars are the measurements obtained from the AGN-dominated IRS/{\it Spitzer} spectra of the Si-s sample. The colours of the stars indicate different range of bolometric luminosities ($L_{bol}$). White stars are for $logL_{bol}<42$, light-green stars are for $43\leq logL_{bol}<44$, blue stars for $44\leq logL_{bol}<45$, green stars for $45\leq logL_{bol}<46$, and black star for $logL_{bol}>46$. Small stars indicate lower luminosities, while larger stars are for higher luminosities.}
\label{fig:models_obs_lambda}
\end{figure*}

\begin{figure*}
\includegraphics[width=1.\columnwidth]{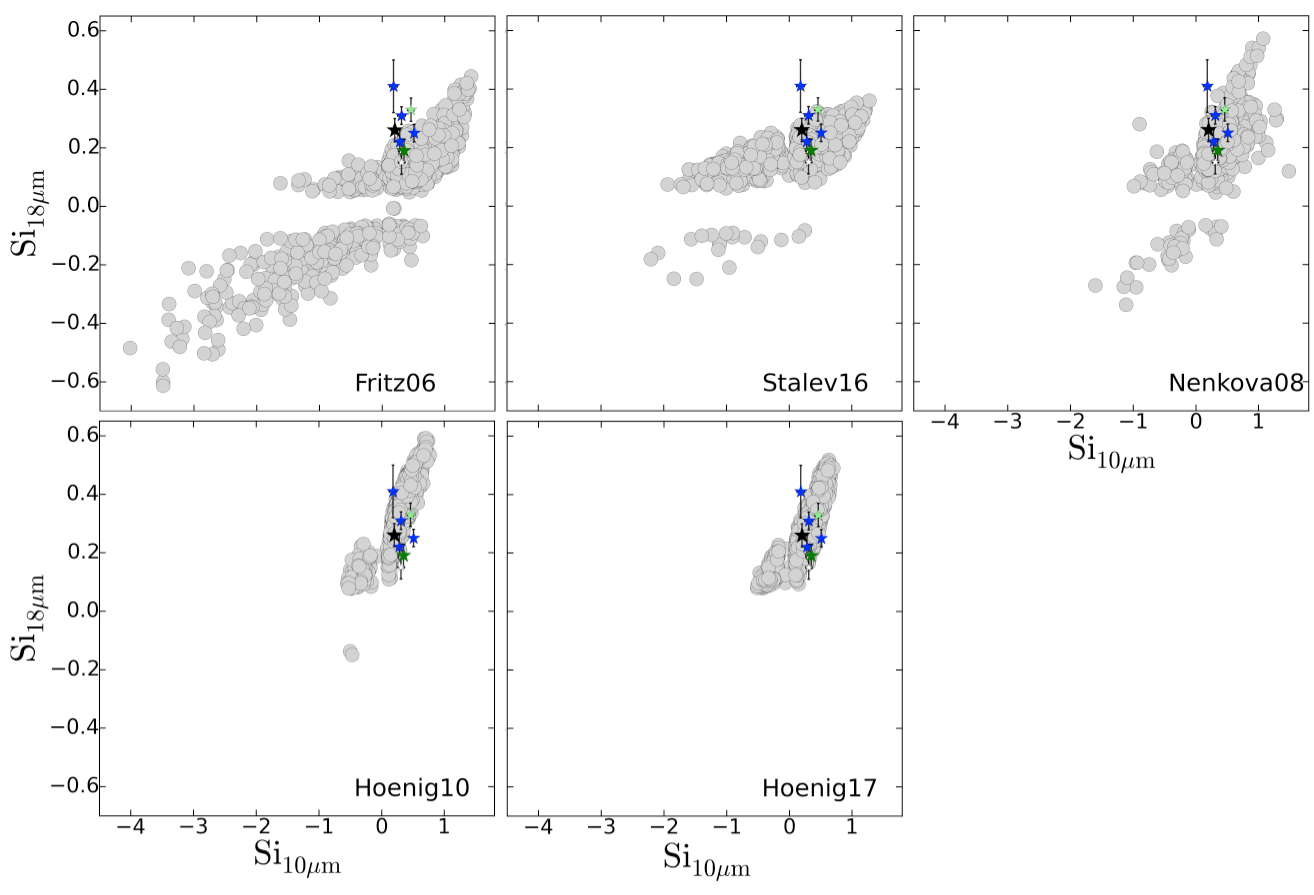}
\caption{{\bf Silicate strength of the 10 and 18$\mu$m silicate features}. Symbols and colours are as in Figure~\ref{fig:models_obs_lambda}.}
\label{fig:models_obs_silicates}
\end{figure*}

\begin{figure*}
\includegraphics[width=1.\columnwidth]{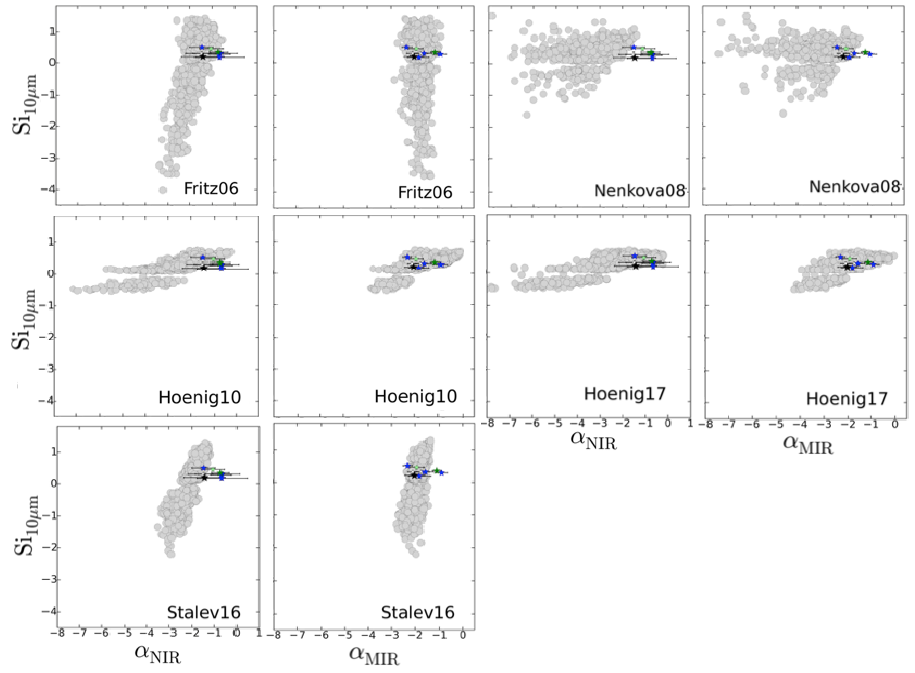} 
\caption{{\bf Synthetic and observed silicate strength of the 10$\mu$m silicate feature, and the NIR $\alpha_{NIR}$ and MIR $\alpha_{MIR}$, spectral indexes. Symbols and colours are as in Figure~\ref{fig:models_obs_lambda}.}}
\label{fig:models_obs_silicate_indx}
\end{figure*}

\section{Analysis}
\subsection{Synthetic and observed silicate peak wavelengths and strengths}
In this section we explore how well the dusty torus models 
reproduce
the central wavelength and strength of both 10 and 18$\mu$m silicate features, and the NIR ($\alpha_{NIR}$) and MIR ($\alpha_{MIR}$) spectral indexes. 
In order to make 
a proper comparison the synthetic and observed central wavelength
and strength of both the 10 and 18$\mu$m silicate features are measured following the same methodology described in Section 2.3, and fixing 
the bands to the sides of the silicate 
features 
between 7-7.5$\mu$m, 14-15$\mu$m, and $25-26\mu$m. The synthetic and observed 
spectral indexes $\alpha_{NIR}$ and $\alpha_{MIR}$ are measured between 5.5-7.5$\mu$m, and 
between 7.5-14.0$\mu$m, respectively, 
according to the following definition $\alpha_{2,1}=-log(f_{\nu}(\lambda_{2})/f_{\nu}(\lambda_{1}))/log(\lambda_{2}/\lambda_{1})$, with $\lambda_{2}>\lambda_{1}$ \citep[see e.g.,][]{Buchanan06}.

In Figures \ref{fig:models_obs_lambda} and
\ref{fig:models_obs_silicates} 
we plot the wavelength of the peak of the both 10 and 18$\mu$m silicate features and the strengths of both silicate features
as predicted by the models, and as observed in the AGN-dominated IRS/{\it Spitzer} spectra of the Si-s sample. Additionally, we color-coded the objects according to their bolometric luminosity, which we estimate using the hard X-ray luminosity and the relation derived by \citet{Marconi04} and \citet{Alexander12}.

Since the models are probabilistic in nature, we compare the envelope of measurements performed on the models to the measurements of the Si-s AGN sample. While the Fritz 06, Stalev16 and Nenkova08 models envelope covers the wavelength space (Figure~\ref{fig:models_obs_lambda}) where our Si-s AGN measurements lie, the Hoenig10 and Hoenig17 models leave miss one and two objects, respectively, with the lowest value central wavelength of the 18$\mu$m silicate feature. Curiously, these objects are of low bolometric luminosity (OQ~208 and NGC~4258). 


For the silicate strengths (Figure~\ref{fig:models_obs_silicates})
we note that the range of synthetic 
values sampled by the models mostly 
match the observations. However, we also note that Hoenig10 and Hoenig17 models show a 
narrower range of values respect to the other models. Additionally, the Hoenig17 model never predicts both silicates in absorption. Similar results were reported 
for a larger sample of AGN \citep{Gonzalez-Martin19b}. Finally, 
all the 
models tend to produce extremely prominent silicate emission features that have not been observed.

In Figure~\ref{fig:models_obs_silicate_indx} we compare the synthetic and observed silicate
strength of the 10$\mu$m silicate feature 
with the $\alpha_{NIR}$ and $\alpha_{MIR}$ spectral indexes.
We note that the range of synthetic values sampled by the Fritz06, Hoenig10, and Hoenig17
models mostly match the observations, independently of the bolometric luminosity. However, the Nenkova08 and Stalev16 models miss several of the NIR and MIR spectral indexes observed. Note that NGC~3998 is excluded from plots in Figures~~\ref{fig:models_obs_lambda}, ~\ref{fig:models_obs_silicates} and ~\ref{fig:models_obs_silicate_indx} because the short spectral range ($\sim7.5-14.5\mu$m) cover by the low resolution IRS/{\it Spitzer} spectrum of this object.

\begin{figure*}
\begin{center}
\includegraphics[width=\textwidth]{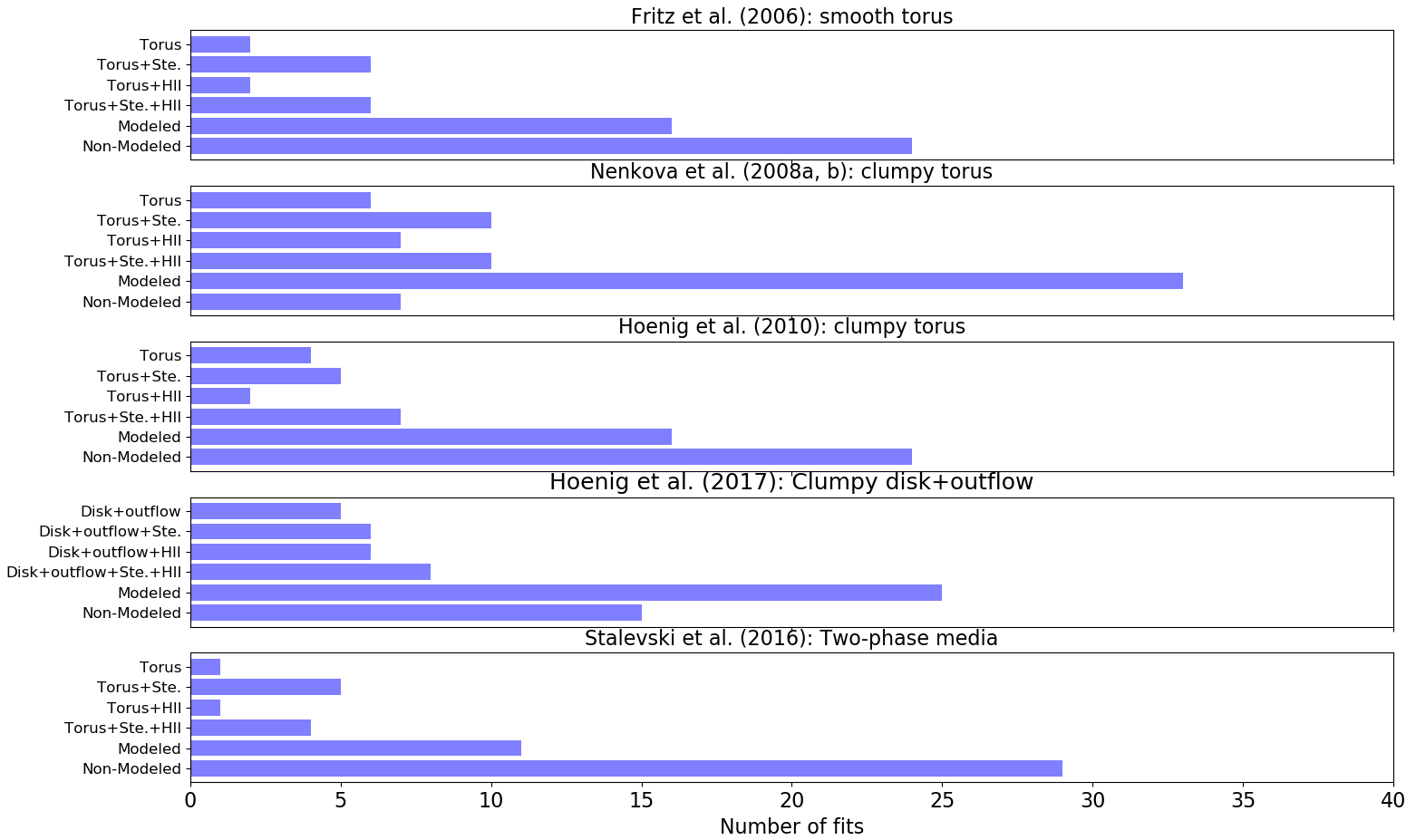} 
\caption{{\bf Fitting components for each torus model}: from top to bottom the horizontal bars show the number of fits with one (AGN), two (AGN + stellar or AGN + HII), and/or three (AGN + stellar + HII) components for each dusty torus model. Additionally, for each torus model, the two last horizontal bars represent the total number of cases modeled and not modeled ($\chi_{\rm red}^{2}>2$), respectively.}
\label{fig:silicates_models}
\end{center}
\end{figure*}
\subsection{Modeling}
We apply the torus models using the 
computational spectral fitting tool {\sc XSPEC}, 
which is part of the {\sc HEASOFT}\footnote{\url{https://heasarc.gsfc.nasa.gov}} 
software. These models were recently converted into 
the {\sc XSPEC} format in order to fit data in a similar way \citep[see section 2 in][]{Gonzalez-Martin19a}. We also use the set of synthetic 
stellar and empirical HII components (ISM) previously converted into {\sc XSPEC} format \citep[][]{Gonzalez-Martin19a}. The former corresponds to a 
stellar population of 10$^{10}$ years and solar 
metallicity from the stellar spectral libraries of \citet{Bruzual_Charlo03}, while the empirical HII components are 
average starburst templates from \cite{Smith07}.

We model the AGN-dominated IRS/{\it Spitzer} spectrum of each 
object in the Si-s sample, using each one of the four torus family models and 
the disk+outflow models. Additionally, we add a 
stellar synthetic and/or HII empirical component to investigate in which cases adding one or both components really improves the fitting. We also add foreground extinction to the 
torus models by using the {\sc ZDUST} component \citep{Pei92}. 

In detail, we start converting the IRS/{\it Spitzer} spectra into the {\sc XSPEC} format, loading the data into {\sc XSPEC}, removing the parts of the spectra dominated by emission lines, 
and modeling the spectra  
assuming the following component combinations: 
\begin{itemize}
\item 1: AGN dust emission: torus models or disk+outflow model
\item 2: AGN + stellar
\item 3: AGN + HII
\item 4: AGN + stellar + HII
\end{itemize}
For each step, we save the reduced $\chi_{\rm red}^{2}$, the 
parameters of the model with their uncertainties, and 
the emission contribution of each component to the 
total emission of the IRS/{\it Spitzer} spectrum between 5.5 to 30 $\mu$m. In 
those cases where $\chi_{\rm red}^{2}>2$ we 
reported the case as ``non-modeled". We perform this 
procedure for each torus model, resulting in 20 spectral fits per 
object. Figure~\ref{fig:silicates_models} shows, for each 
torus model, the number of spectral fits obtained, using one, two, or three 
components. Those 
cases in which none of the component combinations (1,2,3, or 4) are able to model the spectrum with a $\chi_{\rm red}^{2}<2$ are called ``non-modeled" spectral fits (see Table \ref{tab:all_fitting} in Appendix C). In the next section we investigate which model best fits the peak and strength of both silicate emission features 
in the Si-s sample.

\begin{figure*}
\begin{center}
\includegraphics[width=1.1\textwidth, angle=90]{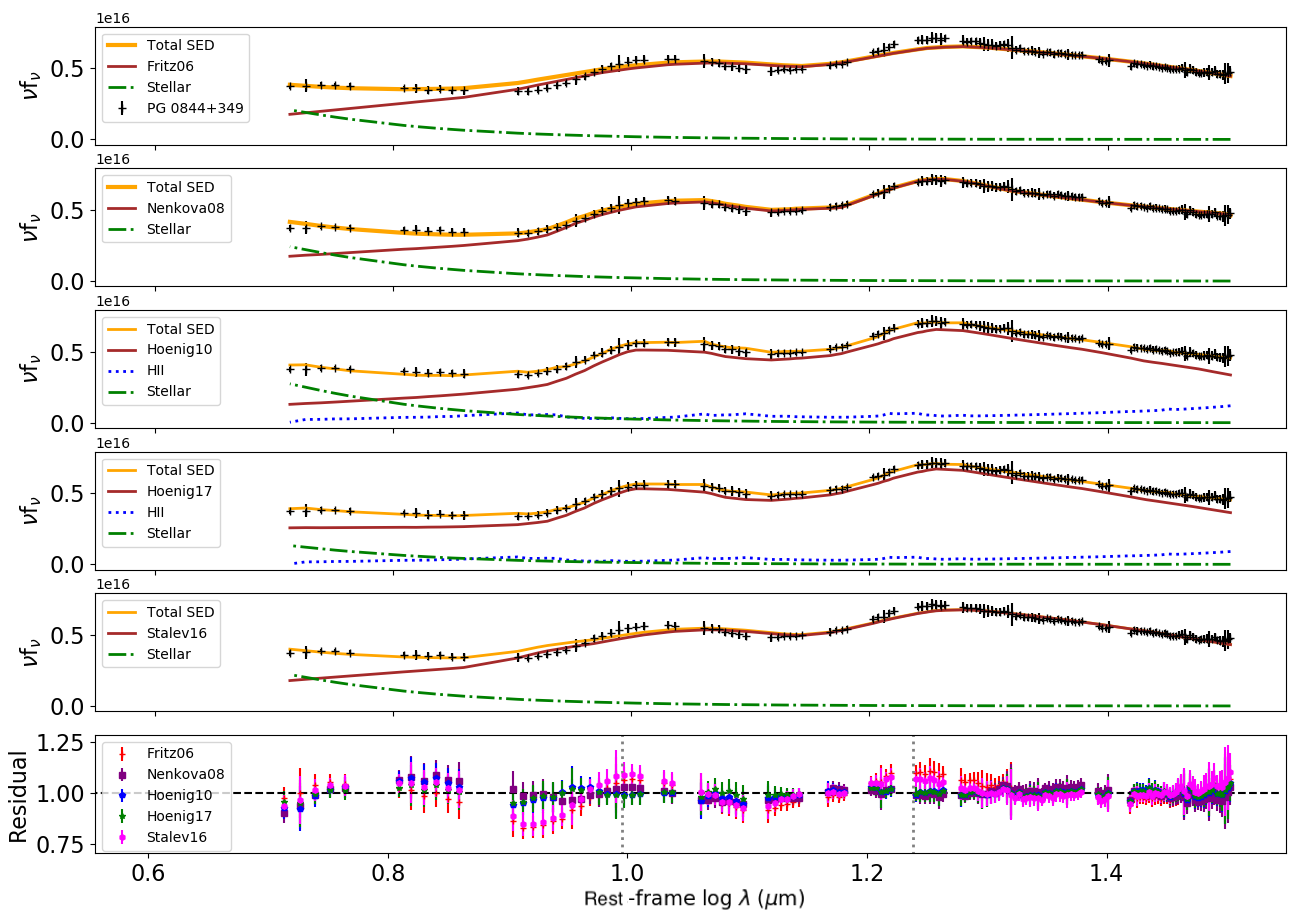}
\caption{{\bf Modeling and residuals of the IRS/{\it Spitzer} spectrum of PG~0844+349}. From top to bottom 
we assume models Fritz06 ($\rm{\chi_{red}^{2}\sim1.20}$), Nenkova08 ($\rm{\chi_{red}^{2}\sim0.32}$), Hoenig10 ($\rm{\chi_{red}^{2}\sim0.20}$), Hoenig17 ($\rm{\chi_{red}^{2}\sim0.13}$), and Stalevski16 ($\rm{\chi_{red}^{2}\sim0.81}$). The $\chi_{\rm red}^{2}$ values refer to fits with both torus models and other components. The last panel shows the residuals defined as the ratio between the data and model. In all panels the black points are the IRS/{\it Spitzer} spectrum and its error in erg s$^{-1}$cm$^{-2}$, and the red solid line is the fitted torus model. The orange line is the total SED that results when 
the stellar (green dot-dashed line) and/or the HII (blue dotted line) components are added to model the spectrum.}
\label{fig:fitted_models}
\end{center}
\end{figure*}

\subsection{Model comparison}
In most cases, we find that the same object can be 
modeled assuming one (AGN), two (AGN+stellar or AGN+HII), or 
three components (AGN+Stellar+HII) with a $\chi_{\rm red}^{2}\sim1$. For each object and torus model we use 
the statistical f-test, which allows us to evaluate in 
which cases the addition of a new component improves the fit from a statistical point of view. We add a new component when the f-test probability is $<10^{-4}$. The computational tool {\sc xspec} 
includes the {\sc ftest} command line, which allows us 
to calculate the f-statistic and its probability, 
when new and old values of $\chi^{2}$ and the degrees of freedom 
(dof) are provided.
As an example, 
in Figure~\ref{fig:fitted_models} we show the best fit of
%
the IRS/{\it Spitzer} spectrum of PG~0844+349 for each torus model. For this 
particular object 
the Fritz06, Nenkova08, and Stalev16 models need an additional stellar 
component
to fit the spectrum, while 
the Hoenig10 and Hoenig17 
models 
need also the HII component. Additionally, 
for this 
object, we observe 
that clumpy and disk+outflow models produce the flattest residuals within the uncertainties and smaller $\chi_{\rm red}^{2}$ 
(see bottom panel in 
Figure~\ref{fig:fitted_models} and Table~\ref{tab:best_fitting}). 

\begin{figure}
\centering
\includegraphics[scale=0.5]{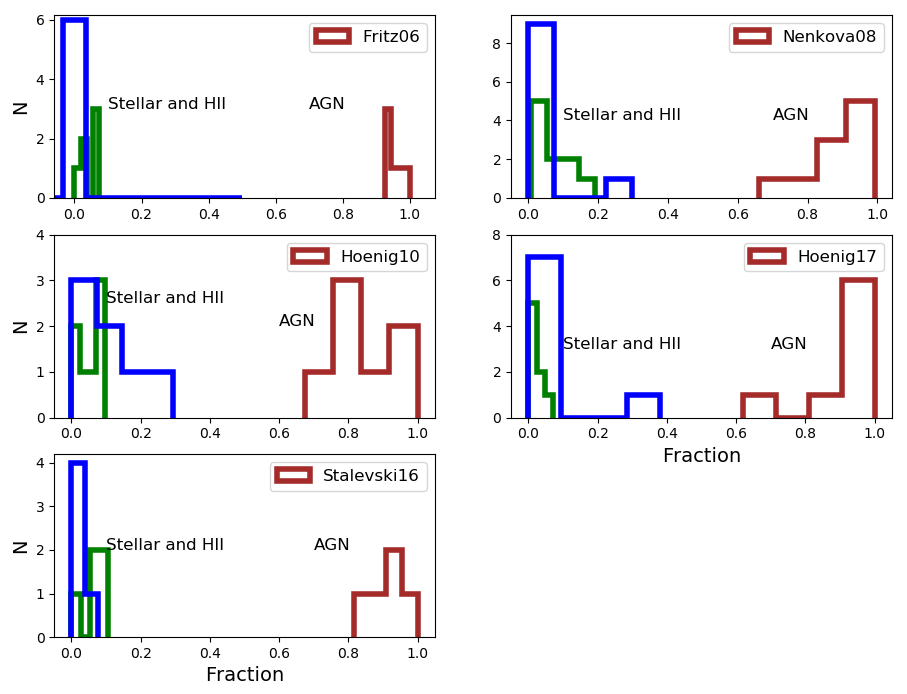}  
  \caption{Fractional contributions of the AGN (brown histogram), stellar (green histogram) and HII (blue histogram) components for Fritz06, Nenkova08, Hoenig10, Hoenig17, and Stalevski16 models.}
 \label{fig:fraction_components}
\end{figure}

\begin{table*}
\begin{minipage}{1.\textwidth}
\caption{Best fitting from smooth models.
Column 1 lists the name of the object, column 2 the $\chi_{\rm red}^{2}$, 
column 3 the combination of components that best fit the IRS/{\it Spitzer} spectra, and columns 4, 5, 6 the percentage contribution of each component. \label{tab:best_fitting}}
\resizebox{18cm}{!}{
\begin{tabular}{l|ccccc|l|ccccc}
				\hline
    &  &  & \multicolumn{3}{c}{Component contributions} &   & & & \multicolumn{3}{c}{Component contributions}\\
Name  & $\chi_{red}^{2}$   &Best combination & AGN & Stellar & HII & Name  &$\chi_{red}^{2}$ &Best combination & AGN & Stellar & HII\\
      &  &     & ($\%$) & ($\%$) & ($\%$)       &      &        &    & ($\%$) & ($\%$) & ($\%$)  \\       
\hline
\multicolumn{6}{c}{Smooth torus models of \citet{Fritz06}}  & \multicolumn{6}{|c}{Disk+outflow models of \citet{Hoenig17}}\\
\hline
PG2304+042 & 1.92 & Torus+Stellar & 93 & 7 &0 & PG2304+042 & 0.57 & Disk+outflow+Stellar & 93 & 7 &0\\
PG~0844+349 & 1.20  & Torus+Stellar & 93 & 7 & 0 & PKS0518-45 &0.31 & Disk+outflow & 100 & 0 & 0\\
PG~1351+640 & 1.52  & Torus+Stellar & 97 & 3 & 0 & PG~0844+349 &   0.13& Disk+outflow+Stellar+HII & 87 & 5 & 8\\
PG~2214+139 &  1.52 & Torus+Stellar & 96 & 4 & 0 & PG~2214+139 & 0.20  & Disk+outflow+Stellar+HII & 95 & 1 & 4\\
PG~0804+761 & 0.63 & Torus+Stellar & 94 & 6 & 0 & PG~0804+761 & 0.49 & Disk+outflow+HII & 95 & 0 & 5\\ 
NGC~3998 &  1.91& Torus+Stellar & 100 & 0 & 0 & OQ~208 &  1.78 & Disk+outflow+Stellar & 98 & 2 & 0\\
  &   &  & &   &   & NGC~4258 & 1.88 & Disk+outflow+HII & 62 & 0 & 38\\
         &.     &.          &.   &.  &.   & NGC~3998 & 0.39 & Disk+outflow & 100 & 0 & 0\\
\hline
\multicolumn{6}{c}{Clumpy torus models of \citet{Nenkova08a, Nenkova08b}}& \multicolumn{6}{|c}{Two-phase media torus models of \citet{Stalev16}}\\
\hline
NGC7213   &  1.11    & Torus+Stellar & 96 &4 & 0 & PG2304+042 & 1.99 & Torus+Stellar & 93 & 7 &0\\
PG2304+042 & 1.25 & Torus+Stellar & 92 & 8 &0 & PKS0518-45 & 1.65& Torus & 100 & 0&0\\
PKS0518-45 &1.29 & Torus+Stellar & 97 & 3&0 & PG~0844+349 &  0.81 & Torus+Stellar & 92 & 8 & 0\\
PG~0844+349 &  0.32 & Torus+Stellar & 91 & 9 & 0 & PG~2214+139 &  0.91 & Torus+Stellar+HII & 82 & 10 & 8\\
PG~1351+640 & 1.92  & Torus+Stellar & 99 & 1 & 0 & PG~0804+761 & 1.11 & Torus+Stellar & 90 & 10 & 0\\
PG~2214+139 & 1.92  & Torus+Stellar & 87 & 13 & 0 &   &  &   &   &  &  \\
PG~0804+761 & 1.60 & Torus+Stellar & 89 & 11 & 0  &.         &.    &.                  &.   &.  &.  \\ 
OQ~208 & 0.32  & Torus+Stellar+HII & 91 & 3 & 6 &.         &.    &.                  &.   &.  &.  \\
NGC~4258 &0.78  & Torus+Stellar+HII & 66 & 4 & 30&.         &.    &.                  &.   &.  &.  \\
NGC~3998 & 0.32 & Torus+Stellar & 81 & 19 & 0&.         &.    &.                  &.   &.  &.  \\
\hline
\multicolumn{6}{c}{Clumpy torus models of \citet{Hoenig10}}&.         &.    &.                  &.   &.  &.  \\
\hline
PG2304+042 & 1.18 & Torus & 100 & 0 &0&.         &.    &.                  &.   &.  &.  \\
PKS0518-45 &0.46 & Torus+Stellar+HII & 89 & 6&5 &.         &.    &.                  &.   &.  &.  \\
PG~0844+349 & 0.20  & Torus+Stellar+HII & 80 & 10 & 10 &.         &.    &.                  &.   &.  &.  \\
PG~2214+139 &  0.71 & Torus+Stellar+HII & 76 & 9 & 14 &.         &.    &.                  &.   &.  &.  \\
PG~0804+761 & 1.79 & Torus+Stellar+HII & 76 & 9 & 15 &.         &.    &.                  &.   &.  &.  \\ 
NGC~4258 &1.64  & Torus+Stellar+HII & 67 & 3 & 29&.         &.    &.                  &.   &.  &.  \\
NGC~3998 & 0.40 & Torus & 100 & 0 & 0&.         &.    &.                  &.   &.  &.  \\
				\hline
			\end{tabular}}\\
		\end{minipage}
	\end{table*}

In Figure~\ref{fig:fraction_components} we plot the histogram of the fractional 
contribution of each spectral component. The AGN component dominates the 
emission in all cases, which is expected due to our selection of 
AGN-dominated IRS/{\it Spitzer} spectrum sources.

Fritz06 models are able to reproduce the IRS/{\it Spitzer} 
spectra in seven of 10 objects. The AGN component contributes more than $93\%$. 
The Nenkova08 models reproduce the IRS/{\it Spitzer} spectra of 10 
objects, in most cases with an AGN contribution $>81\%$, 
except for NGC4258 for which the AGN component 
contributes $66\%$.
For Hoenig10 models we find that four of the 
objects can be modeled with an AGN contribution $>80\%$, and three other
(PG2214+139, PG0804+761, and NGC4258) with a 
contribution between 67 and 76 $\%$. The Hoenig17 models are able to fit seven of the objects  
with an AGN contribution $>87\%$, and NGC 4258
with an AGN component that contributes $>62\%$. 
Finally, in the case of two-phase medium models 
of \citet{Stalev16} only five of the 10 objects can be reproduced 
with a contribution of the AGN $>82\%$. 

Irrespective of the model used, NGC4258 always need a large ($\sim20-38\%$) contribution from the HII component. It is possible that in this case the spectral decomposition tool {\sc DeblendIRS} is not perfectly separating the different spectral contributions at MIR.
We also note that in all cases the stellar component is necessary to take into account the emission in the bluer extreme of the spectrum, while the HII component is necessary to take into account the emission in the redder extreme of the spectrum. A similar result was also found by \citet{Gonzalez-Martin19a} for a large sample of AGN. 

We obtain upper (or lower) limits of the resulting  free parameters and covering factors. The covering factor is defined as $1-P_{\rm esc}$, where $P_{\rm esc}$ 
is the probability that a 
photon emitted in the central engine is able to escape without being absorbed by the torus (in the case of the smooth torus 
and two-phase medium torus models) or by a dusty cloud in clumpy torus models (Nenkova08, Hoenig10). In the case of Hoenig17 models, the 
covering factor derived is the sum of the geometrical covering 
factors of the disk and the covering factor of the outflow 
\citep[see][]{Gonzalez-Martin19b}.

To obtain well constrained parameters a detailed modeling that includes NIR and far-infrared (FIR) data is necessary \citep[see e.g.,][]{Ramos_Almeida14}. However, modeling simultaneously the NIR and MIR components of the IRS/{\it Spitzer} spectrum of type 1 AGNs has been a challenge \citep[e.g,][]{Mor09, Martinez-Paredes17, Hernan-Caballero15}. In this work we assume the NIR as a stellar component. Our purpose is to show if any of the proposed models are able to explain both the peak 
and shape of the strongest silicate emission features. For this purpose, we only 
need to check that the range of values of the covering 
factors obtained from modeling the IRS/{\it Spitzer} 
spectra with each model are within the range of values 
expected for type 1 AGNs \citep[see e.g.,][]{RamosAlmeida11, Alonso-Herrero11, Ichikawa15, Mateos16, Martinez-Paredes17, Gonzalez-Martin17}.

We note that on average the smooth models produce a dusty torus with a small angular widths, with low and high viewing angles, 
although the angular width is poorly constrained. The dusty clumpy torus models of Nenkova08 produce both large and low viewing angles and a range of angular width from  low (15 degrees) to high (70 degrees) values, and a number of clouds along the equatorial ray that are in general $\lesssim7$ clouds, resulting in escape probabilities 
$\gtrsim40\%$. The clumpy models of Hoenig10 produce values of the viewing angle that range from nearly 30 to 80 degrees, angular widths around 55 degrees, and number of clouds in the range 2.5 to 10.0. 
The viewing angles in disk+outflow models of Hoenig17 that we find is between 0 and 50 degrees and the angular widths between 30 and 45 degrees. Although in three cases we obtain lower limits of the angular width, indicating that this could be larger. The two-phase models of Stalev16 produce on average large viewing angles ($\sim80$), only in one case we obtain a viewing angle around 10 degrees, and the angular width is around 80 degrees, resulting in very obscured AGNs (see Tables~\ref{tab:parameters_Fritz06},\ref{tab:parameters_Nenkova08},\ref{tab:parameters_Hoenig10}, \ref{tab:parameters_Hoenig17}, and Figure \ref{fig:parameters} in Appendix C).

In general, 
Fritz06 and Hoenig17 models produce lower covering factors than 
Nenkova08, Hoenig10 and Stalevski16 models. Fritz06 and Hoenig17 models produce  
covering factors around 0.2, although they range from 0 to 1. The Nenkova08 model produces covering factors around 0.6 with a range from 0.3 to 1.0. These values are 
consistent with the range of values obtained by \citet{Martinez-Paredes17} for a sample of PG QSOs using Nenkova08 models. \citet{Feltre12} compare smooth 
and clumpy torus models of Nenkova08 and found that both torus 
model families produce similar MIR continuum shapes for different model 
parameters. In the Hoenig10, and Stalevski16 models the 
covering factors are large (around 0.8), probably due the fact that in these 
models the angular width of the torus tends to be larger, which leads to a more obscured AGN.

\subsection{Residuals}
\begin{figure*}
\begin{center}
\includegraphics[width=\textwidth]{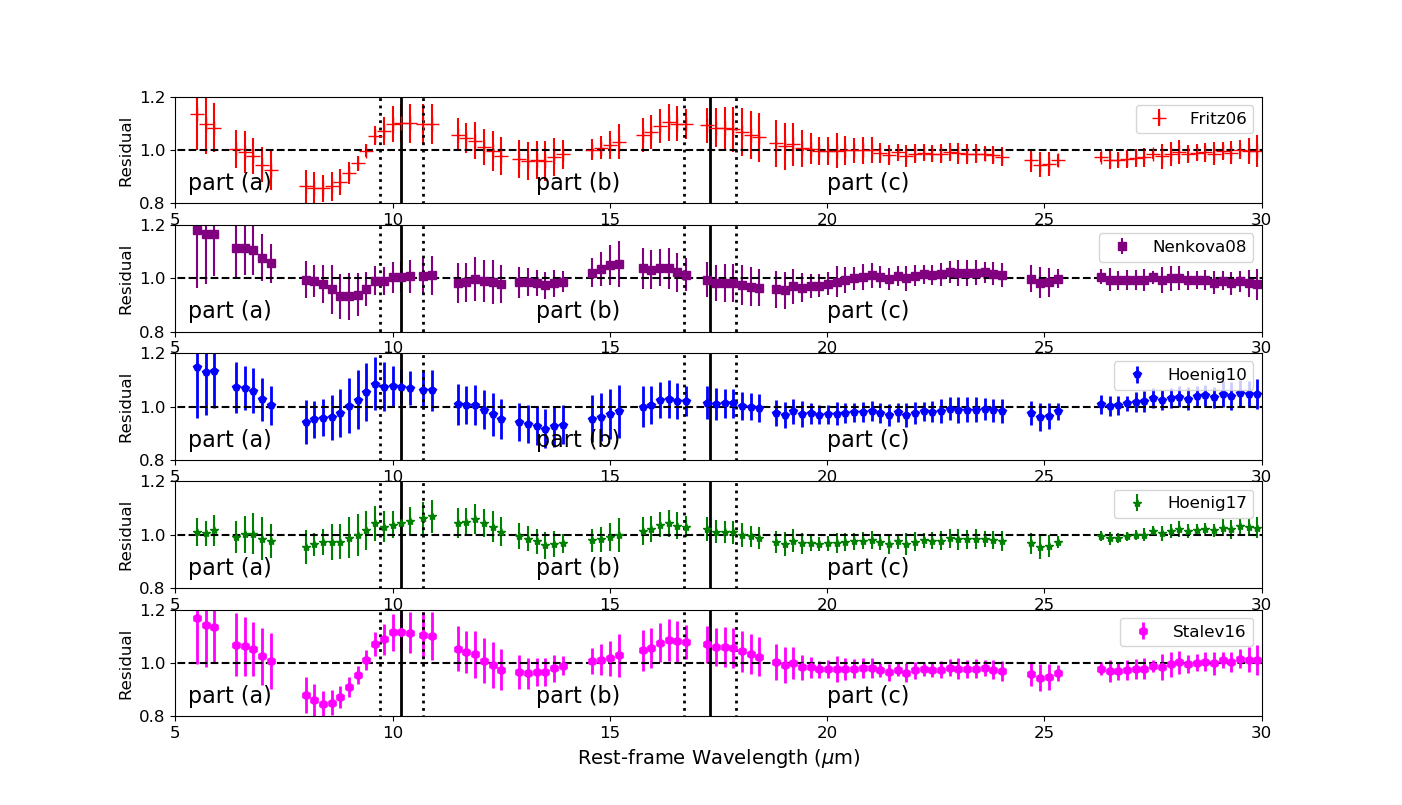} 
\caption{Average residuals (data/model) for all modeling combination including the modeled and non-modeled cases. The red points represent the average residuals of objects fitted with the smooth models. The purple and blue points represent the average residuals obtained from those objects  fitted with the clumpy models of Nenkova08 and Hoenig10, respectively. The green points represent the average residuals obtained from the objects fitted using the disk+outflow (Hoenig17) models, and the magenta points are the residuals obtained from using the two-phase dusty torus model of Stalevski16. The vertical grey solid line indicates the mean wavelength where the 10 and 18$\mu$m silicate features peak. The grey dashed lines are the $1\sigma$ confidence intervals.}
  \end{center}
    \label{fig:residuals}
\end{figure*}

Using the spectral residuals for all modeled and non-modeled cases we calculate the average spectral residuals for each torus model. The vertical black solid lines in Figure~\ref{fig:residuals} represents the mean wavelength where the silicate features peak in the IRS/{\it Spitzer} spectra, and the black dotted lines their $1\sigma$ intervals.
 
In order to discuss qualitatively the similarities and differences between the residuals of the models we divide the spectral range into three parts in Figure~\ref{fig:residuals}. These parts are the region bluewards of 10$\mu$m, between $10\mu$m and $18\mu$m, and redwards 18$\mu$m. In part (a) we observe that on average,  around 5$\mu$m, Fritz06, Nenkova08, Hoenig10, Stalev16 models are the worst at reproducing the bluer extreme of the spectra within the uncertainties. However, at longer wavelengths the Fritz06 and Stalevski16 models are the least accurate at reproducing the shape of the spectra. The Hoenig17 models best reproduce the spectra at all wavelengths within the uncertainties. In the part (b) Nenkova08 show the flattest residual, while in the redder extreme the Fritz06 and Stalevski16 models show the largest residuals, while the Hoenig10 and Hoenig17 models shows flatter residuals. In part (c) all models show similar residuals, although Nenkova08 show the flattest residual within the uncertainties. 

We also observe that the Fritz06 and Stalevski16 models underestimate the strength of the 10 and 18$\mu$m silicate features, while the Nenkova08, Hoenig10, and Hoenig17 models best reproduce the peak and the shape of both features. In general we note that Hoenig17 {and Nenkova08} models show flatter residuals, resulting in the models that best reproduces the shape and the peak of the strong silicate features observed in these objects.

Some objects deserve particular attention. For instance, in Figure~\ref{fig:modeling2} we can see that for NGC~7213 we require the stellar component to fit the bluer extreme of the spectrum, but that both silicate peaks are still underestimated by the Nenkova08 models. 

PG~1351+640 is modeled only by the smooth and clumpy models of Nenkova08. In both cases it is necessary to add the stellar component, which contributes $3\%$ and $1\%$ in the case of smooth (Fritz06) and clumpy models (Nenkova08 and Hoenig10), respectively. 
OQ208 is modeled only by the Nenkova08 and Hoenig17 models. In this case the residuals from the Nenkova08 model are flatter than the residuals from the Hoenig17 models, although in the case of the Nenkova08 model it is necessary to include the HII component with a contribution of $6\%$, in addition to the stellar component ($3\%$) in order to fit the redder extreme of the spectrum. In contrast, the Hoenig17 model is able to reproduce the entire spectral range requiring only a small contribution from the stellar component ($2\%$). The remaining objects are modeled successfully by the Nenkova08, Hoenig10, and Hoenig17 models.

\begin{figure*}
\begin{center}
\includegraphics[width=\textwidth]{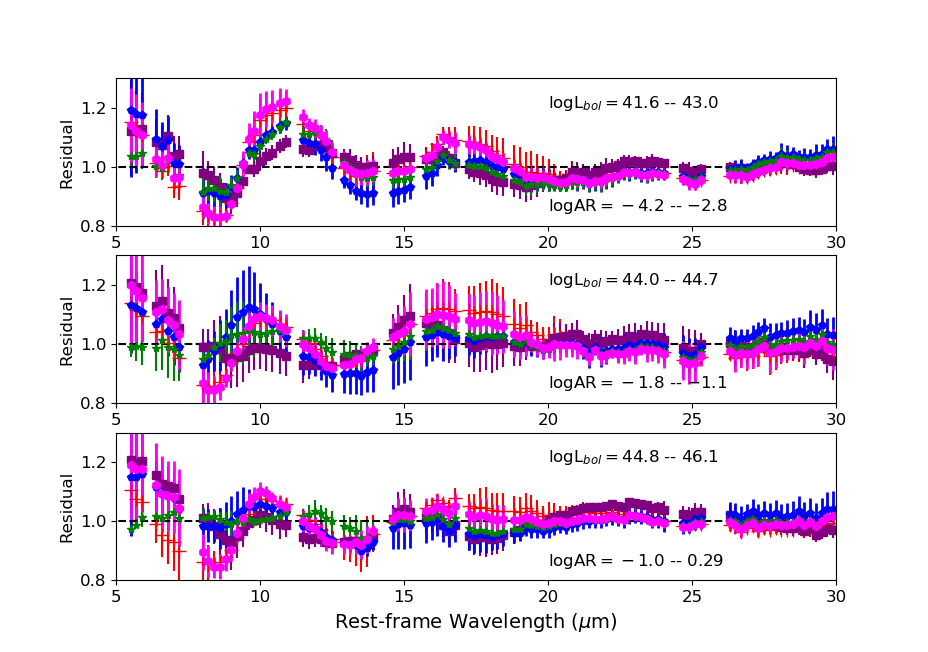} 
\caption{Average residuals for three ranges of bolometric luminosities and BH accretion rates. Colours are as in Figure~\ref{fig:residuals}.}
\end{center}
\label{fig:residuals_luminosity}
\end{figure*}
We divide the objects in the Si-s sample into three groups according to their bolometric luminosities and BH accretion rates. For each group and torus model we combine the residual obtained from fitting the AGN-dominated {\it Spitzer} spectrum with the components C1 (AGN), C2 (AGN+Stellar), C3 (AGN+HII), and C4 (AGN+Stellar+HII). In Figure \ref{fig:residuals_luminosity} we plot the average residual for each group. We note that for the first group, all models are unable to produce a flat residual around the 10$\mu$m silicate feature. But, for the second group, which covers a larger range of bolometric luminosities and BH accretion rates, all the residuals become flatter. At the largest bolometric luminosites of the third group all models show the flattest residuals. These results 
show that all models fail in reproduce 
the central wavelength 
of the 10$\mu$m silicate feature in the objects
with lower bolometric luminosities, as we see in Figure~\ref{fig:models_obs_lambda} in Section 4.1. Additionally, we note that 
the Hoenig17 models always produce 
the flattest residual around 5$\mu$m for
low, intermediate, and high 
luminosities.

\section{Discussion}
In the previous section we used four different torus models (smooth, clumpy, two-phase medium, and disk+outflow) to reproduce the strongest silicate features observed in type 1 AGNs. These features appear broader and shifted with respect to the silicate emission features observed in the standard ISM, suggesting a different dust composition (or geometry) of the torus or surrounding dust of AGNs. We discuss below the dust distribution (Section 5.1) and composition (Section 5.2) of the dust. We also discuss the deficiencies of the models to reproduce the strong silicate emission features in Section 5.3.
\subsection{Dust distribution}

Based on our analysis the clumpy models of Nenkova08 ($\chi_{\rm red}^{2}\sim1.03$), Hoenig10 ($\chi_{\rm red}^{2}\sim0.89$), and Hoenig17 ($\chi_{\rm red}^{2}\sim0.70$) produce a better fitting of the IRS/{\it Spitzer} spectra than the smooth ($\chi_{\rm red}^{2}\sim1.36$) and two-phase medium ($\chi_{\rm red}^{2}\sim1.20$) models. This result is in agreement with the previous evidence that the surrounding medium around the AGN should be clumpy \citep[see][and references therein]{Ramos_Almeida17}. Indeed, \citet[][]{Mendoza-Castrejon15} found that isolated type 1 AGN have a clumpy dust distribution, while interacting type 1 AGN can have both clumpy or smooth dusty distributions. Additionally, we find that
all models produces flatter residual in high luminosity AGN than in low luminosity AGN, 
probably due to all models better sample 
the central wavelength and both 10 and 18$\mu$m 
silicate features of high luminosity AGNs.
However, a high luminosities Hoenig17 models 
produce the flattest residual along all the spectral range.

\citet{Mor09} used IRS/{\it Spitzer} spectra of a sample of 26 nearby QSOs to constrain the clumpy models of Nenkova08. They argue that in order to  model the entire spectral range between 5-30$\mu$m they need to add two more components, one that takes into account the emission produced by the dust in the narrow line region (NLR) \citep{Schweitzer08}, and another one that takes into account the emission produced by dust close to the AGN, not directly related with the dusty torus \citep[e.g.,][]{Minezaki04, Riffel09, Kishimoto07}. However, \citet{Alonso-Herrero11} showed, for a sample of local Seyferts, that when high angular resolution data at NIR and MIR is used, it is not necessary to add any additional components in order to constrain the clumpy models of Nenkova08, although for some type 1 Seyferts galaxies they found that the NIR emission is underpredicted by the fitted SED. In a previous work, we used the starburst-subtracted IRS/{\it Spitzer} spectra from $\sim7.5-15\mu$m plus NIR high angular resolution data from the Near Infrared Camera and Multi-Object Spectrometer (NICMOS) on {\it HST} for a sample of 20 nearby QSOs in order to constrain the clumpy torus models of Nenkova08. \citet{Martinez-Paredes17} found that including the spectral range between $5-8\mu$m resulted in a poor fitting of the 10$\mu$m silicate emission feature. In this work we find that the AGN-dominated {\it Spitzer} spectra of the Si-s type 1 AGN can be fitted by Nenkova08 and/or Hoenig10 models by adding a stellar component that takes into account the bluer spectral range, and in some cases the HII component, in order to improve the fitting in the redder spectral range. A similar result was also found for a large sample of AGNs \citep{Gonzalez-Martin19a, Gonzalez-Martin19b}.

\citet{Hernan-Caballero17} found, for a sample of 85 QSOs, that a superposition of two blackbodies between 1.7-8.4$\mu$m, with temperatures around 1000 K for the hot blackbody component, and 400 K for the warm blackbody component, can fit the {\it Spitzer} spectra between 0.1 and 10$\mu$m. They argue that an additional hotter component of dust is necessary to reproduce the excess emission at 1-2$\mu$m. On the other hand, \citet{Lyu17} argues that the strong silicate emission features observed in dust-deficient PG QSOs can be explained assuming a reduced height scale of the warm dust, i.e., the dust between the sublimation zone and the outer region of cold dust, allowing for less interception of the radiation from the accretion disk with the inner dust, resulting in a decrease of the MIR continuum, but keeping  the heating of the outer dust responsible for the silicate emission features invariant \citep[see Figure 20 in][]{Lyu17}.

\subsection{Dust composition}
In general, the Nenkova08 models produce a better fit than the Hoenig10 ones. They assume the same geometry, but the former uses a standard dust composition, while the latter includes standard ISM dust plus standard ISM with large grains, and also a composition of dust mostly dominated by graphite. The Hoenig10 models assume this dust composition in order to take into account the observational suggestion that the dust composition in AGN deviates from standard ISM dust \citep[][]{Suganuma06, Kishimoto07}. One of the clear improvements in the Hoenig17 models is that they allow the existence of different dust compositions in different parts of the surrounding dusty structure (disk+outflow).

The chemical composition of the dust has been largely studied for the ISM \citep[see][and references therein]{Henning10}, and to a lesser extent in AGN \citep[see][and references therein]{Lyu14}. However, determining the exact chemical composition of the dust has been challenging. \citet{Srinivasan17} studied the dust composition of a large sample of PG QSOs with a redshift $z<0.5$, that showed the 10$\mu$m silicate feature in emission. They found that the dust is mostly composed of amorphous oxides and silicates, plus a small fraction in crystalline form. However, this small fraction is nearly four times larger than the last upper limit ($\lesssim2.2$) reported by \citet{Kemper05} for the ISM, and more similar to the upper limit ($\lesssim5$) previously reported by \citet{Li_draine01} for the ISM.

\subsection{Deficiencies of the models}
In general we find that neither the peak nor the shape of the silicate features of the AGN-dominated IRS/{\it Spitzer} spectra of the Si-s type 1 AGN are perfectly reproduced by the models. However, we note that for each object either the Hoenig17 or the Nenkova08 models produce the smallest $\chi_{\rm red}^{2}$ and flattest residuals. \citet{Hoenig17} point out that their dust composition explains both the observed small NIR reverberation mapping and interferometric sizes with dust sublimation physics. They argue that this combination of disk+outflow clumpy models is able to reproduce the $3-5\mu$m bump in type 1 AGN, and preserve the MIR bump produced by the wind. Indeed, according to this model a standard ISM composition of the dust in the wind would be the responsible for the emission of the silicate features. 
\citet{Garcia-Gonzalez17} found that the Hoenig17 (disk+outflow) models predict MIR slopes (between $8.1-12.5\mu$m) and silicate strengths at 10 $\mu$m that are in agreement with the values observed in type 1 AGN. Particularly, they noted that when clouds are more concentrated towards the inner region of the torus \citep[see e.g.,][]{Hoenig10, RamosAlmeida11, Ichikawa15, Martinez-Paredes17}, the MIR spectral indices are flatter and the silicate features are stronger than those observed in Seyferts and QSOs.

\section{Summary and conclusions}
\label{conclusion}

In order to investigate which model better reproduces the shape and peak of the strongest 10$\mu$m silicate emission features observed in type 1 AGN, we measure the 10$\mu$m silicate emission strength for a sample of local ($z<0.1$) type 1 AGN, for which their IRS/{\it Spitzer} spectra is mostly dominated by the emission of non-stellar processes
($>80\%$).

We find that 
these objects show silicate features in emission. On average the 10$\mu$m silicate feature has a strength of $0.13^{+0.15}_{-0.36}$ that peaks at 10.3$\mu$m, and a 18$\mu$m silicate strength of $0.14^{+0.06}_{-0.06}$ that peaks at $17.3\mu$m. We find that 10 objects are among the AGN with the largest 10$\mu$m silicate strengths ($\sigma_{Si_{10\mu\text{m}}}>0.28$, Si-s sample), and that some of them have been previously classified as objects with prominent silicate features.  

We use four different torus models, Fritz06 \citep{Fritz06}, Nenkova08 \citep{Nenkova08a, Nenkova08b}, Hoenig10 \citep{Hoenig10}, Stalevski16 \citep{Stalev16} and a disk+outflow \citep[Hoenig17,][]{Hoenig17} model
to fit the IRS/Spitzer spectra of the Si-s sample, and investigate which model better reproduces the peak and shape of both silicate emission features. The models assume different dust distributions, namely, smooth, clumpy, and a two-phase medium, as well as different dust compositions. We find that in most cases it is necessary to add a stellar or HII component in order to improve the fit. In most cases we find that the contribution of these components is $<<20\%$, in agreement with our selection requirement that the spectra be dominated by the emission of the AGN. The exceptions is NGC~4258 for which the spectral decomposition seems to underestimate the HII component.

We find that in general Fritz06 and Hoenig17 models produce lower covering factors than the Nenkova08, Hoenig10, and Stalevski17 models. The values are consistent with those reported in previous works for type 1 AGN. We find that the individual and average spectra are reproduced better with clumpy torus models than smooth models. Moreover, the Hoenig17 model shows the flattest residuals along all the spectral range between $\sim5-35\mu$m, while the rest of models fail to reproduce the bluer extreme of the spectrum. However, on average none of the models are able to exactly reproduce the peak and shape of the silicate features. 

In the near future the {\it Mid-Infrared Instrument} (MIRI) onboard of the {\it James Webb Space Telescope} will provide high angular resolution with higher sensitivity and spectral resolution observations which will allow an in-depth investigation of the dust properties in active galaxies. Additionally, new models that include a better description of the properties of the dust will be required.

\section*{Acknowledgements}
MM-P acknowledges support by the KASI and UNAM-DGAPA postdoctoral fellowships. This work is partially supported by the KASI project 2019184100 and Conacyt project CB-2016-281948. O.G.-M. acknowledges support by the PAPIIT projects IA100516. A.A.-H. acknowledges support through grant PGC2018-094671-B-I00 (MCIU/AEI/FEDER,UE). AAH’s work was done under project No. MDM-2017-0737 Unidad de Excelencia "Mar\'ia de Maeztu"- Centro de Astrobiolog\'ia (INTA-CSIC). YK acknowledges	support	from	grant	DGAPA-PAPIIT	106518,	and	from	program	DGAPA-PASPA. T.H acknowledges the support from the National Research Foundation of Korea (NRF) grant funded by the Korea government (MSIT) (2019R1A2C1087045). CRA acknowledges financial support from the Spanish Ministry of Science and Innovation (MICINN) through project PN AYA2013-47742- C4-2-P. CRA also acknowledges the Ramon y Cajal Program of the Spanish Ministry of Economy and Competitiveness. This work is based on observations obtained with the \emph{Spitzer Space
  Observatory}, which is operated by JPL, Caltech, under NASA contract
1407. This research has made use of the NASA/IPAC Extragalactic
Database (NED) which is operated by JPL, Caltech, under contract with
the National Aeronautics and Space Administration. CASSIS is a
product of the Infrared Science Center at Cornell University,
supported by NASA and JPL.



\newpage
\section{Appendix A: Spectral Decomposition}

In order to select those type 1 AGNs in which the IRS/{\it Spitzer} spectrum is mostly ($>80\%$) dominated by emission from dust heated by the AGN we use the spectral decomposition tool {\sc deblendIRS} from \citet{Hernan-Caballero15}. 

{\sc deblendIRS} is a spectral decomposition tool that uses a set of starburst, stellar and AGN templates with IRS/{\it Spitzer} spectrum. This spectral decomposition assumes that the spectral shape of the AGN and its host galaxy are found in others sources where the emission from the AGN or host galaxy completely dominate the spectral emission. The spectral emission of the host galaxy is composed by the stellar emission (passive stellar population) and the emission from the interstellar medium (ISM), which is call PAH, from Polycyclic Aromatic Hydrocarbon, since the emission of these molecules is relate with the presence of young star forming regions (starburst, SB). The spectral decomposition is carry on trying every possible combination between stellar, PAH, and AGN, according to the following linear spectral combination:
\begin{equation}
f_{i,j,k}(\lambda)=af_{i}^{stellar}(\lambda)+bf_{i}^{PAH}+cf_{k}^{AGN}(\lambda),
\end{equation}
where the $i$,$j$ and $k$ indices range all the stellar, PAH, and AGN templates, respectively. The a, b and c coefficients are obtain through the $\chi^{2}$ minimization.  
In Figure~\ref{fig:decomp} we show an example of the spectral decomposition. After find the combination of components that best reproduce the IRS/{\it Spitzer} spectrum {\sc deblendIRS} estimates the fractional contribution of each component to the integrated 5 - 15 $\mu$m luminosity, the luminosity of the starburst at 12$\mu$m, and the luminosity of the AGN at 12, and 6 $\mu$m. Additionally, it gives for the AGN component the silicate strength measured at the wavelength where the silicate feature peaks, and the MIR spectral index measured between 8.1 and 12.5 $\mu$m.
\begin{figure*}
\centering
\includegraphics[scale=0.50]{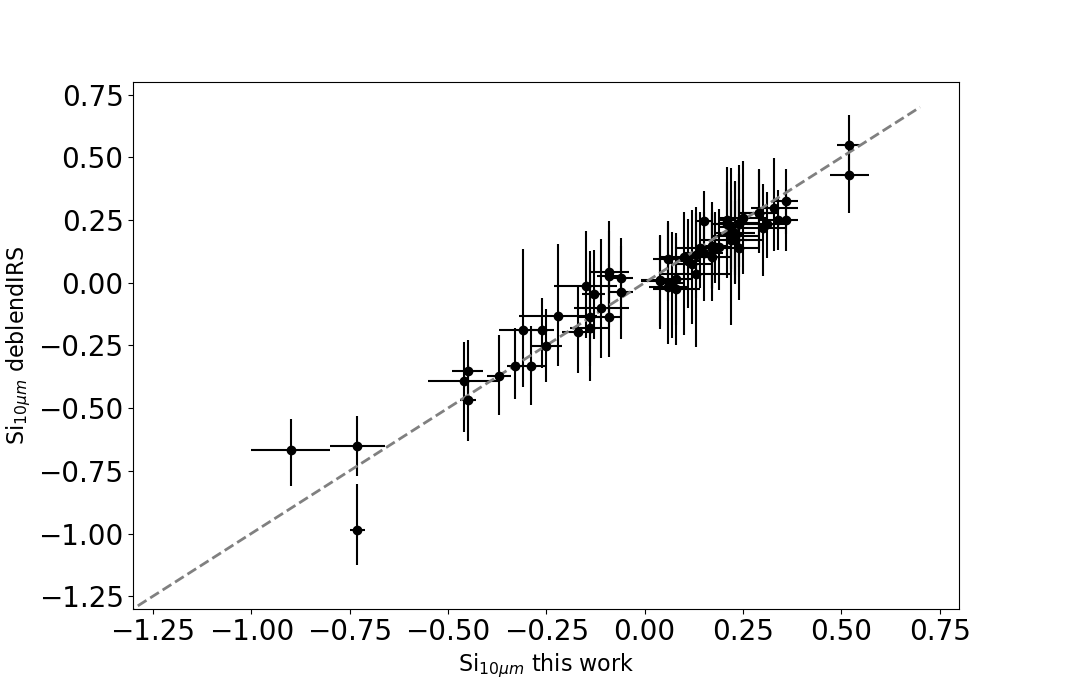}
\caption{10$\mu$m silicate strength as measured using our own methodology (see Section 2.3) and deblendIRS. The grey dashed line represent the 1:1 comparison.}
\label{fig:comparison}
\end{figure*}

\begin{figure*}[htbp]
\centering
\includegraphics[scale=0.8]{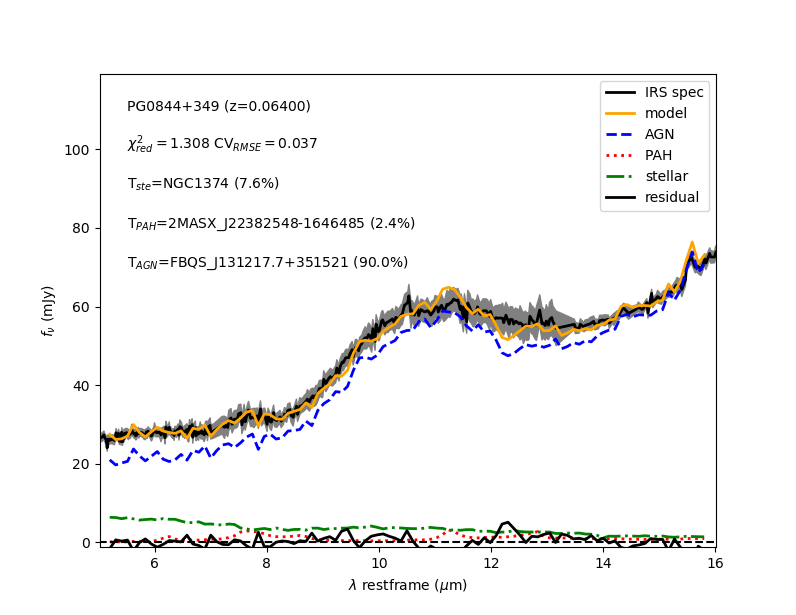}\\
\includegraphics[scale=0.3]{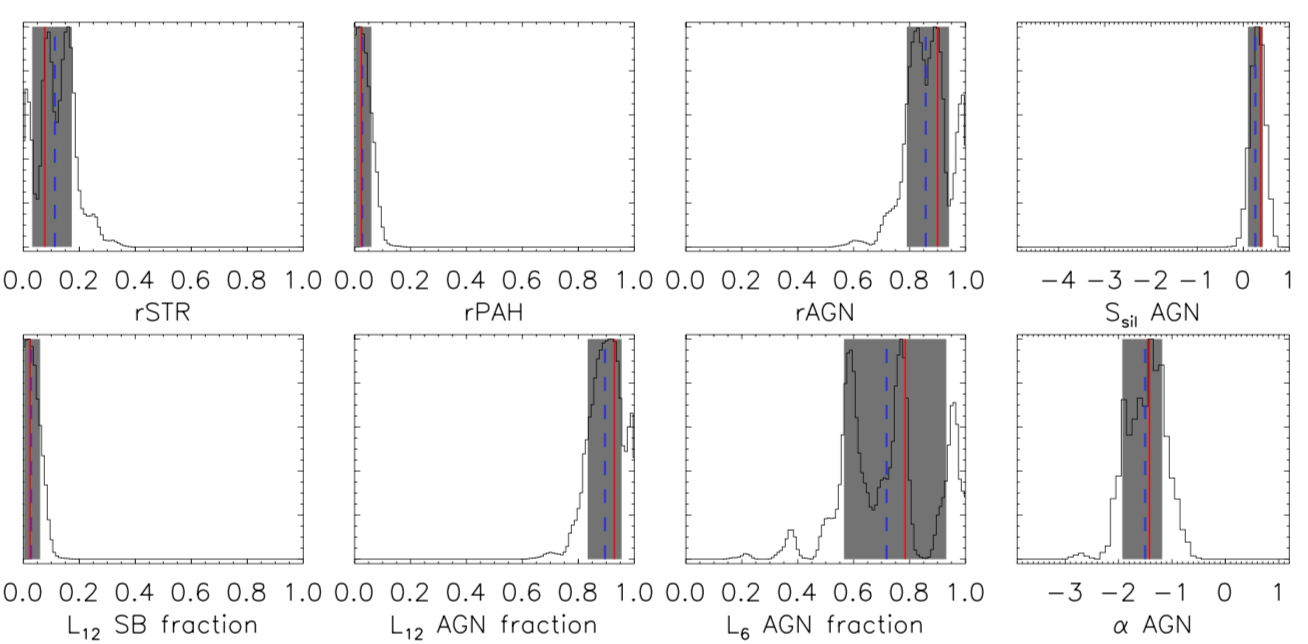}
\caption{{\bf Upper. Spectral decomposition:}the green dot-dashed, blue dashed, and red dotted lines represent the stellar, starburst, and AGN components respectively. The orange solid line is the sum of the three (stellar, starburst, and AGN) components. The black solid line is the resampled IRS/{\it Spitzer} spectrum with their errors (grey shadow). The horizontal solid black line around zero is the residual. {\bf Bottom. Probability distributions:} rSTR, rPAH, and rAGN are the fractional contribution of the stellar, PAH, and AGN components. $L_{12}$ SB fraction, $L_{12}$ AGN fraction, and $L_{6}$ AGN fraction are the monochromatic luminosity of the SB and AGN at 12 and 6 $\mu$m respectively. $S_{sil}$ and $\alpha$ AGN are the silicate strength and the spectral index of the AGN component.}
 \label{fig:decomp}
\end{figure*}
\newpage
\section{Appendix B: Silicate strength measurements}
Here we report the silicate feature strengths measurements for the full sample.
\label{Measurements}
\begin{table*}[htbp]
	\begin{minipage}{1.\textwidth}
		\caption{{\bf Silicate feature strengths measured from the IRS/{\it Spitzer} spectrum}. Columns 1 lists the name. Columns 2 and 3 list the wavelength where the $10\mu$m silicate emission feature peaks, and the $10\mu$m silicate strength. Columns 4 and 5 are like columns 2 and 3 but for the $18\mu$m silicate emission feature. Columns 6, 7, 8, and 9 are like columns 2, 3, 4, and 5 but for the values previously reported in the literature. Column 10 list the reference.\label{tab:silicates_IRS_all}}
\centering
\resizebox{15cm}{!}{
\begin{tabular}{l|ccccccccc}
				\hline
Name &  $\lambda_{p}$& Si$_{10\mu\text{m}}$ &$\lambda_{p}$ & Si$_{18\mu\text{m}}$ & $\lambda_{p}$&Si$_{10\mu\text{m}}$ & $\lambda_{p}$& Si$_{18\mu\text{m}}$ & Ref.\\
     & ($\mu$m)&   & ($\mu$m) & & & & & \\
\hline
NGC7213 &  $10.7\pm0.1$ & $0.52\pm0.05$&    $17.3\pm0.1$ & $0.23\pm0.04$ & 10.1 & 0.6 & 17.9 & 0.16 & 1 \\
3C321 &$9.8\pm0.1$ &$-0.9\pm0.1$ & $18.1\pm0.1$ & $-0.09\pm0.05$  & ...&... & ...& ...& ...\\
3C405 &$9.8\pm0.1$ &$-0.73\pm0.07$ & $18.2\pm0.1$ & $-0.06\pm0.04$  &... & ...& ...& ...& ...\\
3C33	&$9.7\pm0.1$ & $-0.17\pm0.04$ & $17.4\pm0.1$ & $0.12\pm0.04$  &... & ...&... & ...& ...\\
ESO434-G40 &$9.9\pm0.1$ & $-0.37\pm0.03$ & $17.8\pm0.1$ & $0.07\pm0.02$  &... & ...&... & ...& ...\\
IIIZw2 &$10.4\pm0.8$ & $0.04\pm0.05$ & $16.6\pm0.1$ & $0.11\pm 0.03$  & ...&... & ...& ...&... \\
IRAS03450+0055 &$10.3\pm0.1$ & $0.15\pm0.05$ & $16.7\pm0.1$ & $0.08\pm0.05$  &... &... &... &... &... \\
IRAS05218-1212 &$9.2\pm0.1$& $-0.06\pm0.03$& $17.8\pm0.1$& $0.13\pm0.04$  & ...& ...& ...& ...&... \\
MCG-6-30-15	& $9.0\pm0.1$ &$-0.13\pm0.03$ & $17.3\pm0.1$ & $0.07\pm0.04$&10.2 &0.02 &18.3& 0.10& 1\\
MRK1218 & $9.2\pm0.1$ & $-0.22\pm0.1$ & $16.5\pm0.4$ & $0.24\pm0.07$  & ...&... & ...& ...& ...\\
Mrk10	&$9.4\pm0.1$ &$-0.14\pm0.05$ & $17.4\pm0.1$ & $0.1\pm0.06$  & ...& ...&... & ...& ...\\
Mrk110 &$10.3\pm0.1$ & $0.19\pm0.03$ & $18.0\pm0.1$ & $0.25\pm0.03$  & ...&  ...& ...& ...& ...\\
Mrk176 & $11.6\pm0.2$ & $0.07\pm0.03$ & $17.3\pm0.1$ & $0.18\pm0.04$  & ...&... & ...& ...& ...\\
Mrk231 & $9.8\pm0.1$ & $-0.73\pm0.02$ & $17.3\pm0.1$ & $-0.18\pm0.02$ &9.8 &-0.62 & 17.9&-0.23 & 1\\
Mrk3	  &$9.8\pm0.1$ & $-0.45\pm0.04$ & $17.9\pm0.1$ & $0.19\pm0.02$  & ...&... &... &... & ...\\
Mrk348	& $9.4\pm0.1$ & $-0.29\pm0.04$ & $16.6\pm0.1$ & $0.18\pm0.03$  & ...& ...& ...& ...& ...\\
Mrk463E & $9.8\pm0.1$ & $-0.45\pm0.02$ & $18.0\pm0.1$ & $0.05\pm0.03$  & 9.8&-0.4 &18.3 &0.09 & 2 \\
Mrk50	& $10.8\pm0.1$ & $0.22\pm0.08$ & $17.9\pm0.1$ & $0.27\pm0.26$  &10.8 &0.24 & 18.3& 0.37&3 \\
Mrk573	& $9.3\pm0.1$&$-0.14\pm0.03$& $17.3\pm0.1$ & $0.11\pm0.02$  & 9.4&-0.10 &17.2 & 0.04&3 \\
Mrk734	& $10.7\pm0.1$ & $0.06\pm0.04$ & $18.0\pm0.1$ & $0.09\pm0.04$  &... &... & ...&... &... \\
Mrk915 & $9.4\pm0.1$ & $-0.11\pm0.07$ & $17.3\pm0.1$ & $0.18\pm0.05$  &10.4 &0.05 &17.1 &0.27 &3 \\
NGC3081 &$9.3\pm0.1$ &$-0.26\pm0.03$ & $17.3\pm0.1$ & $0.12\pm0.06$  & ...& ...& ...& ...& ...\\
NGC7212 & $9.9\pm0.1$ &$-0.46\pm0.09$ & $17.3\pm0.1$ & $0.18\pm0.04$  & ...&... &... & ...& ...\\
NGC788	& $9.4\pm0.1$ & $-0.25\pm0.04$ & $16.7\pm0.1$ & $0.14\pm0.03$  & 9.5& -0.08& 20.5&0.02 & 3\\
PG1149-110 & $11.6\pm0.1$ & $0.06\pm0.05$ & $17.3\pm0.1$ & $0.13\pm0.03$  &... & ...& ...&... & ...\\
PG1244+026 & $10.9\pm0.1$ &$0.14\pm0.06$ & $17.3\pm0.1$ & $0.06\pm0.04$  & ...&... &... &... & ...\\
PKS2048-57 & $9.9\pm0.1$ & $-0.33\pm0.02$ & $15.3\pm0.1$ & $0.02\pm0.02$  & ...&... & ...& ...& ...\\
TON1542 & $10.9\pm0.1$ & $0.08\pm0.06$ & $17.8\pm0.1$ & $0.12\pm0.03$ & ...& ...& ...& ...& ...\\
UGC3601 & $11.6\pm0.1$ & $0.1\pm0.06$ & $16.6\pm0.1$ & $0.20\pm0.08$  & ...& ...& ...& ...& ...\\
UM614 & $10.9\pm0.1$ & $0.12\pm0.05$ & $17.6\pm0.1$ & $0.19\pm0.1$  & 10.7&0.17 &17.2 &0.16 &3 \\
B3-0754+394& $10.1\pm0.2$ & $0.13\pm0.09$ & $17.7\pm0.1$ & $0.13\pm0.17$  & ...& ...& ...& ...& ...\\
F9	 & $10.7\pm0.1$ & $0.13\pm0.02$ & $17.4\pm0.1$ & $0.13\pm0.03$  & ...&... & ...& ...& ...\\
IIZw136 & $9.2\pm0.1$ &$-0.06\pm0.03$ & $17.9\pm0.1$ & $0.09\pm0.04$  & ...& ...& ...& ...& ...\\
IZw1	& $9.7\pm0.1$ & $0.22\pm0.03$ &$17.8\pm0.1$ & $0.08\pm0.03$  & ...& ...& ...& ...& ...\\
PG0007+106/Mrk1501 & $11.6\pm0.1$ & $0.04\pm0.05$ & $16.3\pm0.3$ & $0.10\pm0.03$  & ...& ...& ...&... &... \\
PG0923+129/Mrk705$^{a}$ & $9.2\pm0.1$ & $-0.09\pm0.03$ & $17.3\pm0.1$ & $0.16\pm0.03$  & ...& ...& ...& ...& ...\\
PG1211+143 & $10.7\pm0.1$ & $0.23\pm0.03$ & $17.6\pm0.1$ & $0.15\pm0.04$  & ...&... & ...& ...&... \\
PG1229+204/Mrk771 & $9.2\pm0.1$ & $-0.15\pm0.08$ & $19.1\pm0.2$ & $0.07\pm0.03$  & ...& ...&... & ...&... \\
PG1411+442 & $10.8\pm0.1$ & $0.17\pm0.04$ & $18.4\pm0.2$ & $0.1\pm0.03$  & ...& ...& ...& ...& ...\\
PG1426+015/Mrk1383 & $10.7\pm0.1$ & $0.17\pm0.04$ & $17.4\pm0.1$ & $0.15\pm0.03$  & ...& ...& ...& ...&... \\
PG1440+356/Mrk478 & $9.3\pm0.1$ & $-0.09\pm0.05$ & $17.3\pm0.1$ & $0.07\pm0.02$  & ...& ...& ...& ...&... \\
PG1448+273 & $11.6\pm0.1$ & $0.08\pm0.04$ & $17.7\pm0.1$ & $0.14\pm0.03$  & ...&... & ...& ...& ...\\
PG1501+106/Mrk841 &$9.4\pm0.1$  & $-0.09\pm0.03$ & $18.2\pm0.1$ & $0.14\pm0.03$ & ...& ...& ...&...  & ...\\
PG1534+580/Mrk290 & $10.7\pm0.1$ & $0.11\pm0.03$ & $17.9\pm0.1$ & $0.15\pm0.03$  & ...&... &... &... & ...\\
ESO198-G24 & $9.9\pm0.1$ & $0.23\pm0.05$ & $16.3\pm0.1$ & $0.25\pm0.06$  &... &... &... &... & ...\\
ESO548-G81 & $10.7\pm0.1$ &$0.21\pm0.04$ & $16.8\pm0.1$ & $0.22\pm0.04$  & 10.7& 0.31&18.0 &0.17 & 3\\
NGC2110 & $10.7\pm0.1$ & "$0.22\pm0.03$ & $16.5\pm0.1$ & $0.16\pm0.02$  & ...& ...& ...& ...& ...\\
Mrk486/PG1535+547 & $10.9\pm0.1$ &$0.17\pm0.05$ & $17.4\pm0.1$ & $0.10\pm0.06$  & ...&... &... & ...& ...\\
3C120$^{b}$ & $10.7\pm0.1$ & $0.19\pm0.03$ & $17.5\pm0.1$ & $0.13\pm0.04$  &10.2 & 0.26& 17.8&0.15 & 1\\
NGC1275 & $10.6\pm0.3$ & $0.15\pm0.02$ & $17.9\pm0.1$ & $0.09\pm0.03$  & ...& ...& ...& ...&... \\
NGC7603 & $11.0\pm0.1$ &$0.21\pm0.02$ & $16.5\pm0.1$ & $0.17\pm0.03$  &10.2 &0.12 &17.7 &0.13 & 1\\
NGC7603$^{c}$ &  & &  &  &11.7 &0.16 &17.1 &0.13 & 3\\
MRK1018 & $10.2\pm0.1$ & $0.23\pm0.06$ & $17.7\pm0.1$ & $0.16\pm0.05$  & ...&... &... & ...& ...\\
3C382 & $10.7\pm0.1$ & $0.24\pm0.05$ & $17.5\pm0.1$     &$0.20\pm0.03$& ... &... & ...&... & ...\\
PG1404+226 & $10.0\pm0.1$ & $0.24\pm0.06$& $16.6\pm0.1$ & $0.2\pm0.08$& ...  & ...& ...& ...& ...\\
PG2209+184$^{a}$ & $11.0\pm0.1$ & $0.25\pm0.06$ & $16.4\pm0.1$ & $0.21\pm0.10$&  ... &... & ...&... & ...\\
UGC12282 &Sy1.9  & 0.0169     &   42.8$^{*}$   & 10& $9.2\pm0.1$ & $-0.31\pm0.06$ & $17.3\pm0.1$ & $0.15\pm0.07$  \\
				\hline
				\hline
			\end{tabular}}\\
			Note.- Ref.: (1): \citet{Thompson09}, (2): \citet{Sirocky08}, (3): \citet{Mendoza-Castrejon15}. $^{a}$The band 1 range from $8.0-8.5\,\mu$m. $^{b}$The band 1 range from $8.0-8.8\,\mu$m. $^{c}$The band 1 and 2, range from $8.0-8.8\,\mu$m, and $14.0-14.5\,\mu$m, respectively. 
		\end{minipage}
	\end{table*}

\newpage
\begin{figure*}[htbp]
\begin{tabular}{cc}
\includegraphics[scale=0.3]{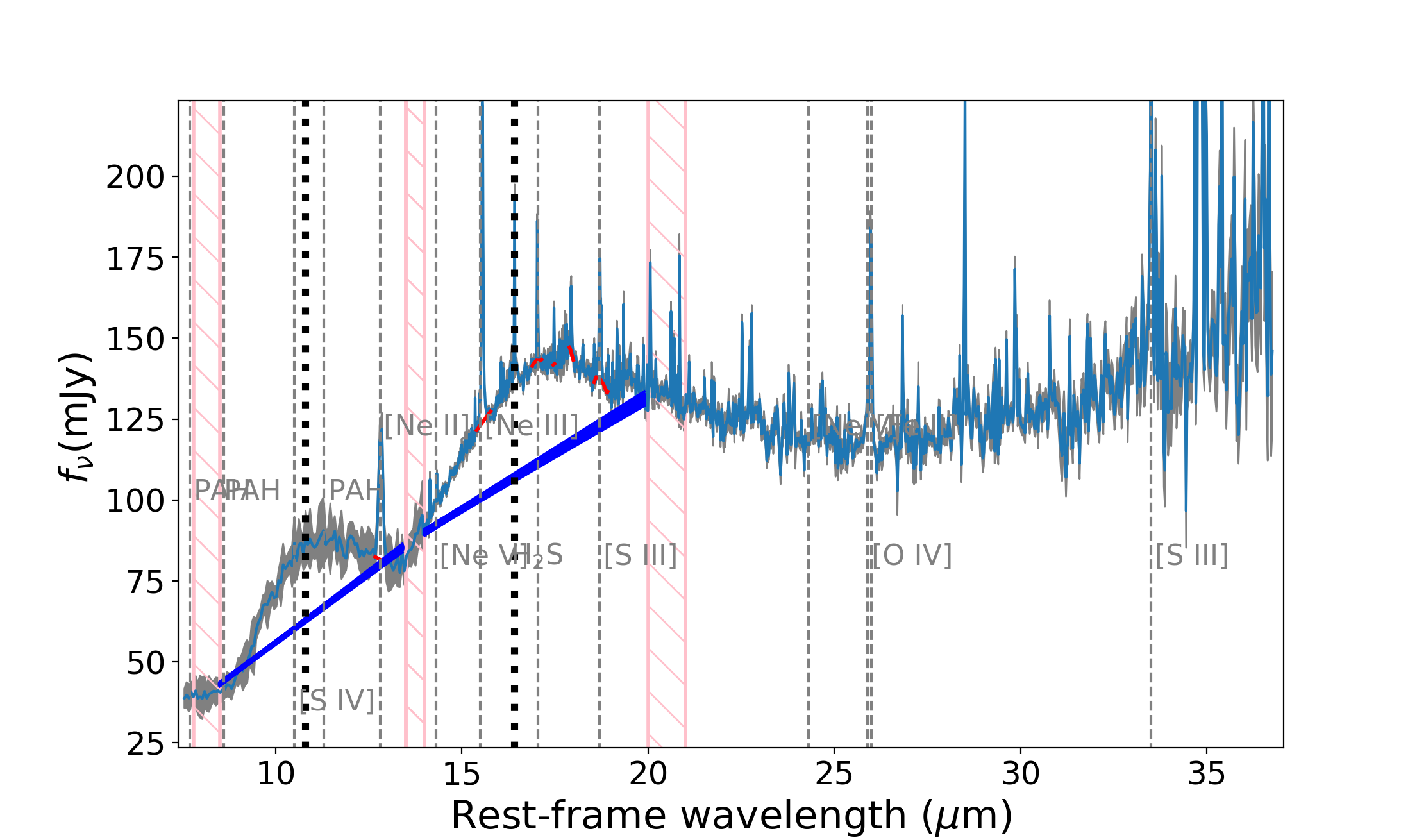} &  \includegraphics[scale=0.3]{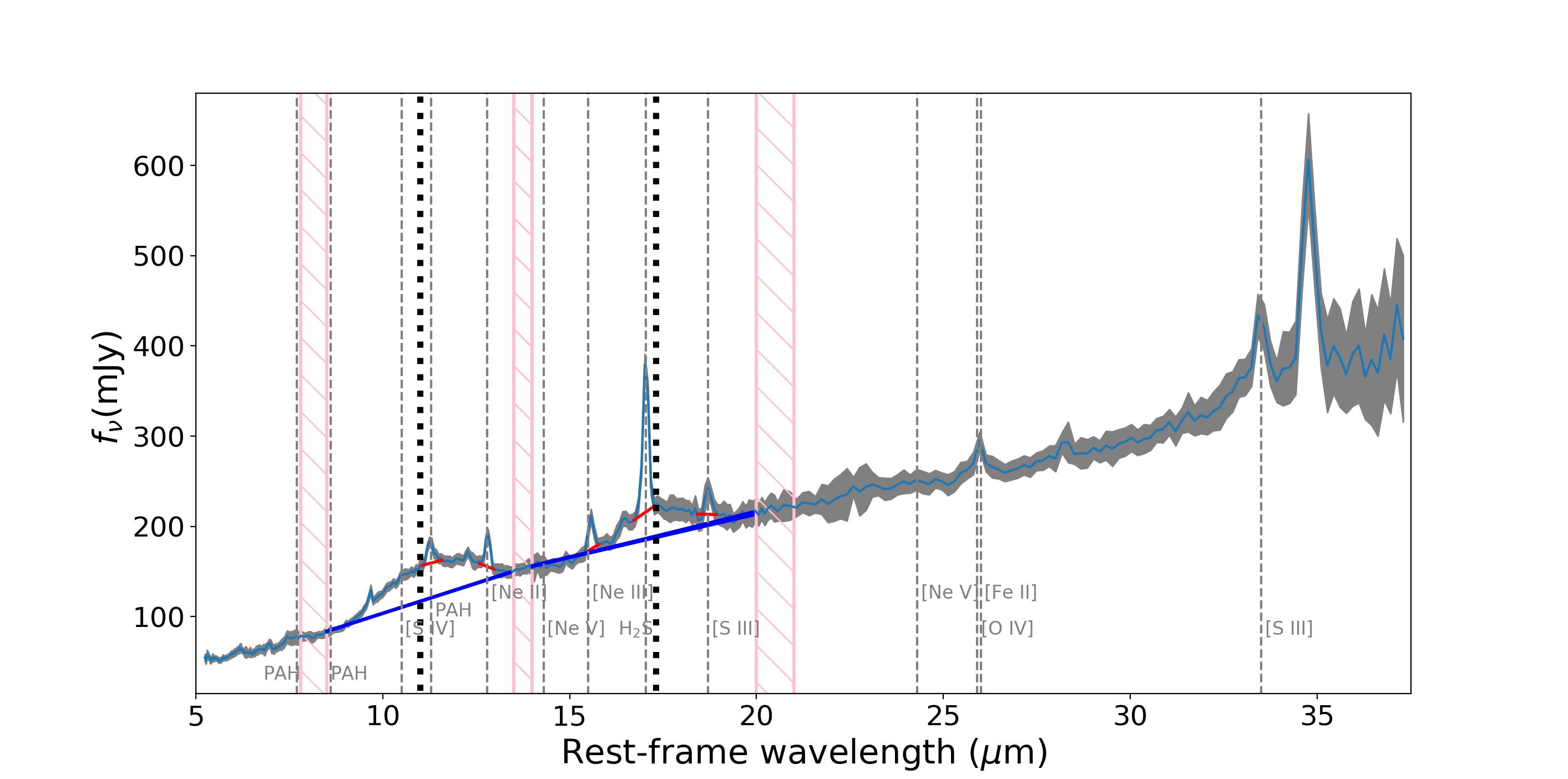}\\
 
\includegraphics[scale=0.3]{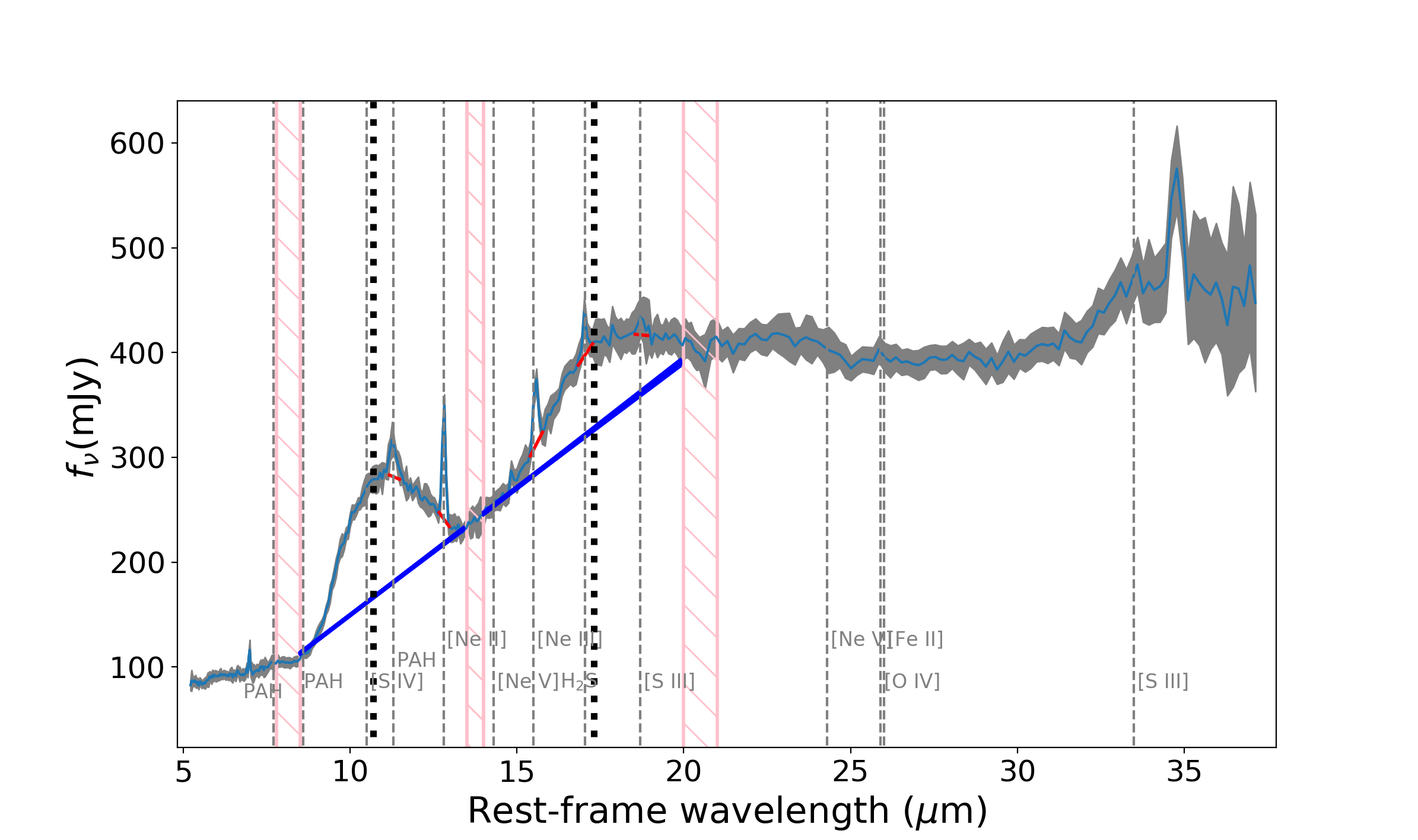} &
\includegraphics[scale=0.3]{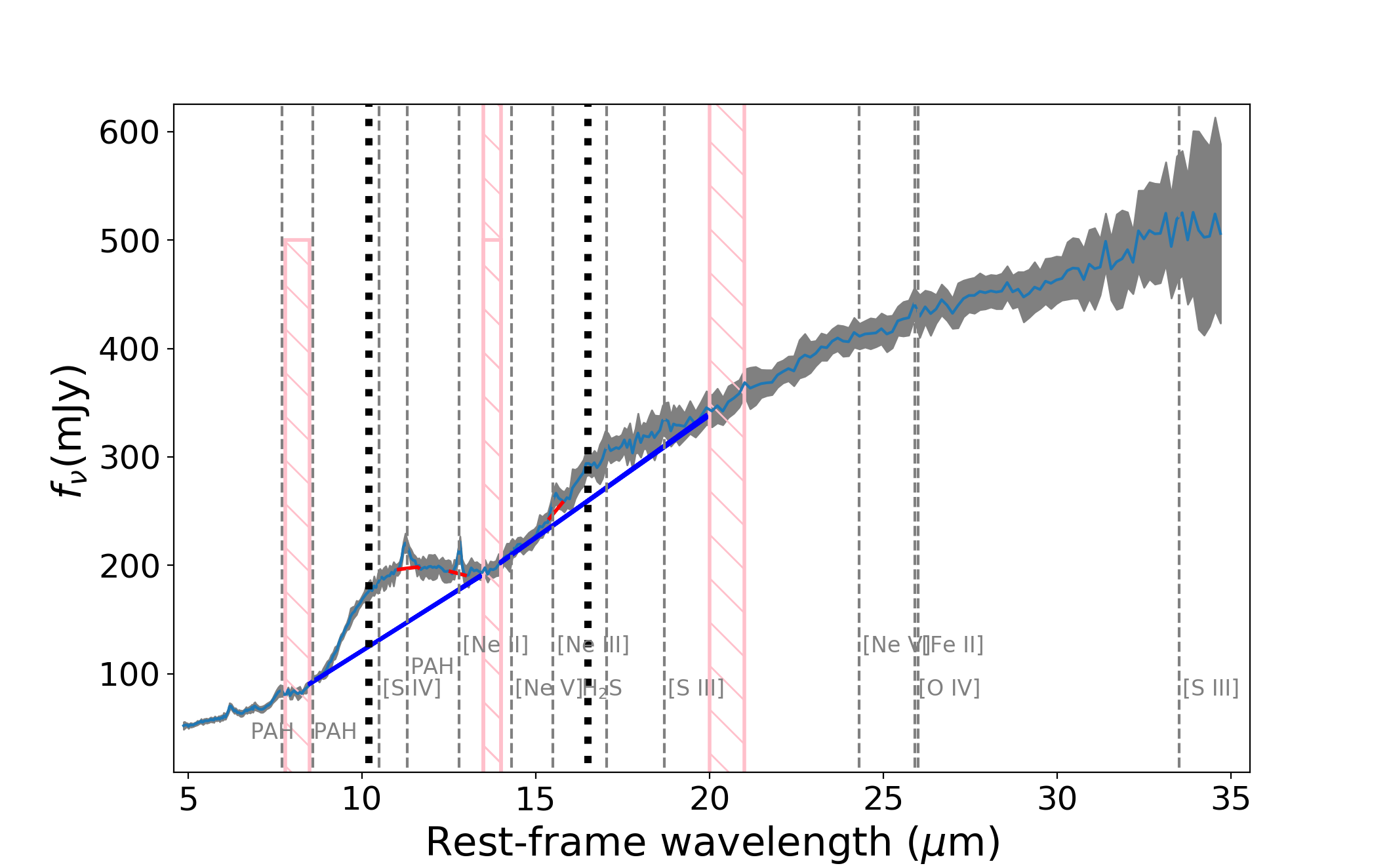} \\
 
\includegraphics[scale=0.3]{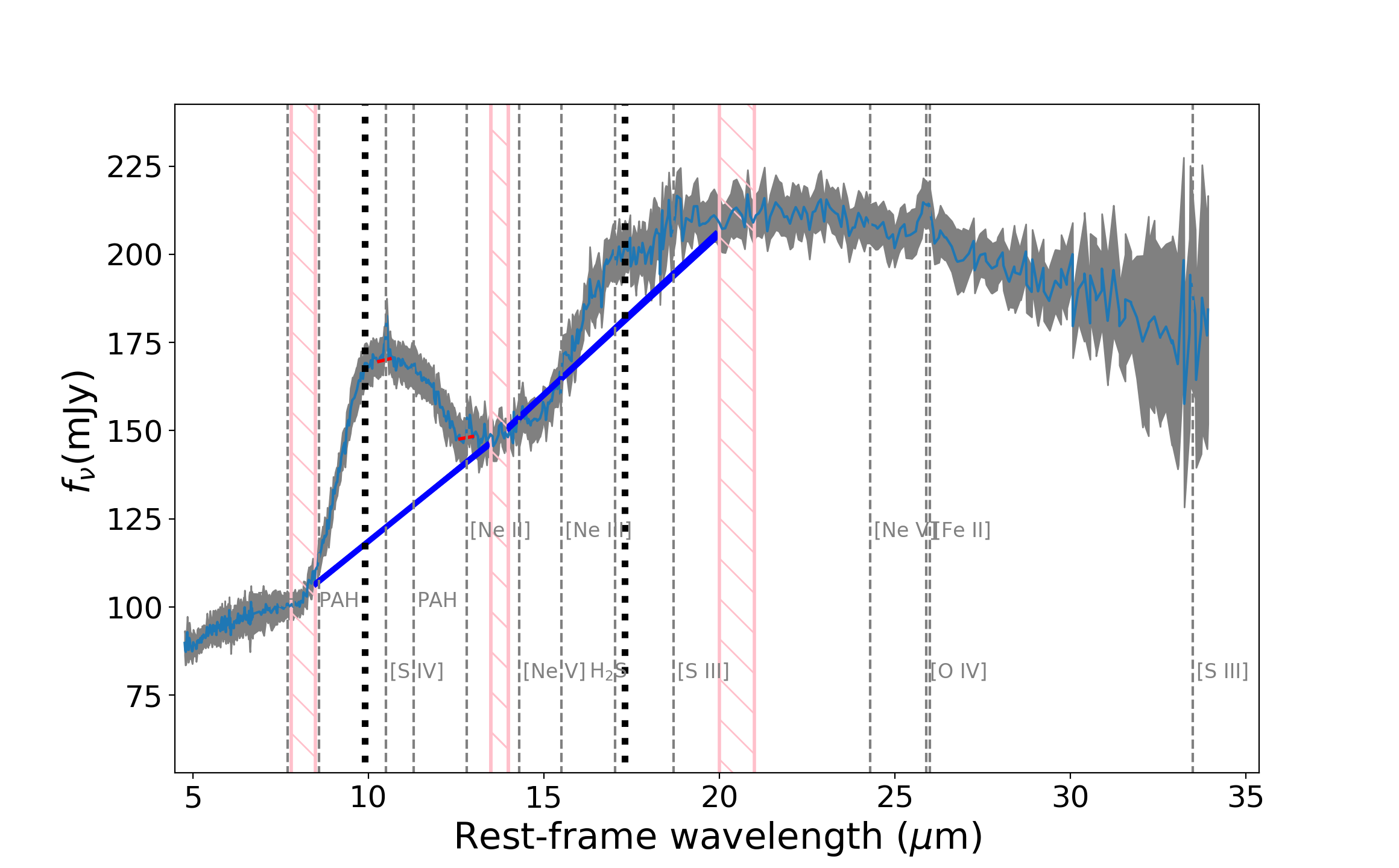} &  \includegraphics[scale=0.3]{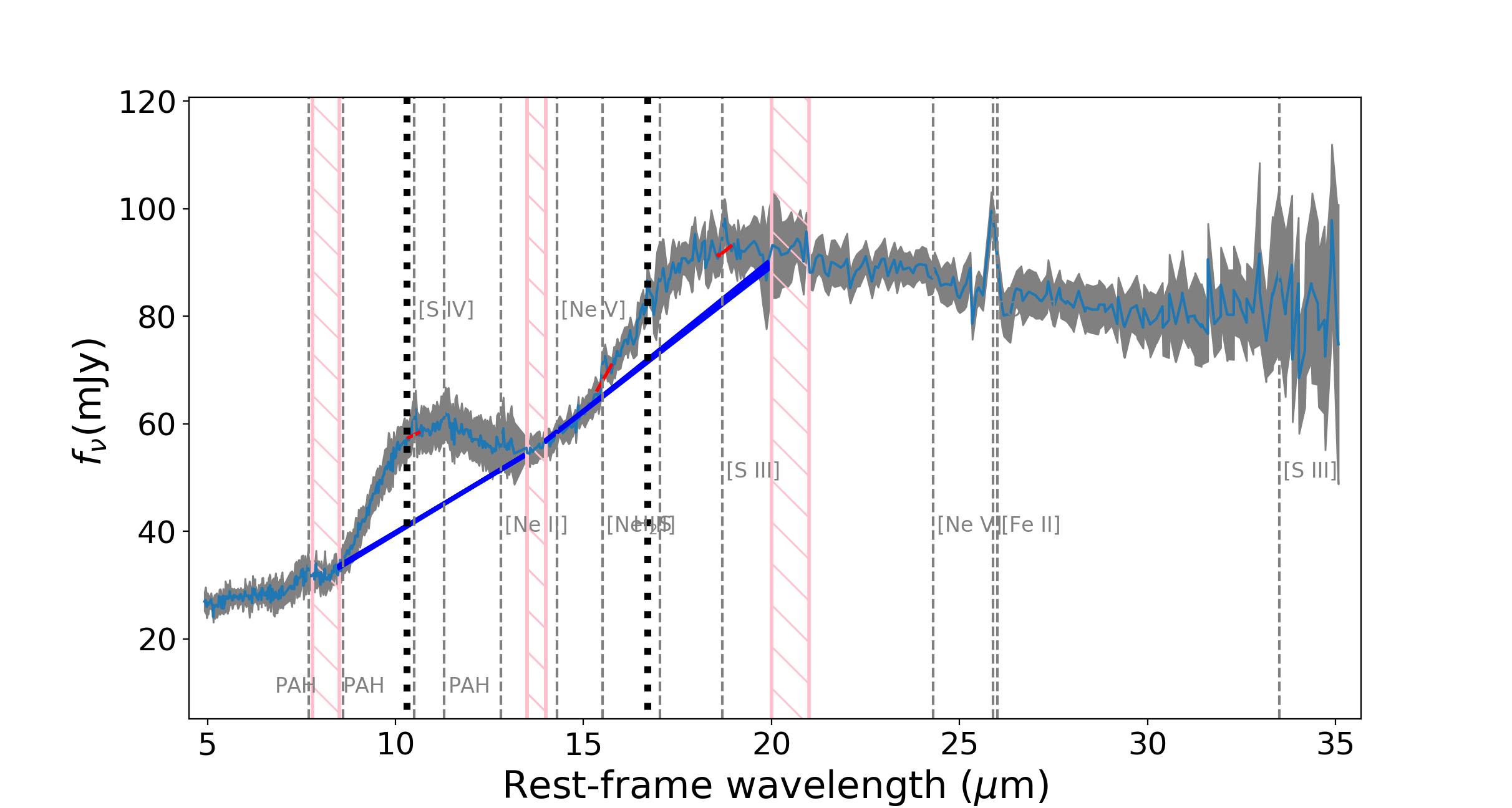}\\

 \end{tabular}
  \caption{IRS/{\it Spitzer} spectrum (light blue solid line). Upper panel: NGC3998 (left), NGC4258 (right). Middle panel: NGC7213 (left), OQ 208 (right). Bottom panel: PG0804+761 (left), PG0844+349 (rigth). The red line is the local continuum that follows the broad features of the IRS/{\it Spitzer} spectrum. The blue solid lines are the bootstrapped local continua and the vertical pink dashed bars are the bands used to fit the continua around the features. The vertical black dashed-lines indicate the wavelength where the silicate strength is measured. The vertical grey dashed-lines mark other emission lines. For NGC3998 the spectrum between 14 and 35$\mu$m is the high angular resolution spectrum from {\it Spitzer}.}
 \label{fig:silicate_measurements}
\end{figure*}

\begin{figure*}[htbp]
\begin{tabular}{cc}
\includegraphics[scale=0.3]{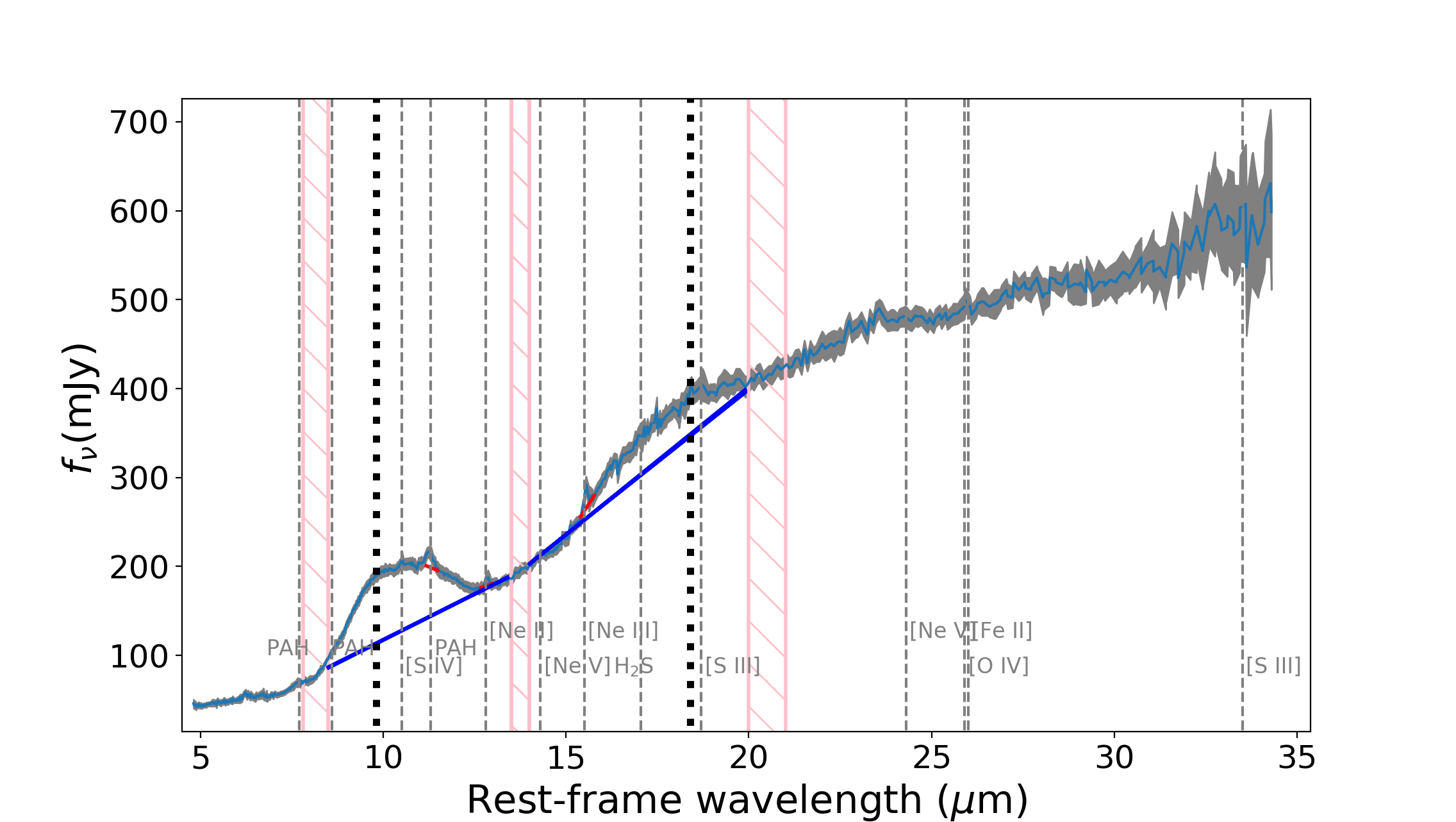} &
\hspace{-1cm}\includegraphics[scale=0.27]{PG2214+139}\\

\includegraphics[scale=0.3]{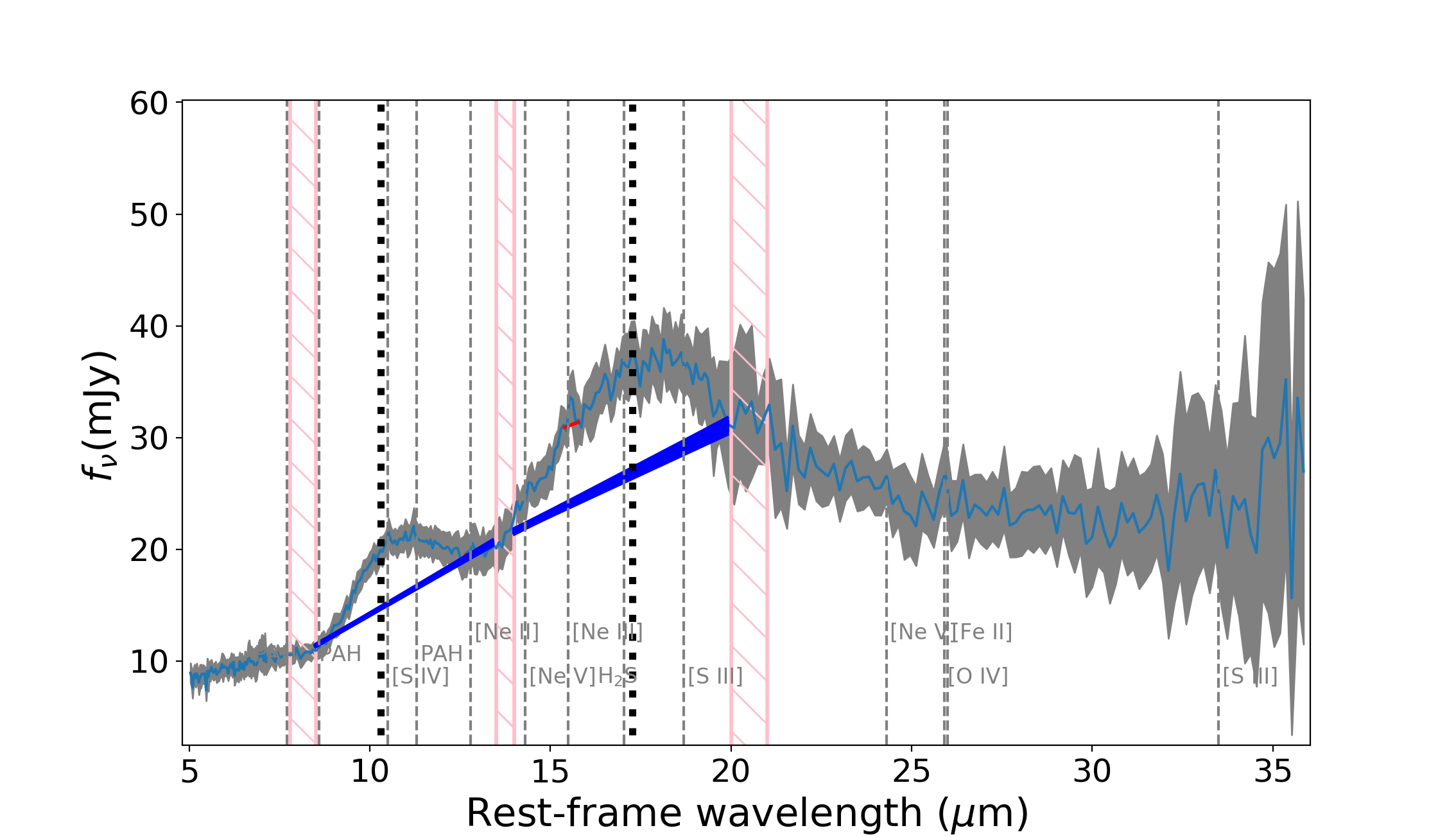}&\hspace{-1cm}\includegraphics[scale=0.3]{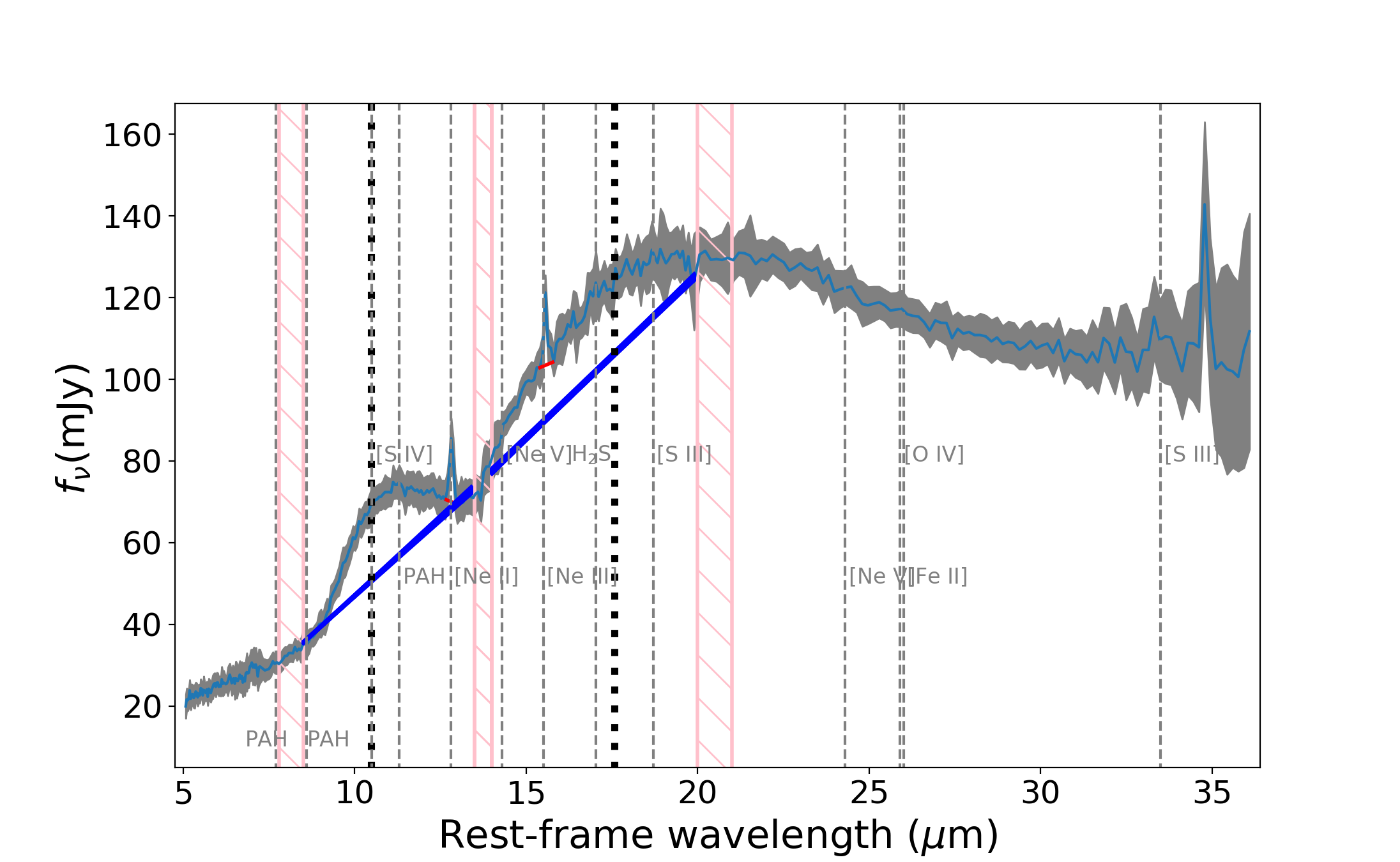}\\
 \end{tabular}
  \caption{IRS/{\it Spitzer} spectrum. Upper panel: PG1351+640 (left), PG2214+139 (right). Bottom panel: PG2304+042 (left), PKS0518-45 (right). Lines as in Figure \ref{fig:silicate_measurements}.}
\end{figure*}

\newpage
\section{Appendix C: Tables of parameters}
Here we report the $\chi^{2}$ and degree of freedom for each object in the Si-s sample fitted with the model combination: AGN, AGN+Stellar, AGN+HII, and AGN+Stellar+HII. We also report the set of parameters obtained from the best fit, and the plots from modeling the non-stellar IRS/{\it Spitzer} spectrum of all objects in the Si-s sample.

\begin{table*}[htbp]
	\begin{minipage}{1.\textwidth}
		\caption{All fitting from smooth models of \citet{Fritz06}, clumpy models of \citet{Nenkova08a,Nenkova08b} and \citet{Hoenig10}, and disk+outflow models of \citet{Hoenig17}. Column 1 lists the name and models. Column 2, 3, 4 and 5 the $\chi^{2}$ and dof of each model combination: C1 indicate AGN model (torus or disk+outflow), C2  AGN+Stellar, C3 AGN+HII, and C4 AGN+Stellar+HII . Column 6 lists the combination of components that best fit the IRS/{\it Spitzer} spectrum.  \label{tab:all_fitting}}
\centering
\resizebox{16cm}{!} {
\begin{tabular}{l|ccccc|l|ccccc}
				\hline
   &    \multicolumn{4}{c}{$\chi^{2}$/dof}   &   &    \multicolumn{4}{c}{$\chi^{2}$/dof}\\
Name  &  C1            & C2  & C3 & C4 &Best & Name  &  C1            & C2  & C3 & C4 &Best\\
\hline
\hline
{\bf NGC7213} &  &  &  &  & &{\bf PG2214+139}&  &  &  &  & \\ 
Fritz+06	&NM&	NM &NM&NM&...& Fritz06 & NM & 187.3/123 & NM & 187.3/122 & C2\\
Nenkova08 & 169.6/114 & 125.1/113 & 169.6/113 & 125.0/112 & C2 & Nenkova+08&NM&236.4/123&NM&236.4/122&C2\\
Hoenig10 & NM & NM & NM & NM & ...& Hoenig+10&NM&NM&NM&84.6/120&C4\\
Hoenig17 & NM & NM & NM & NM & ...& Hoenig+17&NM&NM&NM&22.9/117&C4\\
Stalevski16 & NM & NM & NM & NM & ...& Stalevski16 & NM &  181.6/120 &  NM&  107.9/119& C4 \\
{\bf PG2304+042}	 &  &  &  &  & & {\bf PG0804+761}&  &  &  &  & \\
Fritz+06				& NM & 224.7/117 & NM & 224.7/116 & C2& Fritz+06&124.0/125 & 77.9/124& 124.0/124& 77.9/123&C2\\
Nenkova+08			& 184.9/118 & 146.6/117 & 185.0/117 & 146.5/116 & C2 & Nenkova+08&NF&198.8/124&NF&190.2/123&C2\\
Hoenig+10				& 140/119 & 82.5/68 & NM & 75.9/117 &C1 & Hoenig+10&NM&NM&NM&222.5/124&C4\\
Hoenig+17				& 77.6/116 & 65.7/115 & 76.9/115 &  65.7/114&C2 & Hoenig+17 &89.0/123& 89.0/122&59.6/122&59.6/121&C3\\
Stalevski16        & NM       & 232.2/117 & NM      & NM &C2&Stalevski16 & NM &137.5/124 & NM & 136.0/123 & C2 \\
{\bf PKS0518-45}			&  &  &  &  & & {\bf OQ208} &  &  &  &  & \\
Fritz+06	& NM & NM & NM & NM & ...& Fritz+06 &NM& NM& NM& NM&...\\
Nenkova+08	& 178/9118&	150.8/117&178.9/117&151.0/116&C2& Nenkova+08 &152.4/124&45.7/123&184.4/123&39.0/122&C4\\
Hoenig+10	&87.4/119&	59.7/118&87.4/118&54.0/117&C4& Hoenig+10 &NM &NM &NM &NM&...\\
Hoenig+17	&36.1/116&	36.4/115&35.7/115&35.5/114&C1& Hoenig+17& NM& 215.5/121 & NM & 211.1/120&C2\\
Stalevski16 & 194.8/118 &191.1/117 & 196.4/117 &191.1/116 & C1&Stalevski16& NM &NM &NM &NM&...\\
{\bf PG0844+349}	 &  &  &  &  & & {\bf NGC4258}& &  &  &  \\
Fritz+06&	241.4/121	&144.2/120& NM&144.2/119&C2& Fritz+06 &NM &NM &NM &NM&...\\
Nenkova+08&162.5/121&37.9/120&162.5/120&37.4/119&C2& Nenkova+08& NM& 122.4/110& 82.8/110& 85.0/109&C4\\
Hoenig+10	&NM&	57.7/121&	NM&24.2/120&C4& Hoenig+10 &NM& NM& NM & 180.1/110& C4\\
Hoenig+17&	29.2/119&	27.1/118&20.3/118&14.9/117&C4& Hoenig+17 &NM& NM& 202.5/108& 202.8/107&C3\\
Stalevski16 &NM&97.2/120&NM&97.1/119 & C2&Stalevski16 &NM &NM &NM &NM &... \\
{\bf PG1351+640}				&  &  &  &  & & {\bf NGC3998}&  &  &  &  & \\
Fritz+06&	NM&	187.3/123	&NM	&187.3/122&C2& Fritz+06 &NM &137.8/72 &137.8/72 &137.8/71&C2\\
Nenkova+08&	NM	&236.4/123&	NM	&236.4/122&C2& Nenkova+08 &30.9/73 &23.3/72 &27.1/72 &21.8/71&C2\\
Hoenig+10	&NM&	NM	& NM&	NM&...& Hoenig+10 &29.5/74 &27.2/73 &28.8/73 &28.7/72&C1\\
Hoenig+17	&NM &	NM&NM	&NM&...& Hoenig+17 &27.9/71 &25.3/70 &24.6/70 &23.0/69&C1\\
Stalevski16 &NM &	NM&NM	&NM&...& Stalevski16 & NM& NM& NM& NM& ...\\
\hline
\end{tabular}}\\
\end{minipage}
Note.-NM = Non-modeled, indicate that the spectrum is not fitted by this component or combination of components.
\end{table*}

\begin{table*}[htbp]
\begin{minipage}{1.\textwidth}
\caption{{\bf Derived parameters from smooth models of \citet{Fritz06}}. Column 1 lists the name of the object modeled. Columns from 2 to 7 list the torus parameters that best reproduce the AGN-dominated IRS/{\it Spitzer} spectrum: the viewing angle $i$, the torus angular width $\sigma$ deg, polar index $\gamma$ and radial index $\beta$ of the gas density distribution $\rho(r,\Theta)\propto r^{\beta}e^{-\gamma\times cos(\Theta)}$ within the torus, the radial extend $Y=R_{outer}/R_{inner}$, and the optical depth $\tau_{9.7\,\mu m}$. Column 7 lists the covering factor which is derived using the $i$, $\gamma$, and $\tau$ \citep[see equation in][]{Gonzalez-Martin19b}  \label{tab:parameters_Fritz06}}
\centering
\resizebox{16cm}{!} {
\begin{tabular}{l|cccccc|c}
				\hline
		 & \multicolumn{6}{c}{Parameters}&\\
		 \hline
Name  & $i$ (min;max) &$\Theta$(min;max) & $\gamma$(min;max)&$\beta$(min;max)&Y (min;max)&$\tau$ (min;max) & Covering Factor \\
        & [0 - 90] deg &[20 - 60]& [0 - 6] &[$-1.0$ - 0.0]& [10 - 150] &[0.3 - 10] &[0 - 1]\\
\hline
\hline
PG2304+042  & 0.0$^{*}$ & 20.0$^{*}$ & 6.0$^{*}$ & -0.01$^{*}$& 10.0$^{*}$& 9.8 (9.1;10.0)& 0.2$^{*}$\\  
PG~0844+349 & 12.9 (11.8;14.9) & 20.0$^{*}$& 6.0$^{*}$ & -0.25 (-0.26;-0.20) & 11.8 (11.6;12.0) & 6.0 (5.9; 6.2)&0.1$^{*}$\\
PG~1351+640 & $>84.5$& $<21.0$& 6.0$^{*}$& -0.01$^{*}$& 35.9 (35.6;36.1)& 7.4 (6.9; 7.6)&0.1$^{*}$\\
PG~2214+139 & 70.0 (66.3; 71.6)& $<22.1$& 0.0$^{*}$& -0.5 (-0.6;-0.4)& 10.4 (10.2; 10.7)& 1.0 (0.9; 1.1)&0.6$^{*}$\\
PG~0804+761 & 44.3 (34.1; 51.9)& $<24$& $>5.7$& -0.7 (-0.8;-0.7) & 10.0$^{*}$ & 3.0 (2.7; 3.2)&$>0.1$\\
NGC~3998    & 0.0$^{*}$& $<21$& $>5.9$& -0.01$^{*}$& $>142$& 2.1 (2.1; 2.2)&0.1$^{*}$\\
				\hline
			\end{tabular}}\\
		\end{minipage}
		{\bf Note}.-$^{*}$Parameter unrestricted.
	\end{table*}

\begin{table*}[htbp]
\begin{minipage}{1.\textwidth}
\caption{{\bf Derived parameters from clumpy models of \citet{Nenkova08a, Nenkova08b}}. Column 1 lists the name of the object modeled. Columns from 2 to 7 list the torus parameters that best reproduce the AGN-dominated IRS/{\it Spitzer} spectrum: the viewing angle $i$, the number of clouds along the equatorial ray $N_{0}$, the angular width $\sigma$ deg, the radial extend $Y$, the index of the radial distribution of clouds $q$, and the optical depth $\tau_{V}$. Column 7 lists the covering factor which is derived using the $i$, $N_{0}$, and $\Theta$ \citep[see equation in][]{Gonzalez-Martin19b} \label{tab:parameters_Nenkova08}}
\centering
\resizebox{16cm}{!} {
\begin{tabular}{l|cccccc|c}
				\hline
	 & \multicolumn{6}{c}{Parameters}&\\
	 \hline
Name& $i$ (min;max)& $N_{0}$ (min,max)&$\sigma$ (min,max)& Y (min;max)&$q$ (min,max) & $\tau_{V}$ (min,max) &Covering Factor\\
  & [0 - 90] deg& [1 - 15]&[15 - 70] deg& [5 - 100]&[0.0 - 2.5] & [5 - 300] &[0 - 1]\\
\hline
\hline
NGC7213 &0.01 (0.00;1.47)& 7.0 (6.1;7.3)& $<15.0$& 10.2 (10.0; 10.5)& $<0.01$& 55.1 (51.6; 59.5)& $<0.4$ \\
PG2304+042& 76.3 (74.0;77.4)& 13.1 (10.5; 13.6)& 17.5 (15.0; 21.7)& 10.0 (9.8;10.1) & 0.9 (0.6; 1.0)& 10.0 (10.0;10.3)& 0.5 (04;0.6)\\
PKS0518-45&70.8 (69.0; 75.0)& 12.8 (12.1; 14.1)& 23.6 (20.7; 29.8)& 10.0 (9.9; 10.0)& 0.1 (0.0; 0.5)& 10.0 (10.0 10.6)& 0.6 (0.6;0.8)\\
PG0844+349 & 78.3 (66.4; 90.0) & 6.7 (4.2; 10.9)& 20.3 (15.0; 41.0)& 11.7 (10.5; 14.7)& 0.5 (0.0; 1.2)& 13.9 (10.0; 22.3)& 0.5 (0.3;0.9)\\
PG1351+640 & 0.0$^{*}$& 7.6 (7.3; 10.4)& 45.0 (44.3; 45.3)& 30.3 (29.9; 31.0)& $<0.01$& 15.5 (15.1;15.9)& 0.9$^{*}$ \\
PG2214+139& $>88.6$& 2.7 (2.1; 3.1)& 24.7 (15.1; 39.7)& $>91$& $>2.4$& 36.8 (33.2;39.3)& $>0.5$\\
PG0804+761& $>76.9$& 1.0$^{*}$& 54.5 (27.0; 70.0)& 100.0 (79.2; 100.0)& 2.3$^{*}$& 87.8 (65.9; 100.9)& $>0.5$\\
OQ208 & 25.5 (0.0; 74.0)& 1.3 (1.2; 3.4)& 64.8 (24.4; 70.0)& 11.7 (11.3; 13.5)& 0.01 (0.00; 0.81)& $>265.6$& 0.6 (0.3;0.9)\\
NGC4258& 81.2 (72.8; 86.0)& 1.5 (1.1; 1.8)& $<20.4$& 7.8 (7.2; 8.6)& 0.01 (0.00; 0.05)& $>254.9$& $<0.3$\\
NGC3998& $>87.5$& 1.7 (1.6; 1.8)& $<36.4$& 100.0 (96.8; 100.0)& 0.01 (0.00; 0.02)& 20.1 (16.5; 24.3)& $>0.5$\\
\hline
\end{tabular}}\\
{\bf Note}.-$^{*}$Parameter unrestricted. 
\end{minipage}
\end{table*}

\begin{table*}[htbp]
\caption{{\bf Derived parameters from clumpy models of \citet{Hoenig10}}. Column 1 lists the name of the object modeled. Columns from 2 to 6 list the torus parameters that best reproduce the AGN-dominated IRS/{\it Spitzer} spectrum: the viewing angle $i$, the number of clouds along the equatorial ray $N_{0}$, the angular width $\Theta$, the index of the radial distribution of clouds $a$, and the optical depth per cloud $\tau_{V}$. Column 7 lists the covering factor which is derived using the $i$, $N_{0}$, and $\Theta$ \citep[see equation in][]{ Gonzalez-Martin19b} \label{tab:parameters_Hoenig10}}
\centering
\resizebox{18cm}{!} {
\begin{tabular}{l|ccccc|c}
				\hline
    & \multicolumn{5}{c}{Parameters}&\\
    \hline
Name& $i$ (min;max)&$N_{0}$(min,max)&$\theta$ (min,max)&$a$ (min,max)&$\tau_{V}$(min;max) &Covering Factor\\
  & [0 - 90]&[2.5 - 10.0]&[50 - 60]&[$-2.0$ - 0.0]&[30 - 80] & [0 - 1]\\
\hline
\hline
PG2304+042& 75.4 (49.6;87.6) & $<3.2$& $>55.0$& -0.03 (-0.06; -0.01)& $<32.7$&$>0.6$\\
PKS0518-45& 51.8 (38.5; 55.8)& 5.7 (3.1; 6.9)& 58.5 (47.7; 60.0)& -0.01 (-0.06; -0.01)& 45.9 (41.2; 63.8)&0.9 (0.8-1.0)\\
PG0844+349& 44.9 (0.0; 53.3)& 3.1 (2.6; 4.2)& 55.1 (46.2; 60.0)& -0.08 (-0.17; -0.03)& $>73.9$&0.8 (0.7-0.9)\\
PG2214+139& 30.0 (25.4;33.3)& 8.9 (8.0; 9.8)& 56.7 (54.1; 59.5)& -1.1 (-1.2; -1.1)& $>78.9$&0.97 (0.95, 0.98)\\
PG0804+761&29.9 (25.2; 30.4)& 6.9 (6.4; 7.4)& $>59.7$& -0.7$^{*}$& 80.0 (79.3; 80.0)&$>1.0$\\
NGC4258 & 31.0 (30.0; 32.2)& $>9.7$& $>59.8$& -0.01$^{*}$& $>79.6$&$1.0$\\
NGC3998 & 75.1 (59.1; 90.0)& 2.9 (2.5; 3.6)& 54.7 (32.7; 60.0)& -0.3 (-0.5; -0.2)& $<43.8$&0.8 (0.6;0.9)\\
\hline
\end{tabular}}\\
{\bf Note}.- $^{*}$Parameter unrestricted. 
\end{table*}

\begin{table*}[htbp]
\caption{{\bf Derived parameters from disk+outflow models of \citet{Hoenig17}}. Column 1 lists the name of the object modeled. Columns from 2 to 9 list the disk and outflow parameters that best reproduce the AGN-dominated IRS/{\it Spitzer} spectrum:  Disk: index of the radial distribution of clouds $a$, number of clouds along the equatorial ray $N_{0}$, the scale height in the vertical Gaussian distribution of clouds $h$. Wind: index of the radial distribution of clouds in the wind $a_{w}$; the half-opening angle of the wind $\Theta_{w}$, and its angular width $\sigma_{\Theta}$. The viewing angle $i$ and the ratio between the number of clouds along the cone and $N_{0}$, $f_{wd}$. Column 7 lists the covering factor which is derived using the $i$, $N_{0}$, $\Theta_{w}$ and, $\sigma_{\Theta}$ \citep[see equation in][]{ Gonzalez-Martin19b} \label{tab:parameters_Hoenig17}}
\begin{minipage}{\textwidth}
\resizebox{18cm}{!} {
\begin{tabular}{l|cccccccc|c}
				\hline
    & \multicolumn{8}{c}{Parameters}&\\
    \hline
Name& $i$(min,max)&$N_{0}$(min,max)&$a$(min,max)&$\Theta_{w}$(min,max)&$\sigma_{\theta}$(min,max)&$a_{w}$(min,max)&$h$(min,max)&$f_{wd}$ (min,max)& Covering Factor\\
   & [0 - 90]&[5 - 10]&[$-3.0$ - $-0.5$]&[7 - 15] deg&[30 -45] deg&[$-2.5$ - $-0.5$]& [0.1 - 0.5]&[0.15 - 0.75]& [0 - 1]\\
\hline
\hline
PG2304+042&45.0 (44.1; 45.9)& $<5.3$& -2.5 (-2.7; -2.5)& 10.4 (9.2; 11.2)& $>44.6$& -0.5$^{*}$& $>0.4$& 0.30 (0.30; 0.33)&$>0.3$\\
PKS0518-45 & 29.6 (21.2; 33.9)& $<6.1$& -2.2 (-2.4; -2.0)& $>9.0$ & 38.8 (33.8; 43.0)& -0.5 (-0.9; -0.5)& 0.12 (0.10; 0.16)& $>0.41$& $>0.2$\\
PG0844+349& 18.6 (0.1; 26.9)& $>6.7$& -2.0 (-2.3; -1.6)& $>8.5$& 34.8 (33.5; 42.7)& -0.5 (-0.8; -0.5)& 0.11 (0.10; 0.19)& 0.73 (0.42; 0.75)&$>0.2$\\
PG2214+139&15.6 (0.0; 16.8)& $>6.6$& -3.0 (-3.0; -2.8)& 8.5 (7.0; 13.5)& $>35.5$& -0.5 (-0.9; -0.5)& 0.4 (0.3; 0.5)& $>0.40$&$>0.4$\\
PG0804+761 & 0.0$^{*}$& $>9.7$& -2.4 (-2.4; -1.8)& 10.3 (9.8; 10.8)& $>44.3$& -0.5$^{*}$& 0.18 (0.17; 0.20)& $>0.73$&$>0.2$\\
OQ208& 0.0$^{*}$& 10.0$^{*}$& -0.5$^{*}$& 9.9 (9.7; 10.1)& $>44.9$& -0.5 (-0.5; -0.5)& 0.5$^{*}$& 0.75 (0.74; 0.75)&$>0.4$\\
NGC4258& 30.0 (25.6; 30.4)& $>9.7$& -2.1 (-2.1; -2.0)& 7.1 (7.0; 7.5)& $>44.5$ & -0.5 (-0.5; -0.5)& 0.1$^{*}$& $>0.74$&$>0.1$\\
NGC3998 & $<26.31$& $<6.6$& -0.5 (-1.1; -0.5)& 12.7$^{*}$& 33.4 (31.0; 45.0)& -2.5 (-2.5; -1.5)& 0.1$^{*}$& $>0.27$&0.2$^{*}$\\
\hline
\end{tabular}}\\
{\bf Note}.-$^{*}$Parameter unrestricted.
\end{minipage}
\end{table*}

\begin{table*}[htbp]
\caption{{\bf Derived parameters from two-phase media models of \citet{Stalev16}}. Column 1 lists the name of the object modeled. Column 2 lists the viewing angle $i$, column 2 lists the angular width $\sigma$ of the torus. Columns 3 and 4 list the indices of the radial ($p$) and angular distribution ($q$) of the clouds. Column 5 gives the ratio between the outer and inner radius. Column 6 gives the optical depth $\tau_{9.7}$. Columns 7 list the covering factor. \label{tab:parameters_Stalev16}}
\begin{minipage}{\textwidth}
\resizebox{18cm}{!} {
\begin{tabular}{l|cccccc|c}
				\hline
    & \multicolumn{6}{c}{Parameters}&\\
    \hline
Name& $i$(min,max)&$\sigma$(min,max)&$p$(min,max)&$q$(min,max)&$Y$(min,max)&$\tau_{9.7}$(min,max)& Covering Factor\\
    &[0-90] & [10-80] & [0-1.5] & [0-1.5] & [10-30] & [3-11]&  [0-1]\\   
\hline
\hline 
PG2304+042 & $>83.5$& 80.0 (60.0;79.1)& 0.01 (0.0 ;0.02)& 1.0 (1.0; 1.1)& 10.0$^{*}$& 3.0 (0.1; 3.1)&$>0.7$\\
PKS0518-45& $>66.5$& 80.0 (60.0;79.0)& 0.01 (0.0; 0.01)& 0.7 (0.6; 0.7)& 10.0$^{*}$& 3.0 (0.1; 3.1)&$>0.8$\\
PG0844+349&$>80.5$& 80.0 (60.0;76.2)& 0.01 (0.0; 0.01)& 1.5$^{*}$& 12.3 (11.9; 12.8)& 3.1 (0.1; 3.3)&$>0.6$\\
PG2214+139& 63.0 (58.3; 64.9)& 70.0 (67.8; 74.4)& $>1.5$& 1.5$^{*}$& $<10.1$& 3.8 (3.6; 4.2)&0.7$^{*}$\\
PG0804+761 & 10.0 (0.0; 14.1)& 42.3 (41.6; 44.1)& $>1.5$& 1.5$^{*}$& $<10.2$& 10.8$^{*}$&1.0$^{*}$\\
\hline
\end{tabular}}\\
{\bf Note}.-$^{*}$Parameter unrestricted.
\end{minipage}
\end{table*}

\begin{figure*}[htbp]
\begin{center}
\includegraphics[scale=0.5]{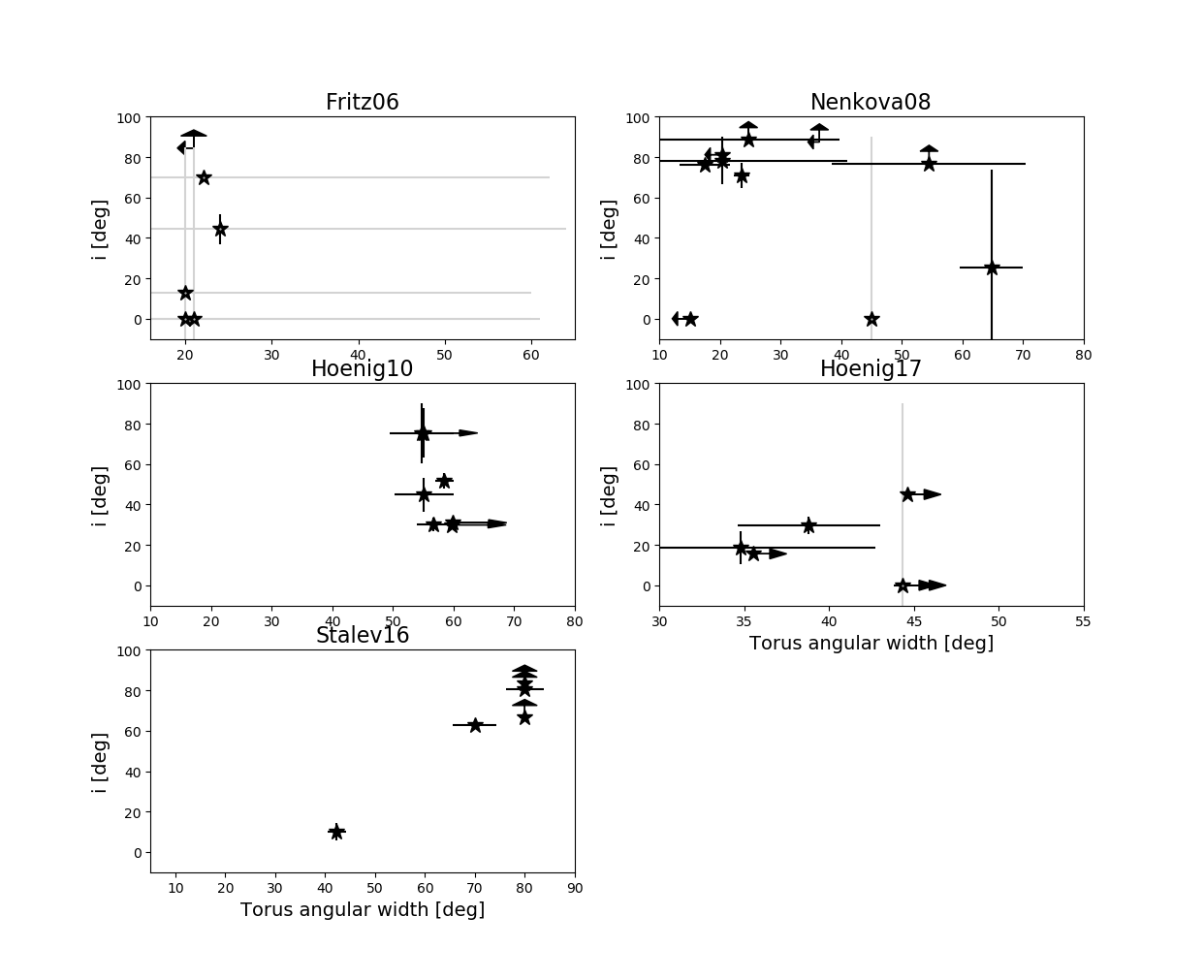}
\caption{{\bf The viewing angle Vs the angular width derived from the Si-s sample for the smooth dusty torus of Fritz06, the clumpy torus models of Nenkova08 and Hoenig10, the disk+outflow torus model of Hoenig10, and the two-phase media models of stalev16.}\label{fig:parameters}}
\end{center}
\end{figure*}

\begin{figure*}[htbp]
\begin{center}
\begin{tabular}{c}
\includegraphics[scale=0.7]{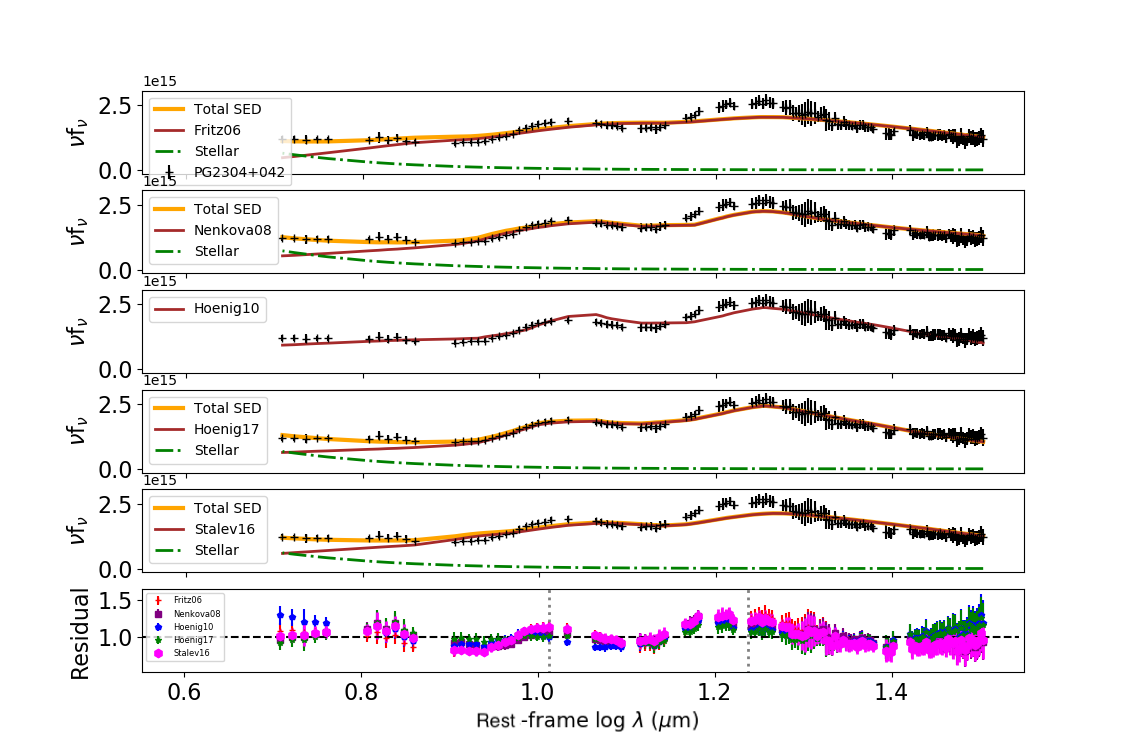} \\ 
\end{tabular}
\caption{{\bf Modeling and residuals of the IRS/{\it Spitzer} spectrum of PG~2304+042}. From top to bottom is the spectrum fitted assuming Fritz06 ($\chi_{red}^{2}\sim1.20$), Nekova08 ($\chi_{red}^{2}\sim0.32$), Hoenig10 ($\chi_{red}^{2}\sim0.20$), Hoenig17 ($\chi_{red}^{2}\sim0.13$), and Stalevski16 ($\chi_{red}^{2}\sim0.81$) model. The last panel show the residuals defined as the ratio between the data and model. In all panels the black points are the IRS/{\it Spitzer} spectrum and its error in (erg s$^{-1}$cm$^{-2}$), and the red solid line is the fitted torus model. The orange line is the total SED that results when more than one component, the stellar (green dot-dashed line) and/or HII (blue dotted line), is added to model the spectrum.}
  \label{fig:modeling1}
\end{center}
\end{figure*}

\begin{figure*}[htbp]
\begin{center}
\begin{tabular}{c}
\includegraphics[scale=0.6]{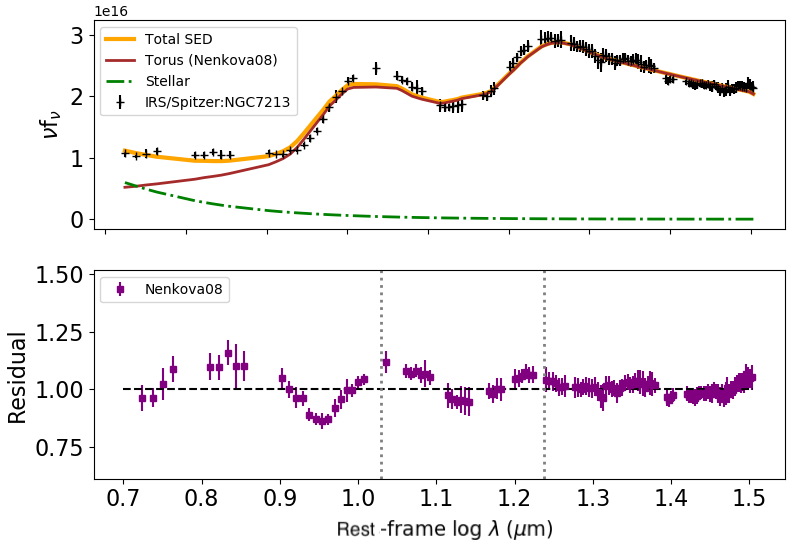} 
\end{tabular}
\caption{{\bf Modeling and residuals of the IRS/{\it Spitzer} spectrum of NGC~7213}. From top to bottom is the spectrum fitted assuming Nekova08 ($\chi_{red}^{2}\sim1.11$). The last panel show the residuals defined as the ratio between the data and model. In all panels the black points are the IRS/{\it Spitzer} spectrum and its error in erg s$^{-1}$cm$^{-2}$, and the red solid line is the fitted torus model. The orange line is the total SED that results when more than one component, the stellar (green dot-dashed line) and/or HII (blue dotted line), is added to model the spectrum.}
\label{fig:modeling2}
\end{center}
\end{figure*}

\begin{figure*}[htbp]
\begin{center}
\begin{tabular}{c}
\includegraphics[scale=0.5]{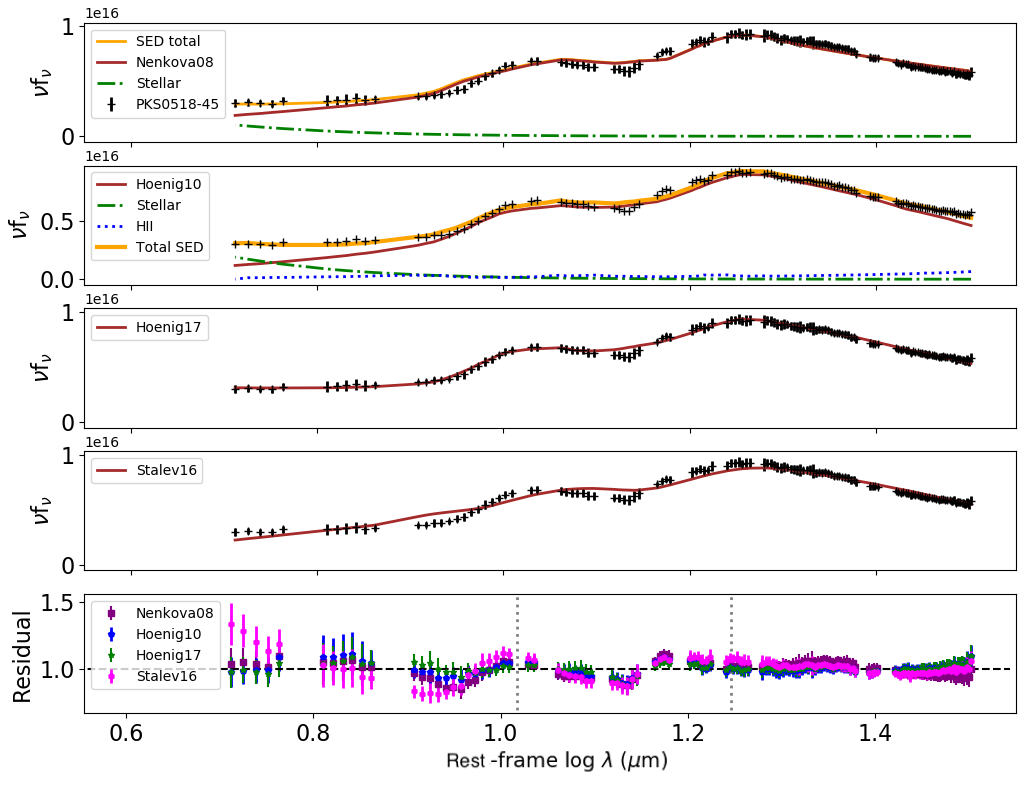}\\
\end{tabular}
\caption{{\bf Modeling and residuals of the IRS/{\it Spitzer} spectrum of PKS~0518-45}. From top to bottom is the spectrum fitted assuming Nekova08 ($\chi_{red}^{2}\sim1.29$), Hoenig10 ($\chi_{red}^{2}\sim0.46$), Hoenig17 ($\chi_{red}^{2}\sim0.31$), and Stalevski16 ($\chi_{red}^{2}\sim1.65$) model. The last panel show the residuals defined as the ratio between the data and model. In all panels the black points are the IRS/{\it Spitzer} spectrum and its error in (erg s$^{-1}$cm$^{-2}$), and the red solid line is the fitted torus model. The orange line is the total SED that results when more than one component, the stellar (green dot-dashed line) and/or HII (blue dotted line), is added to model the spectrum.}
\label{fig:modeling3}
\end{center}
\end{figure*}

\begin{figure*}[htbp]
\begin{center}
\includegraphics[scale=0.6]{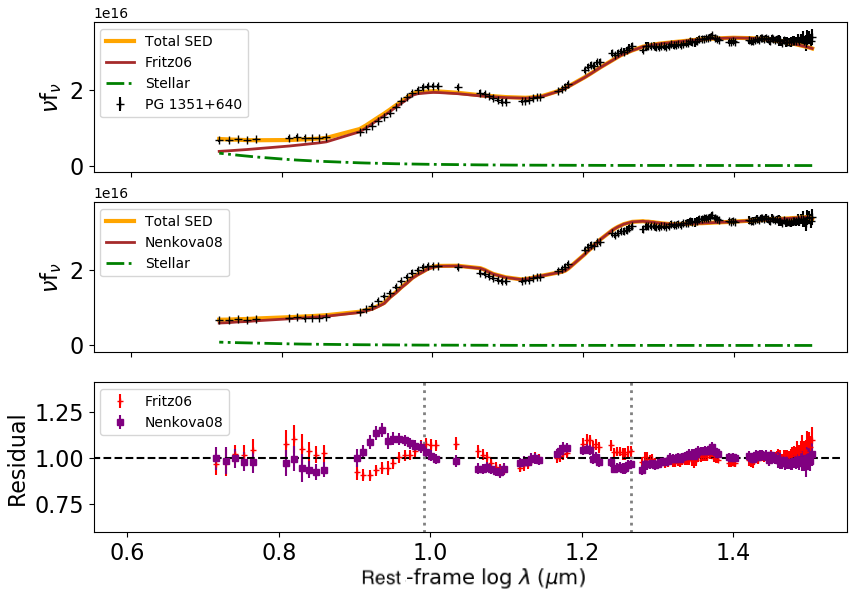}  
\caption{{\bf Modeling and residuals of the IRS/{\it Spitzer} spectrum of PG~1351+640}. From top to bottom is the spectrum fitted assuming Fritz06 ($\chi_{red}^{2}\sim1.52$), Nekova08 ($\chi_{red}^{2}\sim1.92$) model. The last panel show the residuals defined as the ratio between the data and model. In all panels the black points are the IRS/{\it Spitzer} spectrum and its error in (erg s$^{-1}$cm$^{-2}$), and the red solid line is the fitted torus model. The orange line is the total SED that results when more than one component, the stellar (green dot-dashed line) and/or HII (blue dotted line), is added to model the spectrum.}
\label{fig:modeling4}
\end{center}
\end{figure*}

\begin{figure*}[htbp]
\begin{center}
\includegraphics[scale=0.4]{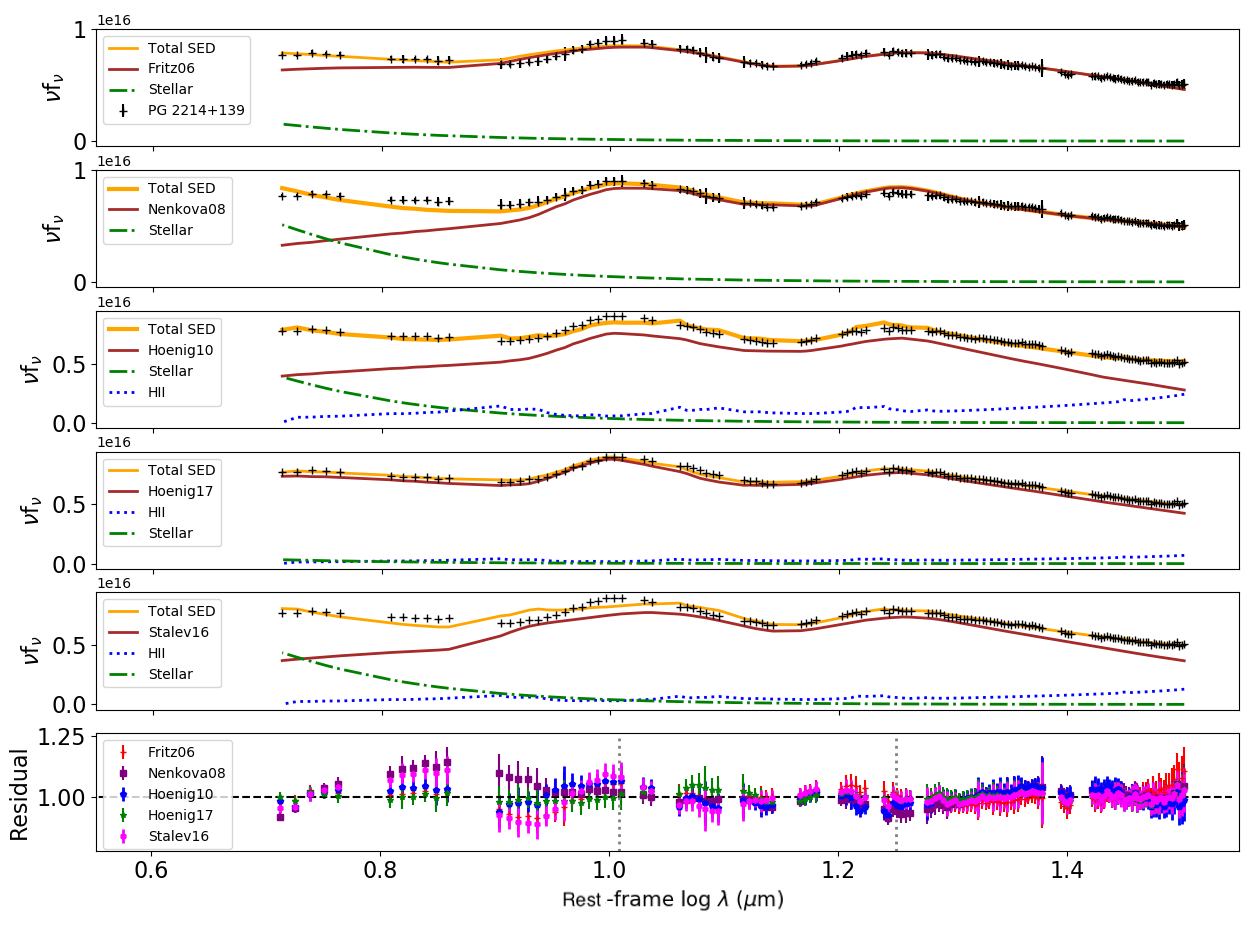}
\caption{{\bf Modeling and residuals of the IRS/{\it Spitzer} spectrum of PG~2214+139}. From top to bottom is the spectrum fitted assuming Fritz06 ($\chi_{red}^{2}\sim1.52$), Nekova08 ($\chi_{red}^{2}\sim1.92$), Hoenig10 ($\chi_{red}^{2}\sim0.71$), Hoenig17 ($\chi_{red}^{2}\sim0.20$), and Stalevski16 ($\chi_{red}^{2}\sim0.91$) model. The last panel show the residuals defined as the ratio between the data and model. In all panels the black points are the IRS/{\it Spitzer} spectrum and its error in (erg s$^{-1}$cm$^{-2}$), and the red solid line is the fitted torus model. The orange line is the total SED that results when more than one component, the stellar (green dot-dashed line) and/or HII (blue dotted line), is added to model the spectrum.}
\label{fig:modeling5}
\end{center}
\end{figure*}

\begin{figure*}[htbp]
\begin{center}
\includegraphics[scale=0.4]{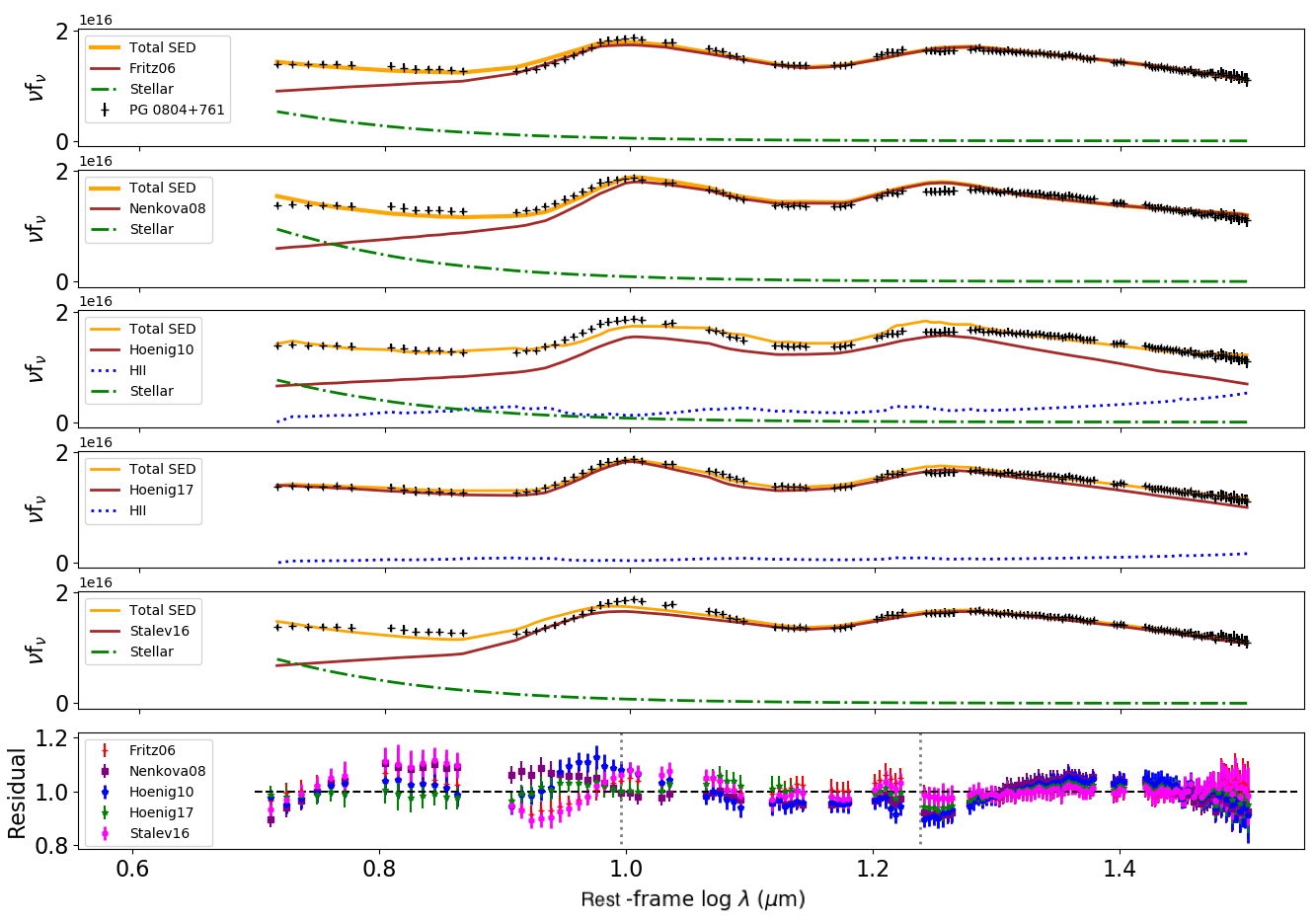}
\caption{{\bf Modeling and residuals of the IRS/{\it Spitzer} spectrum of PG~0804+761}. From top to bottom is the spectrum fitted assuming Fritz06 ($\chi_{red}^{2}\sim0.63$), Nekova08 ($\chi_{red}^{2}\sim1.60$), Hoenig10 ($\chi_{red}^{2}\sim1.79$), Hoenig17 ($\chi_{red}^{2}\sim0.49$), and Stalevski16 ($\chi_{red}^{2}\sim1.11$) model. The last panel show the residuals defined as the ratio between the data and model. In all panels the black points are the IRS/{\it Spitzer} spectrum and its error in (erg s$^{-1}$cm$^{-2}$), and the red solid line is the fitted torus model. The orange line is the total SED that results when more than one component, the stellar (green dot-dashed line) and/or HII (blue dotted line), is added to model the spectrum.}
\label{fig:modeling6}
\end{center}
\end{figure*}

\begin{figure*}[htbp]
\begin{center}
\includegraphics[scale=0.45]{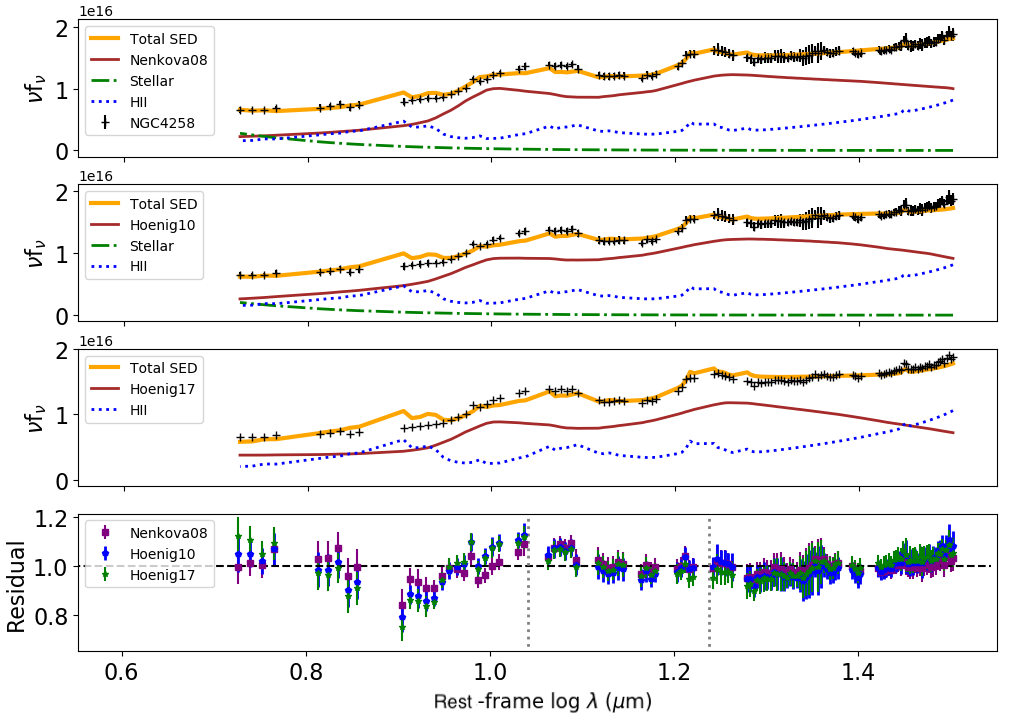}
\caption{{\bf Modeling and residuals of the IRS/{\it Spitzer} spectrum of NGC4258}. From top to bottom is the spectrum fitted assuming Nekova08 ($\chi_{red}^{2}\sim0.78$), Hoenig10 ($\chi_{red}^{2}\sim1.64$), and Hoenig17 ($\chi_{red}^{2}\sim1.88$) model. The last panel show the residuals defined as the ratio between the data and model. In all panels the black points are the IRS/{\it Spitzer} spectrum and its error in (erg s$^{-1}$cm$^{-2}$), and the red solid line is the fitted torus model. The orange line is the total SED that results when more than one component, the stellar (green dot-dashed line) and/or HII (blue dotted line), is added to model the spectrum.}
\label{fig:modeling7}
\end{center}
\end{figure*}

\begin{figure*}[htbp]
\begin{center}
\includegraphics[scale=0.5]{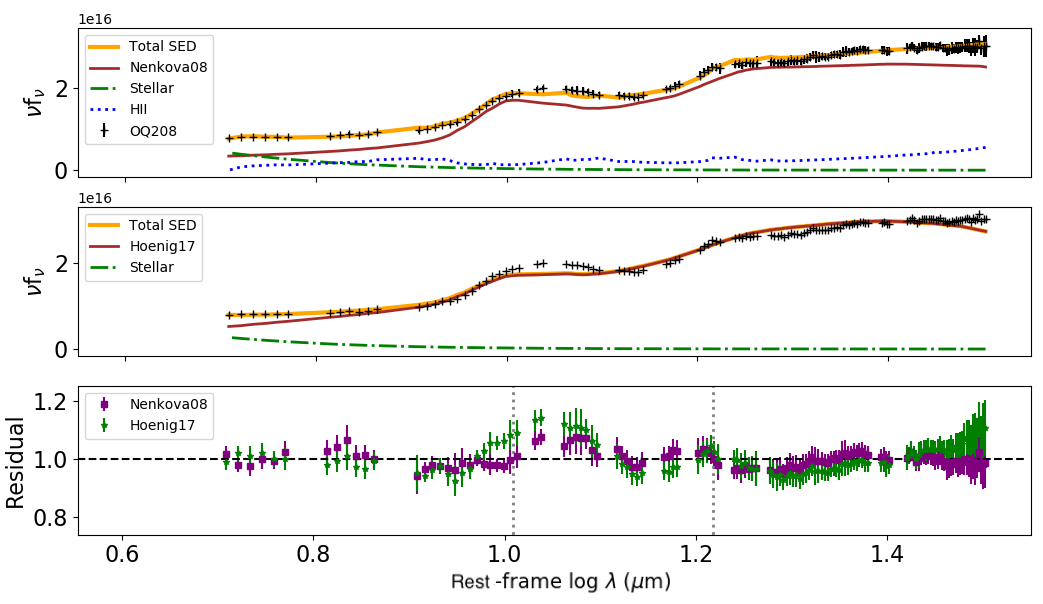}
\caption{{\bf Modeling and residuals of the IRS/{\it Spitzer} spectrum of OQ208}. From top to bottom is the spectrum fitted assuming Nekova08 ($\chi_{red}^{2}\sim0.32$), and Hoenig17 ($\chi_{red}^{2}\sim1.78$) model. The last panel show the residuals defined as the ratio between the data and model. In all panels the black points are the IRS/{\it Spitzer} spectrum and its error in (erg s$^{-1}$cm$^{-2}$), and the red solid line is the fitted torus model. The orange line is the total SED that results when more than one component, the stellar (green dot-dashed line) and/or HII (blue dotted line), is added to model the spectrum.}
\label{fig:modeling8}
\end{center}
\end{figure*}

\begin{figure*}[htbp]
\begin{center}
\includegraphics[scale=0.5]{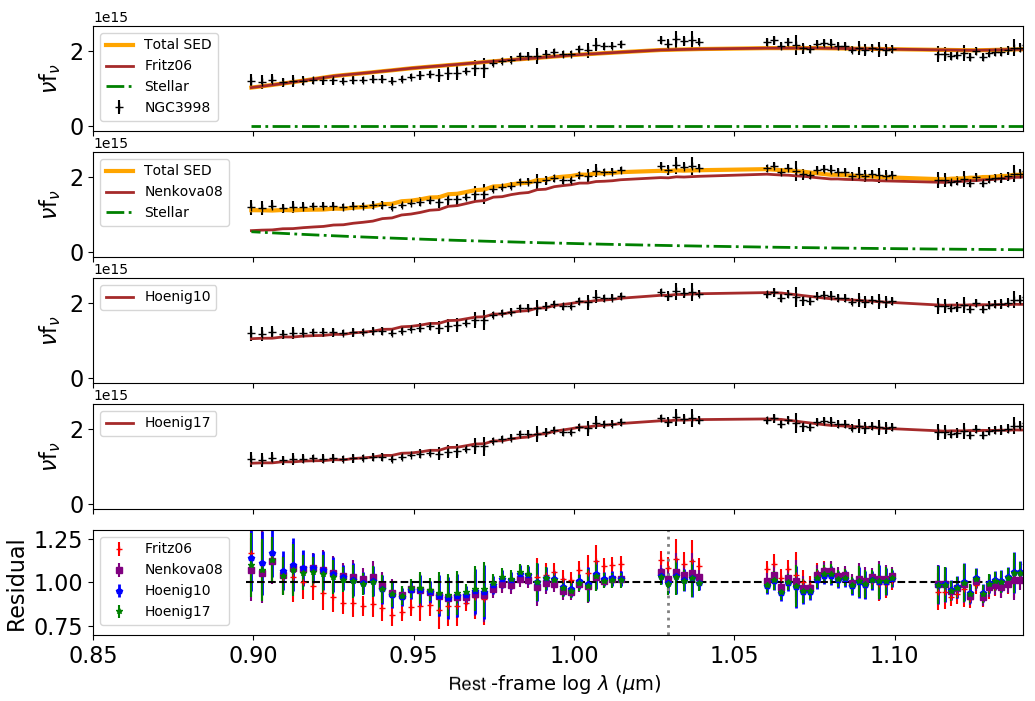}
\caption{{\bf Modeling and residuals of the IRS/{\it Spitzer} spectrum of NGC3998}. From top to bottom is the spectrum fitted assuming Fritz06 ($\chi_{red}^{2}\sim1.91$), Nekova08 ($\chi_{red}^{2}\sim0.32$), Hoenig10 ($\chi_{red}^{2}\sim0.40$), and Hoenig17 ($\chi_{red}^{2}\sim0.39$) model. The last panel show the residuals defined as the ratio between the data and model. In all panels the black points are the IRS/{\it Spitzer} spectrum and its error in (erg s$^{-1}$cm$^{-2}$), and the red solid line is the fitted torus model. The orange line is the total SED that results when more than one component, the stellar (green dot-dashed line) and/or HII (blue dotted line), is added to model the spectrum.}
  \label{fig:modeling9}
  \end{center}
\end{figure*}

\end{document}